\documentclass[a4paper,fleqn,usenatbib]{mnras}

\usepackage[T1]{fontenc}
\usepackage{ae,aecompl}

\usepackage[english]{babel}
\usepackage{graphicx}	% Including figure files
\usepackage{amssymb}	% Extra maths symbols
\usepackage{enumitem}
\usepackage{natbib}
\bibpunct{(}{)}{;}{a}{}{,}

\usepackage{amsmath}	% Advanced maths commands

\def\msun{{\rm M_{\odot}}}

\title[Probing AGN triggering trough clustering analysis]{Constraining AGN triggering mechanisms through the clustering analysis of active black holes}

\author[M. Gatti et al.]{
M. Gatti,$^{1,2}$ 
F. Shankar,$^{3}$ 
V. Bouillot,$^{4}$ 
N. Menci,$^{1}$ 
A. Lamastra,$^{1}$ 
M. Hirschmann,$^{5}$\newauthor
F. Fiore$^{1}$ \\
$^{1}$INAF - Osservatorio Astronomico di Roma, via di Frascati 33, 00040 Monte Porzio Catone, Italy\\
$^{2}$IFAE - Institut de Fisica d'Altes Energies, Universitat Autonoma de
Barcelona, E-08193 Bellaterra (Barcelona), Spain\\
$^{3}$Department of Physics and Astronomy, University of Southampton, Southampton SO17 1BJ, UK\\
$^{4}$Centre for Astrophysics, Cosmology $\&$ Gravitation, Department of Mathematics $\&$ Applied Mathematics, University of Cape Town,\\ Cape Town 7701, South Africa\\
$^{5}$UPMC-CNRS, UMR7095, Institut d' Astrophysique de Paris, F-75014 Paris, France\\}

\date{Accepted 2015 November 19. Received 2015 October 26; in original form 2015 July 25}

\pubyear{2015}

\begin{document}
\label{firstpage}
\pagerange{\pageref{firstpage}--\pageref{lastpage}}
\maketitle

\begin{abstract}
The triggering mechanisms for Active Galactic Nuclei (AGN) are still debated. Some of the most popular ones include galaxy interactions (IT) and disk instabilities (DI). Using an advanced semi analytic model (SAM) of galaxy formation, coupled to accurate halo occupation distribution modeling, we investigate the imprint left by each separate triggering process on the clustering strength of AGN at small and large scales. Our main results are as follows: i) DIs, irrespective of their exact implementation in the SAM, tend to fall short in triggering AGN activity in galaxies at the center of halos with $M_h>10^{13.5} h^{-1}\msun$. On the contrary, the IT scenario predicts abundance of active, central galaxies that generally agrees well with observations at every halo mass.  ii) The relative number of satellite AGN in DIs at intermediate-to-low luminosities is always significantly higher than in IT models, especially in groups and clusters. The low AGN satellite fraction predicted for the IT scenario might suggest that different feeding modes could simultaneously contribute to the triggering of satellite AGN. iii) Both scenarios are quite degenerate in matching large-scale clustering measurements, suggesting that the sole average bias might not be an effective observational constraint. iv) Our analysis suggests the presence of both a mild luminosity and a more consistent redshift dependence in the AGN clustering, with AGN inhabiting progressively less massive dark matter halos as the redshift increases. We also discuss the impact of different observational selection cuts in measuring AGN clustering, including possible discrepancies between optical and X-ray surveys.
\end{abstract}


%%%%%%%%%%%%%%%%%%%%%%%%%%%%%%%%%%%%%%%%%%%%%%%%%%

%%%%%%%%%%%%%%%%% BODY OF PAPER %%%%%%%%%%%%%%%%%%

\section{Introduction}

%Clustering analysis has proven to be a powerful tool to investigate the spatial distribution of billions of objects in the Universe. In the past decades, it has been successfully applied to the study of the galaxy population, greatly favored by large area surveys - e.g. Two Degree Field Galaxy Redshift Survey \citep{Colless2001}; Deep Extragalactic Evolutionary Probe \citep{Davis2003}; zCOSMOS \citep{Lilly2007}; Sloan Digital Sky Survey \citep{Abazajian2009}. Recently, clustering analysis has also been applied to the study of the spatial distribution of Active Galactic Nuclei (AGN). 

AGN are believed to be powered by the accretion of matter onto supermassive black holes (SMBHs)(\citealp{Soltan1982,Richstone1998}). We also know that the masses of SMBHs in the local Universe relate to several properties of their host galaxies \citep{Magorrian1998,Marconi2003,McConnell2013,Kormendy2013,Laesker2014}.  
 However, the physical mechanisms responsible for such intense accretion episodes in the galactic nuclei and their relation with the cosmological evolution of galaxies still remain unclear.

Various pictures have been proposed. One possible scenario envisages galaxy interactions as the main fueling mechanism. Gravitational torques induced by interacting galaxies would be effective in causing large gas inflows in the central region of the galaxy, eventually feeding the central SMBH. Luminous quasars are indeed often found to be associated with systems undergoing a major merger or showing clear signs of morphological distortion \citep{DiMatteo2005,Cox2008,Bessiere2012,Urrutia2012,Treister2012}. Less violent minor mergers or fly-by events have been invoked to account for moderately luminous AGN \citep{Combes2009,Koss2010,Satyapal2014}.

On the other hand, there is mounting observational evidence based on the star-forming properties and on the morphology of AGN hosts, suggesting that moderate levels of AGN activity might not be casually connected to galaxy interactions \citep{Lutz2010,Rosario2012,Mullaney2012a,Santini2012,Rosario2013,Villforth2014}. Theoretically,  \textit{in-situ} processes, such as disk instabilities or stochastic accretion of gas clouds, have also been invoked as effective triggers of AGN activity \citep{Dekel2009,Bournaud2011}.

AGN clustering analysis could help to unravel the knots of this complex situation. Indeed, clustering measurements can provide vital information on the physics of AGN. Through the use of the 2-point correlation function (2PCF, \citealt{Arp1970}) the spatial distribution of AGN can be effectively used to investigate the relation between AGN and their host halos, enabling to pin down the typical environment where AGN live. This in turn could potentially provide new insights into the physical mechanisms responsible for triggering and powering their emission.

 %By constraining the typical environments where a certain class of AGN live, clustering could shed light on the typical conditions that are more likely to trigger central SMBHs. 

For instance, clustering analysis of optically-selected bright quasars carried out at different redshifts (\citealt{Porciani2004,Myers2006,Coil2007,Shen2007,Padmanabhan2009,White2012}) have pointed out a typical host halo mass of the order of $log M_h [h^{-1} \msun]  \sim 12.0-12.5$, corresponding to the typical mass scale of galaxy groups. These observational evidences would add support to the interactions scenario, since in these environments the probability of a close encounter is higher \citep{Hopkins2008,McIntosh2009}.

Otherwise, a number of clustering studies on moderately luminous X-ray selected AGN \citep{Coil2009,Hickox2009,Cappelluti2010,Allevato2011,Krumpe2012,Koutoulidis2013} have obtained higher typical halo masses, in the range $log M_h [h^{-1} \msun]  \sim 12.5-13.5$. These results have usually been interpreted as a possible sign of alternative triggering mechanisms at play \citep{Fanidakis2011}.

However, great attention must be paid in interpreting these observational results. For example, the relatively small number of AGN, typically around a few percent of the whole galaxy population at z<1, requires especially in deep surveys the use of large AGN samples covering wide ranges of redshift, luminosity and host galaxy properties to gather statistically significant samples. This in turns renders the comparison among different data-sets non-trivial. %Secondly, the analysis of the 2PCF provides only an average halo mass, without providing any additional insight about how the AGN population is distributed in DM halos with different masses. 

The halo occupation distribution (HOD) formalism \citep{Cooray2002,Berlind2003,Cappelluti2012} allows to extract from the 2PCF the full distribution of host halo masses for a given sample of AGN. The halo model in fact constraints the AGN HOD function P(N|$M_h$), which provides the probability distribution for an halo of mass $M_h$ to host a number N of AGN above a given luminosity. Only recently a number of observational studies have begun to focus on the AGN HOD (e.g, \citealt{Miyaji2011,Starikova2011,Allevato2012,Krumpe2012,Richardson2012,Richardson2013,Shen2013}), although still facing uncertainties due the degeneracy in the shape and normalization of the HOD (e.g, \citealt{Shen2013}).

Given the uncertain observational situation, a valuable complimentary way to effectively probe the different interpretations concerning clustering analysis, is to rely on a comprehensive cosmological model for galaxy formation. For instance, implementing various physical mechanisms for triggering AGN activity in a semi analytic model (SAM) for galaxy formation (see \citealt{Baugh2006} for a review), it is possible to compare the predicted $P(N|M_h)$ of each different model with a wide range of different AGN 2PCF and HOD measurements, in order to narrow down the efficiency of each separate AGN triggering mechanism included in the SAM. %Particularly, SAMs, by accounting for the evolution of the AGN population along with the evolution of the DM density field, directly provides the AGN HOD for each timestep of the simulation; then, by applying the HOD formalism, it is possible to obtain from the AGN HOD the AGN 2PCF and compare it with present observational constraints.

In \citet{Menci2014} and \citet{Gatti2015} we included in an advanced SAM for galaxy formation two different analytic prescriptions for triggering AGN activity in galaxies. We first considered AGN activity triggered by disk instabilities (DI scenario) in isolated galaxies, and, separately, the triggering induced by galaxy interactions (major mergers, minor mergers and fly-by events, IT scenario).
The analytic prescriptions included in the SAM to describe each physical process are based on hydrodynamical simulations, thus offering a solid background for describing accretion onto the central SMBH. 

%The physical descriptions of the two feeding modes included in our SAM have been worked out and tested by their authors against aimed hydrodynamical N-body simulations, so that they constitute a solid background for describing the accretion of SMBH. Moreover, since they mainly depends on a few general properties of the galaxy population (e.g, galaxy encounter timescale, galaxy disk mass, etc.), our main results little depend on the exact parameterization/modelling for evolving the galaxies within their host dark matter halos adopted in our SAM. 

Relying on this framework, the aim of this paper is to investigate the imprint left by each separate triggering process (IT and DI modes) on the clustering strength of the AGN population, both on small and large scales and at different redshift and luminosity. The final goal is to highlight key features in the clustering properties of the two modes that might constitute robust probes to pin down the dominant SMBH fueling mechanism.

The paper is organized as follows. In Sect. 2, we briefly describe our SAM and the two AGN triggering mechanisms considered. In Sect. 3, we review the HOD formalism and the theoretical model used to obtain the 2PCF. In Sect. 4 and 5, we present our main results concerning the clustering properties of the AGN population as predicted by our two modes and we compare with a wide range of observational constraints (both AGN HOD and 2PCF measurements). We discuss our results in Sect. 6 and summarize in Sect. 7. Throughout the paper we assume standard $\Lambda$CDM cosmology, with $\Omega_M=0.3$,$\Omega_{\Lambda}=0.7$, $\Omega_b=0.045$, $\sigma_8=0.8$, to make contact with the one adopted in most of the observations we will be comparing our models to.

\section{Semi-analytic model}
\subsection{Evolving dark matter halos and galaxies in the model.}
Our analysis relies on the Rome\footnote{Details on the code can be found at http://www.oa-roma.inaf.it/menci/index.html} semi analytic model (SAM) described in \citet[see \citealt{Menci2014} for the latest update]{Menci2006,Menci2008}, which connects the cosmological evolution of the underlying dark matter halos with the processes involving their baryonic content such as gas cooling, star formation, supernova feedback, and chemical enrichment.  

An accurate Monte Carlo method tested against N-body simulations is used to generate the merging trees of dark matter halos up to $z \sim 10$ following the extended Press \& Schechter formalism \citep{Bond1991,Lacey1993}, ultimately providing the merging rates of dark matter halos. 

The SAM takes into account both dark matter halos containing galaxies, i.e. groups and clusters (labeled by their mass $M_h$, virial radius $R_{vir}$ and circular velocity V, defined as $V= \sqrt{M_hG/R_{200}}$), and the dark matter clumps associated to each member galaxy (characterized by mass $m_h$,
virial radius r and circular velocity $v_c$). As cosmic time proceeds, smaller halos are included in larger ones as satellites. A satellite might either merge with another satellite during a binary aggregation or fall into the centre as a result of dynamical friction, contributing to increasing the mass content of the central dominant galaxy. The typical central-satellite merging timescales increase over cosmic time, thus inevitably increasing the number of satellites as the host halos scale up from groups to clusters \citep{Menci2006}. 

We assume initially at $z \sim 10$ one galaxy in each dark matter host halo, with the latter following the Press $\&$ Schechter mass distribution. The baryonic processes taking place in each dark matter halo are computed according to the common procedures also adopted in other SAMs (e.g. \citealt{Cole2000}): for a given galactic halo of mass $M_h$, at the moment of its formation we assign to the hot phase an initial amount $M_h \Omega_b / \Omega_M$  of gas at the virial temperature. As the cosmic time proceeds, a fraction of hot gas $\Delta m_c$ cools due to atomic processes and settles into a rotationally supported disk with exponential profile with mass $M_c$, disk radius $R_d$ and disk circular velocity $V_d$ computed as in \citet{Mo1998}.

The cooled gas $M_c$ is gradually converted into stars at a rate $\dot{M_*} = M_c/\tau_*$ according to the Schmidt-Kennicut law with $\tau_*$ = 1 Gyr.  An additional channel of star formation implemented in the model is provided by the starburst following the triggering of AGN activity (see below). At each timestep, a fraction of the cooled gas is returned to the hot phase at the virial temperature of the halo due to the energy released by the Supernovae following star formation. Our SAM also accounts for the AGN feedback, modeled according to the blast wave model \citep{Cavaliere2002,Lapi2005}. The supersonic outflows generated by AGN during their luminous phase compress the gas into a blast wave that moves outwards, eventually expelling a certain amount of gas from the galaxy and returning it to the hot phase.

The integrated stellar emission produced by the stellar populations of the galaxies in different bands are computed by convolving the star formation histories of the galaxy  progenitors with a synthetic spectral energy distribution from population synthesis models (\citealt*{Bruzual2003}; Salpeter initial mass function is assumed). Ultimately, we have recently included in our SAM  a description of the tidal stripping of stellar material from satellite galaxies,  computed following the procedure introduced by \citet*{Henriques2010} in the Munich SAM.

\subsection{AGN triggering}
Two different AGN feeding modes are implemented in our SAM: 

i) IT mode. The triggering of the AGN activity is provided by galaxy interactions (major mergers, minor mergers and fly-by events). 

ii) DI mode. The accretion onto the central SMBH occurs due to disk instabilities, where the trigger is provided by the break of the axial symmetry in the distribution of the galactic cold gas. 

DI and IT modes have been included and operate in the SAM separately. The full properties of the AGN population are always determined only \textit{by one} of the two feeding modes. In the following, we give a basic descriptions for both IT and DI models.

\subsubsection{IT mode}
Any galactic halo with given halo circular velocity $v_c$ included in a host halo with halo circular velocity V, interacts with other galactic halos at a rate
\begin{equation}\label{int}
\tau_r^{-1}=n_T (V)\,\Sigma (r_t,v_c,V_{rel})\,V_{rel}(V)
,\end{equation}
where $n_T=3 N_T/4\pi R_{vir}^3$ is the number density of galaxies in the host halo, which contains $N_T$  galaxies. $V_{rel} (V)$ represents the relative velocity between galaxies in the host halo, and $\Sigma$ the cross section for such encounters, which is provided by \citet{Saslaw1985} in terms of the galaxy tidal radius $r_t$ \citep{Menci2004}. Any kind of interaction destabilizes a fraction $f$ of cold gas in the galactic disk; the fraction can be expressed in terms of the variation $\Delta j$ of the specific angular momentum $j\approx GM/V_d$ of the gas as \citep{Menci2004}
\begin{equation}\label{fdest}
f\approx \frac{1}{2}\,
\Big|{\Delta j\over j}\Big|=
\frac{1}{2}\Big\langle {M'\over M}\,{R_d\over b}\,{V_d\over V_{rel}}\Big\rangle\, 
,\end{equation}
where $b$ is the impact parameter, evaluated as the greater of the radius $R_d$ and the average distance of the galaxies in the halo, $V_d$ the galaxy disk circular velocity, $M'$ is the mass of the  partner galaxy in the interaction, and the average runs over the probability of finding such a galaxy
in the same halo where the galaxy with mass $M$ is located. %The pre-factor accounts for the probability that half of the inflow rather than of the outflow is related to the sign of $\Delta j$.
We assume that in each interactions $1/4$ of the destabilized fraction $f$ feeds the central BH, while the remaining feeds the circumnuclear starbursts  \citep{Sanders1996}. Hence, the BH accretion rate is equal to %$\dot{M}_{BH} =1/4 \,f\,M_c/\tau_b$, with $\tau_b=r_d/v_d$ the timescale for the AGN to shine.

\begin{equation}
%\label{macc_ID}
 {dM_{BH}\over dt}={1\over 4}{f\,M_c\over \tau_b}  
,\end{equation}
with $\tau_b=R_d/V_d$ the timescale for the AGN to shine.

%where the time scale $\tau_{b}=R_d/V_d$  is assumed to be the crossing time of the destabilized galactic disk. %The duration of an accretion episode, that is, the time scale for the QSO or AGN to shine, is equal to the crossing time of the destabilized galactic disk ($\tau_b$).

\subsubsection{DI mode}
In the DI scenario, disk instability arises in galaxies whose disk mass exceeds a given critical value
\begin{equation}
\label{efstathi}
M_{crit} =  {v_{max}^2 R_{d}\over G \epsilon}
,\end{equation} 
where $v_{max}$ is the maximum disk circular velocity, $R_d$ the scale length of the disk, and $\epsilon$ the stability parameter (we set its value to be $\epsilon=0.75$, in agreement with the value assumed by \citealt{Hirschmann2012} in their SAM).  The above critical mass is provided by \citet{Efstathiou1982} on the basis of N-body simulations. %The value of the stability parameter is slightly uncertain, since different values might (or might not) give satisfactory results once compared to observations and/or N-body simulation (see, e.g., \citealt{Athanassoula2008}). 
We note that slightly changing the value of epsilon has a minor impact on the model predictions, little affecting the number density of AGN (see also \citealt{Hirschmann2012}).

At each time step of the simulation, we compute the critical mass competing to each galaxy following Eq. \ref{efstathi}. % if the criterion is satisfied, the perturbation settles down and triggers a mass inflow from the outskirts of the galaxy onto the central SMBH.
If the criterion is satisfied, then we assume the disk becomes unstable driving a mass inflow onto the central SMBH and a circumnuclear starburst. The mass inflow is computed according to the model proposed by \citet*{Hopkins2011} and is equal to:
\begin{multline}
\label{hopkins}
\dot{M}_{BH}  \approx 
%{\alpha \over (1 + f_0/f_{gas})}
%{f_d^{4/3}\over (1+2.5\,f_d^{-4/3}) }
{\alpha \,
f_d^{4/3}\over 1+2.5\,f_d^{-4/3}(1 + f_0/f_{gas}) } \times \\
\left( \frac{M_{BH}}{10^8 M_{\bigodot}}\right)^{1/6}
\left( \frac{M_d(R_0)}{10^9 M_{\bigodot}}\right)
\left( \frac{R_0}{100 pc }\right) ^{-3/2}
M_{\bigodot} yr^{-1}
,\end{multline}
where 
\begin{equation}
f_0 \approx 0.2 f_d^2 \left[ \frac{M_d(R_0)}{10^9 M_{\bigodot}}\right]^{-1/3}
{\hspace{2cm}}  f_{gas} \equiv {M_{gas} (R_0)\over M_d(R_0)}
.\end{equation} 
%Here $M_{BH}$ is the central black hole mass,  $f_d$ is total disk mass fraction, $M_d$ and $M_{gas}$ the disk and the gas mass calculated in $R_0$ (we take $R_0 = 100$ pc) and $\alpha$ a parameter taken to be $\alpha = 10$ (the constant $\alpha$ parametrizes several uncertainties of the model, see MN14 for more details).
Here $M_{BH}$ is the central black hole mass,  $f_d$ is the total disk mass fraction, $M_d$ and $M_{gas}$ the disk and the gas mass calculated in $R_0$ (we take $R_0 = 100$ pc). The constant $\alpha$ parametrizes several uncertainties related to some of the basic assumptions of the mass inflow model; its value is not completely freely tunable, but physically admissible values are in the range $\alpha = 2-10$ (see \citealt{Menci2014} and \citealt{Gatti2015} for further details). A higher normalization ($\alpha=10$) roughly corresponds to slightly higher AGN luminosity and shorter duty cycle, since there is a faster gas consumption and a faster stabilization of the disk. The opposite is true for lower normalizations.

In what follows, we will show for the DI scenario three different predictions, corresponding to the cases $\alpha =$ 2, 5 and 10, so as to span all the reasonable values of the normalization of the mass inflow predicted by the model.

For both scenarios, we converted the BH mass inflows into AGN bolometric luminosity using the following equation:
\begin{equation}\label{Lagn}
L_{AGN}=\eta\,c^2\,\dot{M}_{BH}
\end{equation}
%where $dM_{BH}/dt$ is taken from Eq. \ref{macc_ID} for the IT mode and from Eq. \ref{hopkins} for the DI scenario. 
We adopted an energy-conversion efficiency $\eta= 0.1$ \citep{Yu2002,Marconi2004,Shankar2004,Shankar2009}. The luminosities in the UV and in the X-ray bands were computed from the above expression using the bolometric correction given in \citet{Marconi2004}. 

In \citet{Menci2014} and \citet{Gatti2015} we already constrained the regimes of effectiveness of the two mechanisms by comparing them with a wide range of different properties concerning the AGN population (AGN luminosity function, Eddington ratio distribution, Magorrian relation, etc.) and AGN host galaxies (host galaxy mass function, colors magnitude diagram, $SSFR-M_*$ relation, etc.). 

%While our IT scenario provided a quite good match to all the observational constraints we compared with, DIs were effective mainly in triggering moderately luminous ($L \sim L_{knee}$) AGN, hosted by medium sized ($M_* < 10^{11} \msun$), disky and gas rich galaxies. These ranges of validity for the DI scenario must be kept in mind in the next sections when comparing with clustering measurements.

DIs were particularly effective in triggering moderately luminous ($L \sim L_{knee}$) AGN, but they failed in producing luminous AGN especially at high redshift and provided a very low abundance of massive ($M_{BH}>10^9 \msun$) SMBH at $z >1$. In light of eq. 4. and 5., which require DI AGN host galaxies to be gas rich and disk dominated, DIs resulted particularly effective in triggering AGN activity in medium-sized galaxies $10^9 \msun<M_*<10^{11} \msun$, generally characterized by an enhanced star-formation activity locating DI AGN mainly in the main sequence or in the starburst region of the $SSFR-M_*$ plane (e.g., \citealt{Daddi2007,Fontanot2009}). Ultimately, our DI AGN hosts were characterized by bluish colors in the color-magnitude plane, populating mainly the blue cloud and the green valley in the redshift range $0<z<2.5$.

On the contrary, the IT scenario spanned a broader range of SMBH and host galaxies properties, providing a better match to all the observations we compared with. Not only were IT able to trigger moderate luminous AGN in medium-sized galaxies, but they were also responsible for the emission of the most powerful AGN (in case of strong interactions), as well as for the triggering of AGN activity in massive ($M_* \sim 10^{12} \msun$), bulge dominated galaxies.  IT were also able to trigger  AGN activity both in actively star forming galaxies and in passive galaxies, spanning the whole $SSFR-M_*$ plane up to $z \sim 2.5$; at $ z <1.5$, IT host galaxies were characterized by a wide range of colors,populating the blue cloud, the green valley and the red sequence simultaneously. These ranges of validity for the two scenarios (especially for the DI scenario) must be kept in mind in the next sections when comparing with clustering measurements.

We stress that the results in this work do not heavily rely on the exact implementation of these two
processes in our SAM, nor on the exact parameterization/modelling adopted for evolving the galaxies within their host dark matter haloes (Sect. 2.1). In fact, the frequency and type of IT mainly depend on the dark matter merger rates, dynamical friction and encounter timescales, which are common features in every galaxy evolution model based on a $\Lambda$CDM cosmology. The DIs are instead $"$in-situ$"$ processes, mainly controlled by the disk instability criterion of eq. \ref{efstathi}. While eq. \ref{efstathi} is clearly a condition tightly linked to the specific features of the SAM (scale radius evolution, maximum circular velocity), we show that our main results are broadly preserved irrespective of the exact threshold chosen for the disk instability. Moreover, a number of comparison studies have shown that our SAM is quite "typical", producing galactic outputs at all epochs in line with several other state-of-the-art SAMs (e.g., \citealt{Gruppioni2015}).

\section{Probing AGN clustering properties: AGN MOF and 2PCF calculation}

To investigate the clustering properties of the AGN population we rely on the halo model \citep{Kauffmann1997,Cooray2002,Tinker2005,Zheng2007} and we make use of two fundamental probes: the mean AGN occupation function (MOF) $\langle N(M_h) \rangle$ and the AGN 2-point correlation function (2PCF) $\xi(r)$.  %, according to which it is possible to reformulate the pieces of information provided by clustering analysis concerning the spatial distribution of AGN into an accurate description of how AGN populate dark matter halos with different mass. 

The key element of the halo model is constituted by the halo occupation distribution (HOD), which is defined as the conditional probability $P(N|M_h)$ that a halo of mass $M_h$ contains $N$ AGN. Observationally, the full P(N|M) could be specified by determining all its moments from AGN clustering at each order; unfortunately, due to the paucity of AGN, it is not possible to accurately measure higher order statistics (such as 3PCF). The problem is solved by using an approximate description of the halo model and considering only the lowest-order moment, namely the mean occupation function (MOF) defined as $\langle N(M_h) \rangle = \Sigma_N N P(N|M_h)$.%(also known as mean occupation function, MOF).% and $\langle N(N-1) \rangle_{M_h}$. 

%Hence, in the following, we make use of two key probes: the mean AGN occupation function (MOF) $\langle N(M_h) \rangle$ and the AGN 2-point correlation function (2PCF) $\xi(r)$. 

Our SAM, by accounting for the evolution of the AGN population along with the evolution of the dark matter density field, directly provides the AGN MOF for each timestep of the simulation. As a standard practice we express the AGN MOF for the DI and IT modes as the sum of a central and a satellite component $\langle N(M_h) \rangle = \langle N(M_h) \rangle_{cen} + \langle N(M_h) \rangle_{sat}$.

Following the halo model we then compute the two-point correlation function $\xi (r)$ from the predicted AGN MOF, %(\citealp{Kauffmann1997,Cooray2002,Tinker2005,Zheng2007}). 
under the assumption that $\xi (r)$ can be faithfully described as the sum of two contributions: the 1-halo term $\xi_{1-h} (r)$, mainly due to the contribution of AGN residing in the same halo, and the 2-halo term $\xi_{2-h} (r)$, due to the correlation of objects residing in different halos. Both terms can be obtained from the AGN MOF once having specified the halo mass function $n(M_h)$ \citep*{Press1974,Sheth1999} and the halo bias factor $b(M_h)$ at a specific redshift (see, e.g., \citealt{Zheng2007}). % More details on the halo model and on how we compute the 1-halo and 2-halo terms are given in Appendix A.

A point that is important to stress is that both the AGN MOF and the 2PCF obtained from our SAM are not affected by any evident degeneracy concerning their shape and normalization. In basic HOD modelling instead, the AGN MOF is usually inferred indirectly from clustering measurements: once having assumed a parametric expression for $\langle N(M_h) \rangle$, the 2PCF is used to constrain the parameters of the AGN mean occupation function. This indirect approach is however limited by the inevitable degree of degeneracy in the different parameterizations of the input MOF that can account for the measured 2PCF (e.g., \citealt{Shen2013}).

 %This approach however has a relevant drawback: the AGN MOF is strongly degenerate, and different shapes and normalization can equally reproduce the AGN 2PCF. This is basically due to the fact that while the 2PCF provides only a characteristic halo mass for the AGN sample, the AGN MOF provides a more detailed description of how AGN populates halos with different masses, which is difficult to constrain with clustering data alone. 

Our direct approach conversely starts from the predicted AGN mean occupation functions provided by our SAM, which in turn yields, within the HOD formalism, a unique expression for the 1-halo and 2-halo terms of the AGN 2PCF.

%which are well determined and not constrained by any kind of parameterization. Within the assumptions of the halo model, a unique expression for the 1-halo and 2-halo terms of the AGN 2PCF is then associated to every AGN mean occupation functions obtained from the SAM. 

\section{Results: General trends}

%In this section we present the results concerning the clustering properties of the two scenarios. Before comparing with observational results (AGN HOD and 2pcf), we discuss the clustering properties from a general point of view, without trying to reproduce any particular observational measurement (and its selection effects).

In this section we present the results concerning the clustering properties of the IT and DI scenarios. Before focusing on a detailed comparison with a number of AGN MOF and 2PCF measurements, we discuss some important, general features of the clustering properties of our two models.

\subsection{AGN duty cycle}

\begin{figure}
\includegraphics[scale=0.25]{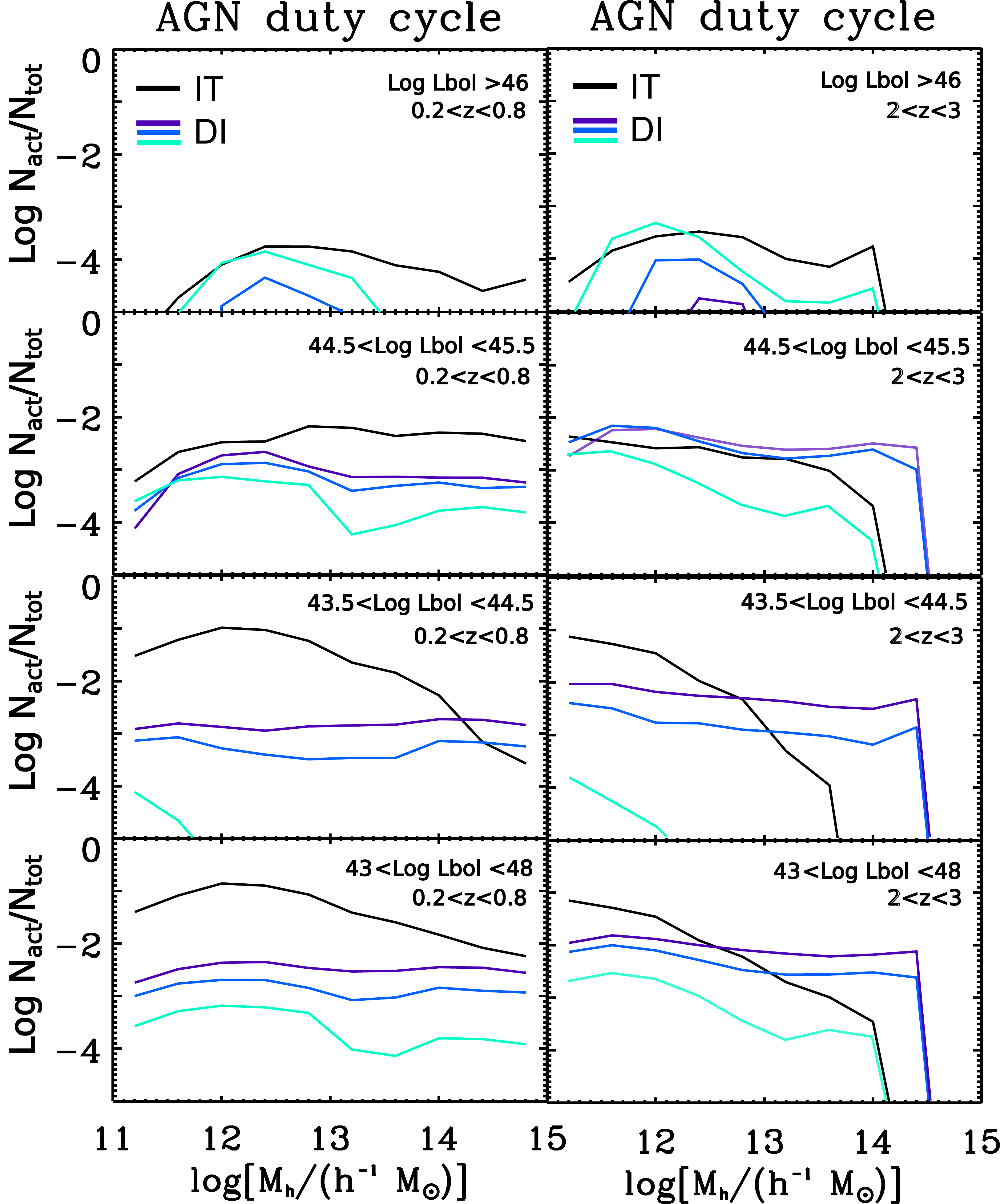}
\caption{AGN duty cycle for the DI and IT scenarios as a function of AGN bolometric luminosity and dark matter host halo mass, computed at redshift $0.2<z<0.8$ and $2<z<3$. Only galaxies (both active and passive) with $M_*>10^{9} \msun$ have been considered. $N_{act}$ refers to the number of active galaxies with bolometric luminosity in the range indicated by the label, while $N_ {tot}$ refers to the total number of galaxies (both active and passive).} The predictions of our model are represented by continuous lines: black line for the IT scenario, light blue, blue and purple for the DI scenario. Light-blue represents the prediction with the highest value of the normalization of the inflow (see Sect. 2.2), blue for the intermediate value, purple for the lowest value.
\label{dutycycle}
\end{figure}

As a first check, fig. \ref{dutycycle} shows the AGN duty cycle for the two scenarios as a function of dark matter host halo mass and AGN bolometric luminosity. This plot is useful to pin down the luminosity intervals and the typical environments where DI and IT are most efficient in triggering AGN activity in galaxies. In computing the AGN duty cycle, we have taken into account galaxies with stellar mass $M_* > 10^{9} \msun$.

At low redshift ($z \sim 0.5$), the IT scenario is in general more efficient than DIs in triggering AGN activity in galaxies, especially for low luminosity AGN ($L_{bol} \sim 44$). Conversely, at higher redshift ($z\sim 2.5$), DIs substantially increase their efficiency, due to the higher gas and disk fraction of AGN host galaxies, becoming more efficient than galaxy interactions in triggering moderately luminous AGN ($Log L_{bol} \sim 45$). 

In the IT scenario the AGN duty cycle depends both on the environment and on the AGN luminosity. Luminous AGN are mainly found in dark matter halos with mass $M_H \sim 10^{12} h^{-1} \msun$, while the AGN duty cycle flattens for moderately luminous AGN ($L_{bol} \sim 45$). Low-luminosity AGN, which dominate by number the total AGN population, are again characterized by a substantial peak around $M_H \sim 10^{12} h^{-1} \msun$. Galaxy interactions are indeed favored in group environments, due to a high density of galaxies with low relative velocity (lower with respect to massive clusters). This differs substantially from the prediction of the DI scenario. Aside for high luminosity AGN, the AGN duty cycle is almost flat, both at low and high redshift. As expected, being $"$in-situ$"$ processes, DIs are in fact not necessarily linked to the large-scale structure environment.

%Being $"$in-situ$"$ processes, DIs show a milder environmental dependence with respect to IT, causing the different duty cycle. 

%At z ∼ 1.25, we showed in §4 and Figure 6 that the cluster and field AGN fractions were comparable and both ∼ 3 − 4% for AGN with hard X-ray luminosities greater thana few times 1043 erg/s
%In the local universe, very luminous AGN are rarely found in the field and very rarely found in clusters. An SDSS study by Kauffmann et al. (2004) found that AGN with L[O III]> 107 L were approximatel
%nd found the field and cluster AGN fractions are the same for low-luminosity AGN (LX > 1041−42 erg/s) Haggard 2010

% In contrast with the local universe,where the luminous AGN fraction is higher in the field than in clusters, the X-ray and MIR-selected AGNfractions in the field and clusters are consistent at 1 < z < 1.5. This is evidence that the cluster AGN populationhas evolved more rapidly than the field population from z ∼ 1.5 to the present.

\begin{figure}
\includegraphics[scale=0.25]{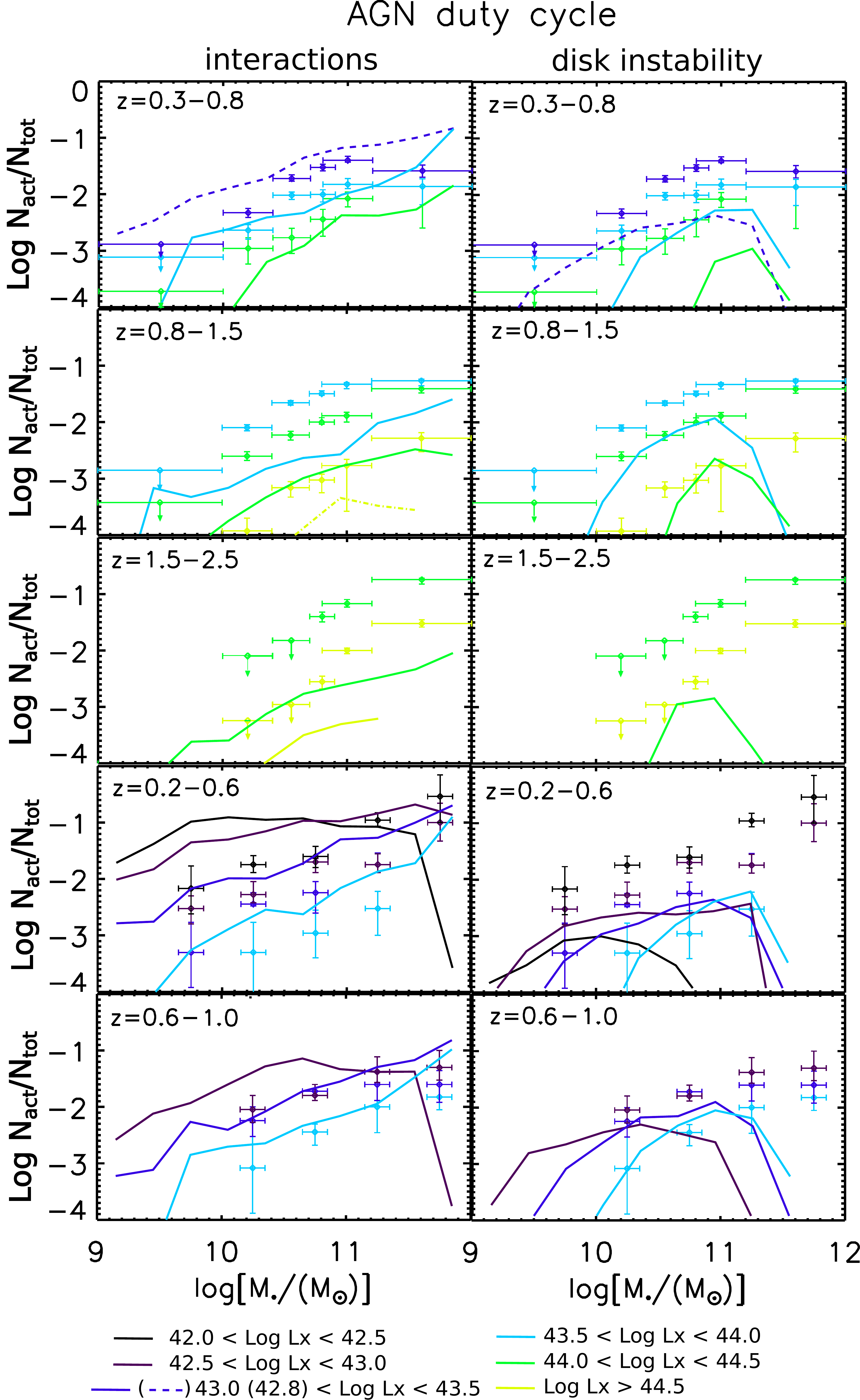}
\caption{AGN duty cycle for the DI and IT scenarios as a function of AGN bolometric luminosity and host galaxy stellar mass. $N_{act}$ refers to the number of active galaxies within a logarithmic luminosity bin, while $N_ {tot}$ refers to the total number of galaxies (both active and passive).
 The three upper rows show the comparison with data from \citet{Bongiorno2012}, while in the two lower ones we compare with data from \citet{Aird2012}. For the DI scenario, only the prediction with $\alpha=2$ is shown. Solid lines represent the SAM predictions, while points represent observational data. The color code refers to different luminosity bin, as shown by the label below the plot}.
\label{dutycyclebonaird}
\end{figure}

Fig. \ref{dutycyclebonaird} shows the AGN duty cycle as a function of host galaxy stellar mass instead of dark matter halo mass. Our predictions are here compared with data from \citet{Aird2012} and \citet{Bongiorno2012}.

First, we note that all the predictions for the IT scenario for the different luminosity/redshift bins are broadly characterized by the same shape, differing only in their normalization. This feature, as also noted by \cite{Bongiorno2012}, indicates that this scenario must be characterized by a broad Eddington ratio distribution, which is what we showed in \cite{Menci2014} for the IT scenario.  Broad Eddington distributions naturally allow for the more massive systems to be active at different modes thus increasing their duty cycle (see also, e.g., \citealt{Shankar2010,Shankar2013}, and references therein). A narrower distribution, instead, would force the most luminous AGN to be preferentially linked to the most massive, less numerous SMBH and host dark matter halos. 

%have to be hosted by the most massive systems, with less luminous AGN hosted by smaller galaxies.
The DI scenario, on the contrary, does show some luminosity dependence. The most luminous AGN are preferentially hosted by the most massive galaxies, with the distribution spreading towards lower stellar masses for lower luminosities. This feature follows from an underlying tighter Eddington ratio distribution characterizing the DI scenario, a trend already noted in \cite{Menci2014}.

We caution that a caveat to the aforementioned statement is that the Eddington ratio distribution is only proportional to $\propto \dot{M}_{BH}/M_{BH}$ and not to $\propto \dot{M}_{BH}/M_{*}$, so the relation between the Eddington ratio distribution and the shape of the AGN duty cycle as a function of stellar mass and AGN luminosity is not one-to-one. Anyway, in case of a sufficiently tight $M_{BH}-M_{*}$ relation, this should be preserved. This is especially the case of DIs: indeed, in \citet{Menci2014} we have showed that they are characterized by a tighter $M_{BH}-M_*$ relation compared to the IT scenario, a fact that further strengthen the link between AGN luminosity and host galaxy stellar mass.

Some relevant discrepancies between our predictions and observational data need to be discussed at this point. 
First of all, we note that the DIs fall short in triggering AGN activity in massive galaxy hosts: regardless of the luminosity and redshift range, the duty cycle drops for galaxies with mass $M_* >10^{11} \msun$. In brief, as outlined by \cite{Gatti2015}, disk instabilities tend to be disfavored  by low gas and disk fractions, a typical condition for massive galaxies.%this is due to a combination of the stabilizing effect of the hot component (represented by the factor $v_c^2$ in eq. 4) and of the low disk and gas fractions that characterize such massive hosts and which disfavors our DI scenario.

Second, there is substantial over production of faint AGN ($42<Log L_{X}< 43$) at $z<1$ for the IT scenario, especially in low-mass hosts. Even if this might partially be due to an underestimate in the data completeness correction for low luminosity AGN, we expect the IT scenario to over-produce faint AGN, owing to the well known excess of small objects produced by SAMs. However, this should little affect the main results of this paper, since the majority of the data we will be comparing our models to hardly ever extends to such low luminosity/low host stellar mass sample.

Finally, a more severe discrepancy occurs at high redshift with respect to data from \cite{Bongiorno2012}. Their data imply a strong redshift evolution of the AGN duty cycle ($\propto (1+z)^4$), which is not reproduced by our predictions, though it is broadly consistent with some continuity equation models (e.g., \citealt{Shankar2013}). From the point of view of our SAM, we can have basically two different explanations: 1) the IT and DI models do not produce enough moderate-to-high luminosity AGN at such redshifts; 2) there is an incorrect correspondence between the AGN luminosity and galaxy stellar mass. %(i.e., the models do not reproduce correctly the

%Eddington distributions of the two models do not evolve enough towards higher values at high redshift.

However, our models show a pretty good match with the AGN luminosity function at redshift $1<z<2$ (especially in the IT scenario). Even accounting for the uncertainties in the LF measurements, this cannot totally explain the one order of magnitude discrepancy shown by the AGN duty cycle in fig. \ref{dutycyclebonaird}. 
%Our SAM also reproduces correctly the galaxy mass function at such redshift, aside for the well-known excess of hierarchical models concerning $M_*<10^{10} \msun$; also this point, hence, should play a minor role. 
It is possible that the AGN duty cycle in the most massive bin might be affected by statistical fluctuations owing to the low abundance of massive hosts. At face value, the most relevant point might concern the correspondence between the AGN luminosity and galaxy stellar mass (in this respect, see \citealt{Lamastra2013}, where we show that the host stellar mass-AGN luminosity distribution predicted by our SAM exhibits a poor match with observational data from \citealt{Bongiorno2012}).

%the Eddington ratio distribution, which does not properly evolve towards higher values at high redshift.

It is also important to stress that observational biases and completeness correction might have a role here, affecting the estimate of the AGN duty cycle and possibly contributing to the discrepancy with our SAM. We note that the data from \citet{Aird2012} and \citet{Bongiorno2012} are not in perfect agreement with each other (if we consider overlapping redshift bins), with our SAM best reproducing the data from \citet{Aird2012}. This tension between the two data sets might depend, for instance, on the AGN host galaxy mass estimate, which is related to the technique used to perform the SED fitting. In this respect, \citet{Bongiorno2012} noted that fitting the optical SED with a galaxy+AGN template produces masses from 10 times smaller to 6 times higher than the estimates obtained using a galaxy template only (which is the case of \citealt{Aird2012}).

Assuming that the discrepancy between our predictions and data at high redshift is a true discrepancy, we note that if the models are failing in reproducing the normalization of the AGN duty cycle for every stellar mass, this should only affect the normalization of the AGN MOF and not the AGN 2PCF, if the relative probability of triggering centrals and satellites remains unaltered. Conversely, if we are assigning at some fixed $L_{bol}$ a lower (higher) stellar mass, we might under (over) predict the average AGN bias factor.

\subsection{Redshift and luminosity dependence of AGN clustering}

A second preliminary investigation concerns any redshift and luminosity dependence of predicted clustering of AGN. %In fact, understanding whether the AGN MOF evolves with redshift and/or with AGN luminosity is essential to correctly interpret 2PCF measurements. 2PCF measurements often probe wide redshift and luminosity ranges, so as to increase the number of sources available. Usually the 2PCF (and the AGN MOF obtained from it)  is assumed to be representative of the true 2PCF of the median redshift and median luminosity of the sample, a practice that might be not completely correct in presence of any redshift and/or luminosity evolution of the AGN clustering strength. 
%Several authors have investigated the luminosity and redshift evolution of clustering measurements. %We summarize the observational results obtained so far in Appendix B.

Several authors have investigated the luminosity and redshift evolution of AGN clustering measurements. 
 \citet{Shen2013}, dividing their $<z>=0.5$ sample in different luminosity bins in the range $-23.5 < m_i < -25.5$, found weak/no sign of luminosity evolution in their clustering measurements, even if they cannot completely rule out stronger evolution, given their uncertainties. Similar results have been found by \citet{Chatterjee2013}, at a slightly lower redshift. \citet{Richardson2012} have shown that dividing their $<z>=1.4$ sample of luminous AGN in two sub-samples above and below the median redshift little affects the AGN 2PCF, hence suggesting no strong redshift evolution. No redshift evolution has been also found by \citet{Allevato2011}, who showed that moderately luminous XMM COSMOS AGN reside in dark matter halos with constant mass up to $z \sim 2$.

On the contrary, \citet{Chatterjee2012}, with their cosmological hydrodynamic simulation, suggested a strong redshift and luminosity evolution of the AGN MOF in the redshift interval $z  \sim 1-3$, but they focused on lower bolometric luminosities in the range $10^{38}-10^{42}$ ergs$^{-1}$. \citet{Krumpe2012,krumpe2015}, dividing their low redshift ($0.07 < z < 0.50$) X-ray selected AGN sample in two luminosity bins, found a mild luminosity dependence at $\sim 2 \sigma$ level, suggesting that $M_{BH}$ might be the driver of such mild dependence. \citet{Eftekharzadeh2015}, studying a sample of spectroscopically confirmed quasars from the BOSS survey, found that quasar clustering remains similar over a decade in luminosity at $z \sim 2.5$, but the typical halo mass decreases with increasing redshift in the range $2.2<z<3.4$. 
\citet{Allevato2014} found that COSMOS AGN at $z\sim3$  inhabit less massive dark matter halos with respect to their low redshift counterparts, suggesting a redshift evolution for $z>2$.  
Ultimately, \citet{Koutoulidis2013}, focusing on moderately luminous X-ray AGN in the redshift range $0<z<3$, noticed no strong redshift evolution up to $z \sim 1.5$, but they found a positive dependence of the bias factor on AGN X-ray luminosity. Particularly, in their picture at redshift $z \sim 1$ moderately luminous AGN with $L_X$ up to $10^{44}$ ergs$^{-1}$ inhabit more massive halos ($M_h \sim 10^{13} \msun$) with respect to less luminous AGN; the authors furthermore suggest that high luminosity AGN ($L_X>10^{44}$ ergs$^{-1}$) might also occupy less massive halos with $M_h \sim 10^{12} \msun$, in agreement with clustering measurements of luminous QSOs.

%As a first check, we show in fig. \ref{bias} the evolution with redshift of the bias factor and of the typical halo mass for all the 2PCF measurements we compared with in the previous sections. The typical halo mass refers to the average halo bias of the sample. The plot exhibits a clear trend: as the redshift increases, the average halo mass moves towards lower values. This trend concerns both optical and X-ray surveys, regardless of the different selection cuts concerning each AGN sample. The main reason of this decrease in the average bias is that at high redshift, halo more massive than $M_h \sim 10^{13} \msun$ becomes progressively rarer, so as AGN populate on average less massive environments.

%Fig. \ref{lumreddep} provides some insights into any redshift and luminosity dependence of AGN clustering for the DI and IT scenario. The contour plots represent the number density of AGN as a function of host halo mass and bolometric luminosity, for two different redshift. At fixed redshift,  dividing the countour plot into horizontal slices provides insights into any AGN HOD (and clustering) luminosity evolution (the AGN HOD could be obtained by dividing the slice by the number density of DM halo). On the other hand, comparing the distributions for the two different redshift considered provides insights into any redshift evolution of the bias factor. %%(even if the median halo mass does not precisely correspond to the mass mapped by the bias factor). 

\begin{figure}
    %\centering
%\includegraphics[scale=0.25]{halo_lum05.pdf}
%\includegraphics[scale=0.25]{halo_lum25.pdf}
\includegraphics[scale=0.16]{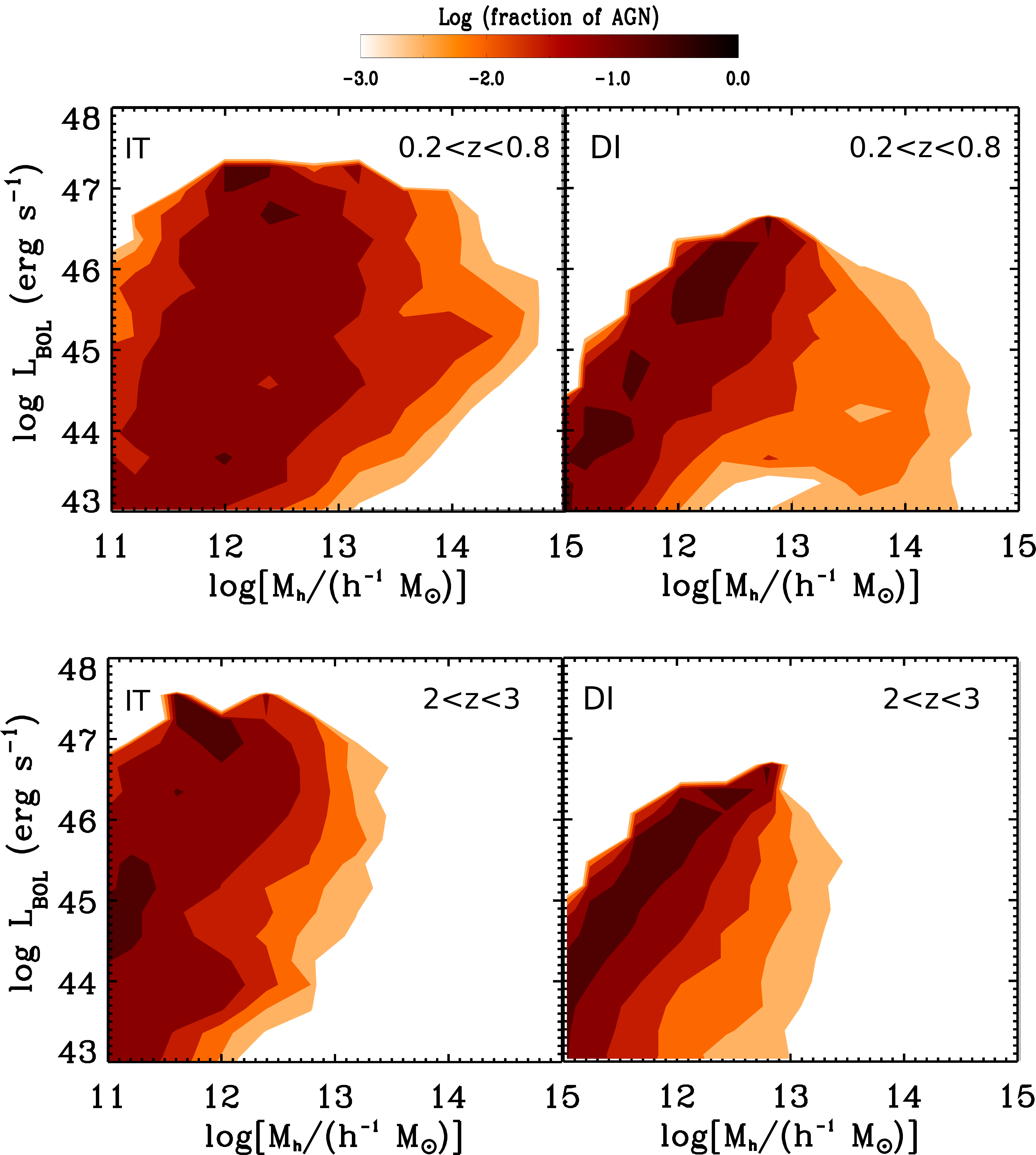}

\caption{Contour plots representing the distribution of AGN as a function of host halo mass and AGN bolometric luminosity, for two different redshift bins ($0.2<z<0.8$ and $2<z<3$). The AGN fraction represents the number density of AGN weighted for the actual number density in a given luminosity bin. Both the predictions for the IT and DI scenario are shown (left and right panels respectively). As for the DI scenario, only the prediction with normalization $\alpha=0.5$ is considered. No additional cut in the AGN host galaxy properties has been considered. } %The panels on the right represent the number density of AGN as a function of host halo mass for three different luminosity cuts. Predictions for both the IT and DI scenario are shown. Vertical dotted lines represent the average halo mass.}
\label{lumreddep}
\end{figure}

%excess at low halo masses for low-luminosity AGN in the IT scenario: might be due to the excess of faint galaxies in SAMs, which increases the number of possible interactions. less effective in the DI scenario, since these faint galaxies might not satisfy the triggering conditions or might have small disk which might not lead to significant inflow (=lower luminosities). 

Fig. \ref{lumreddep} provides some insights into any redshift and luminosity dependence of AGN clustering for the DI and IT scenario. The contour plots represent the distribution of AGN as a function of host halo mass and bolometric luminosity, for two different redshift bins.

At low redshift, fig. \ref{lumreddep} shows that luminous AGN $Log L_{bol} >46$ inhabit mainly halos with mass $\sim 10^{12}-10^{13} \msun$ for both the IT and DI scenarios (even if DIs do not trigger the most luminous AGN with $Log L_{bol} \sim 47$). As we consider intermediate luminosities ($Log L_{bol}\sim 44-45$), the distribution gradually spreads over the full range of halo masses; this is particularly true for the IT scenario, while for the DI scenario the spreading is less pronounced (see the levels of the countour plot, indicating that the majority of DI AGN inhabit low mass environments). Finally, for low luminosity AGN ($Log L_{bol}\sim 43-44$), the distribution becomes slightly steeper towards lower halo masses.%, especially for IT. 

 This non-linear behavior in the number densities of active galaxies can be explained as follows. Luminous AGN are naturally found in halos with mass $\sim10^{12} -10^{13} \msun$ because the most luminous AGN necessarily require large gas reservoirs and massive BHs.
 
 For the IT scenario, luminous AGN are triggered by strong interactions, which are more common in less massive environments than in clusters, due to a higher relative velocity between galaxies. For the DI scenario the SMBH mass inflow is maximized by the presence of an unstable, massive and gas rich disk (see eq. \ref{hopkins}), which cannot be found in massive environments %(disrupted by consequent mergers and tidal stripping). 
 Moderately luminous AGN are less constrained by the above-mentioned requirements and naturally reside in a wide range of dark matter halos. Ultimately, the slight increase towards low halo masses for low luminosity AGN might be due to the tendency of low mass SMBHs to reside on average in less massive dark matter halos.

As for the high redshift ($z\sim2.5$) bin, the most relevant difference with respect to the low redshift case is that all AGN activity is moved towards less massive halos. In short, structures with mass greater than $M_h > 10^{13}$ become significantly rarer, relegating active galaxies to live mainly in less massive environments. This plot alone would suggest a non-negligible redshift evolution of the AGN clustering, calling into question the use of wide redshift interval to compute the AGN 2PCF.

\begin{figure}
    \centering
\includegraphics[scale=0.35]{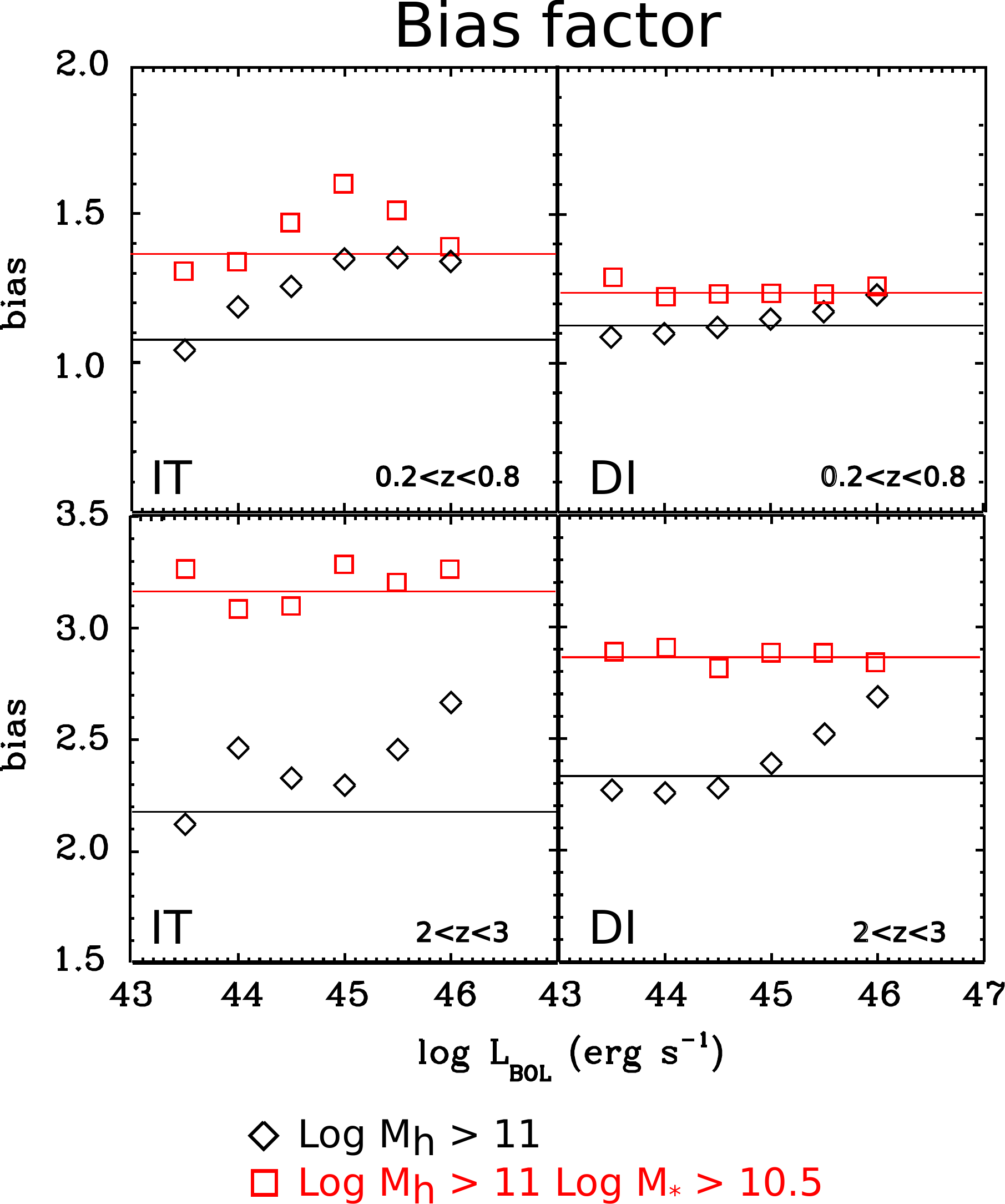}

\caption{Bias factor of the AGN population for the DI and IT scenarios, as a function of redshift and AGN bolometric luminosity. Black diamonds refer to the bias factor computed considering AGN residing in halos with mass $M_h >10^{11} h^{-1} \msun$; red squares represent the bias factor computed considering the additional cut in the host galaxy stellar mass of $M_*>10^{10.5} \msun$. Horizontal lines represent the average bias factor over the full luminosity range.}
\label{bias1}
\end{figure}

In the standard prediction obtained accounting for all the AGN residing in $M_h>10^{11} h^{-1} \msun$, we note that the average bias factor (i.e., the bias factor computed over the full luminosity interval) is very similar for the two scenarios, being dominated by the less luminous AGN and averaging at around $b\sim 1.1$, roughly corresponding to $M_h \sim 10^{12} h^{-1} \msun$ (fig. \ref{bias1}). Luminous AGN are associated with a slightly higher bias factor with respect to faint AGN, corresponding to dark matter halos as massive as $M_h \sim 10^{12.5} h^{-1} \msun$, but in general, no strong evidence for luminosity dependence is observed. At redshift $z \sim 2.5$, the two scenarios are again characterized by similar average bias factors ($b \sim 2.3$, corresponding to $M_h \sim 10^{11.5} h^{-1} \msun$), with the most luminous AGN residing again in slightly more massive dark matter halos ($M_h \sim 10^{11.9} h^{-1} \msun$). With respect to the low redshift case, a consistent evolution towards lower dark matter halo mass is observed, for every luminosity bin. 

The analysis of the bias factor alone as shown in fig. \ref{bias1}, might not be a sufficient discriminator for AGN triggering mechanisms, being the differences relatively small, with also relatively weak luminosity dependence.

Nevertheless, there is an important point that needs to be stressed. In fig. \ref{lumreddep} and \ref{bias1} we have not made any particular cut in the properties of the AGN host galaxy population, simply considering AGN residing in dark matter halos more massive than $M_h > 10^{11} h^{-1} \msun$. However, any additional selection in the host galaxy properties might alter the distribution, affecting the implied bias factor. Among all, the major effect might be represented by the clustering dependence on  the host galaxy stellar mass, since it is known to correlate to the dark matter halo mass \citep{Vale2004,Shankar2006,Moster2013,Shankar2014}.

Fig. \ref{bias1} for example shows that considering only galaxies  with $M_*>10^{10.5} \msun$ considerably alters the bias factor distributions. Irrespective of the exact model, the overall bias as a function of bolometric luminosity in fact increases in normalization and flattens out. This proves that the underlying AGN light curve (e.g., \citealt{Lidz2006}) is not the only responsible for shaping the bias-$L_{bol}$ relation. Particularly, at low redshift the average bias factor corresponds to halos with $M_h \sim 10^{12.5} h^{-1} \msun$, while at high redshift we have $M_h \sim 10^{12} h^{-1} \msun$. Since high redshift observations are generally biased towards bright hosts with high stellar mass, this (often implicit) cut in host galaxy stellar mass might hide a possible intrinsic redshift evolution in the bias factor.

Observationally, a good example of the importance of taking into account various selection effects comes from \cite{krumpe2015}.  The authors, studying the clustering properties of a X-ray selected RASS/SDSS AGN sample and an optical selected SDSS AGN sample in the redshift range $0.12<z<0.36$ find similar dependencies in the clustering strength with $M_{BH}$, $L_{Bol}/L_{Edd}$, FWHM, and $L_{H \alpha}$ in both the X-ray selected and optically selected AGN sample, but only their X-ray sample showed a clear AGN luminosity dependence. The lack of a clear luminosity trend in the optically selected sample has been mainly imputed by the authors to the complexity of the optical SDSS AGN selection, responsible for introducing various selection bias substantially affecting the AGN low-luminosity bin and hampering the detection of any luminosity dependence of AGN clustering.

\subsection{AGN satellite fraction}
Another general feature of our DI and IT models that can be investigated is the AGN satellite fraction $f_{sat}$, i.e. the fraction of AGN that reside in satellite galaxies with respect to the total AGN population. 

From an observational point of view, clustering measurements constrain the AGN satellite fraction %from small scale clustering, once having assumed a parametric form for the central and satellite MOF. Its value is pretty uncertain, as it highly depends on the particular MOF parametric form chosen. 
mainly from small-scale clustering, with an associated systematic uncertainty broadly related to the exact parametric form adopted for the input MOF.
Previous observational estimates both at low (z $\sim 0.5$) and intermediate redshift ($\sim 1.5$) have derived a satellite fraction usually in the range $f_{sat} \sim 0.01 - 0.1$, with $f_{sat} \sim 10\%$ often considered an upper limit (\citealt{Starikova2011,Richardson2013,Shen2013,Kayo2012}), 
though \citealt{Leauthaud2015} claim a satellite fraction as high as $f_{sat} \sim 18\%$.
%but see \citealt{Leauthaud2015}, which infer a higher satellite fraction $f_{sat} \sim 18\%$).

\begin{figure}
    \centering
\includegraphics[scale=0.3]{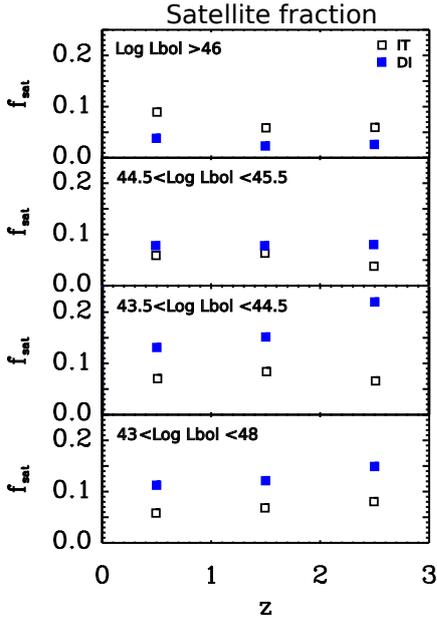}

\caption{AGN satellite fraction, defined as the fraction of AGN in satellite galaxies, as a function of redshift and AGN bolometric luminosity. In computing the satellite fraction, AGN residing in halos with mass $M_h > 10^{11} h^{-1} \msun $ have been considered. The predictions for the DI scenario are represented by filled blue squares, while those for the IT scenario by black empty squares. For the DI scenario, only the prediction with normalization $\alpha=5$ is shown.}
\label{sat_frac}
\end{figure}

\begin{figure}
    \centering
\includegraphics[scale=0.25]{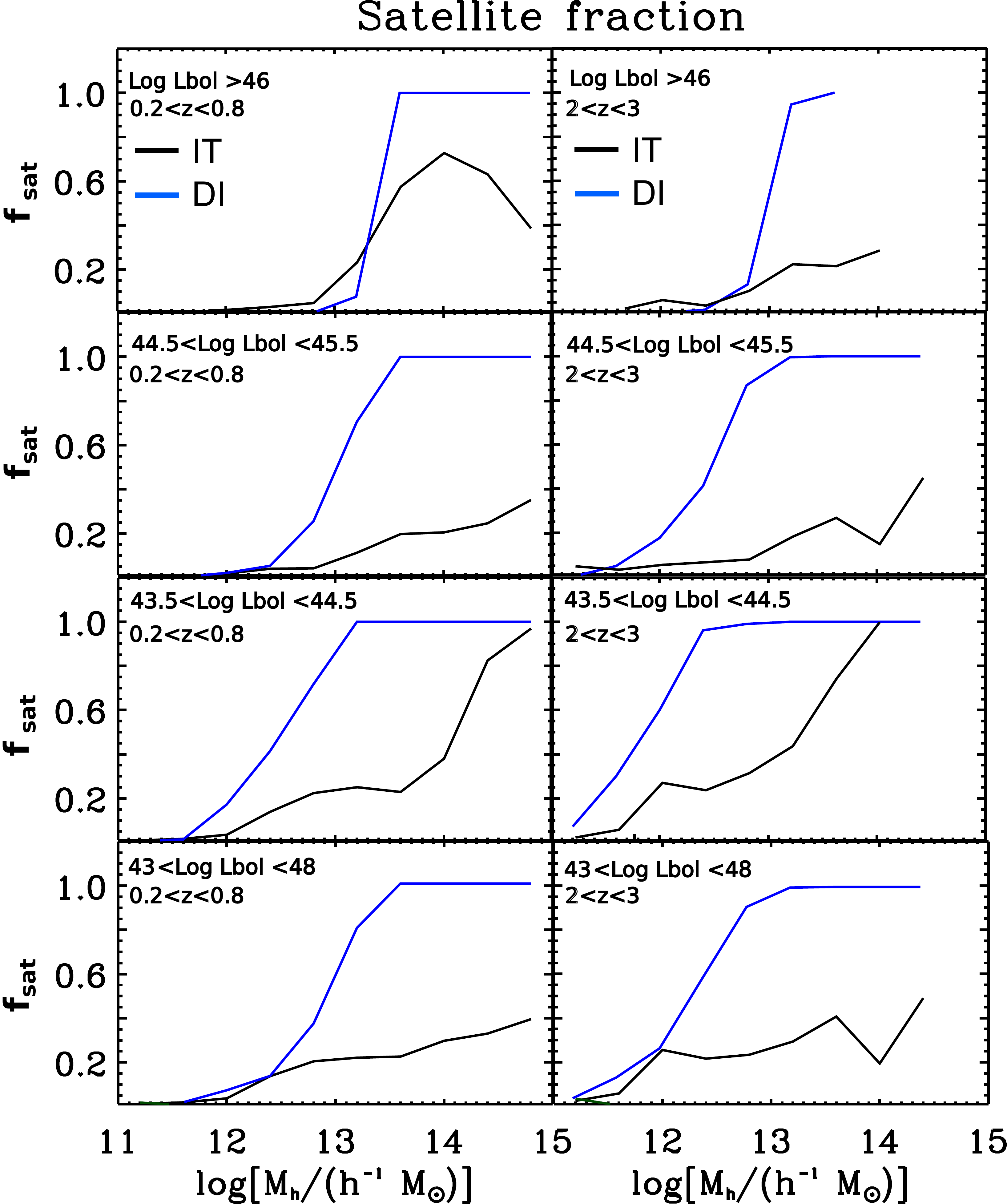}
\includegraphics[scale=0.25]{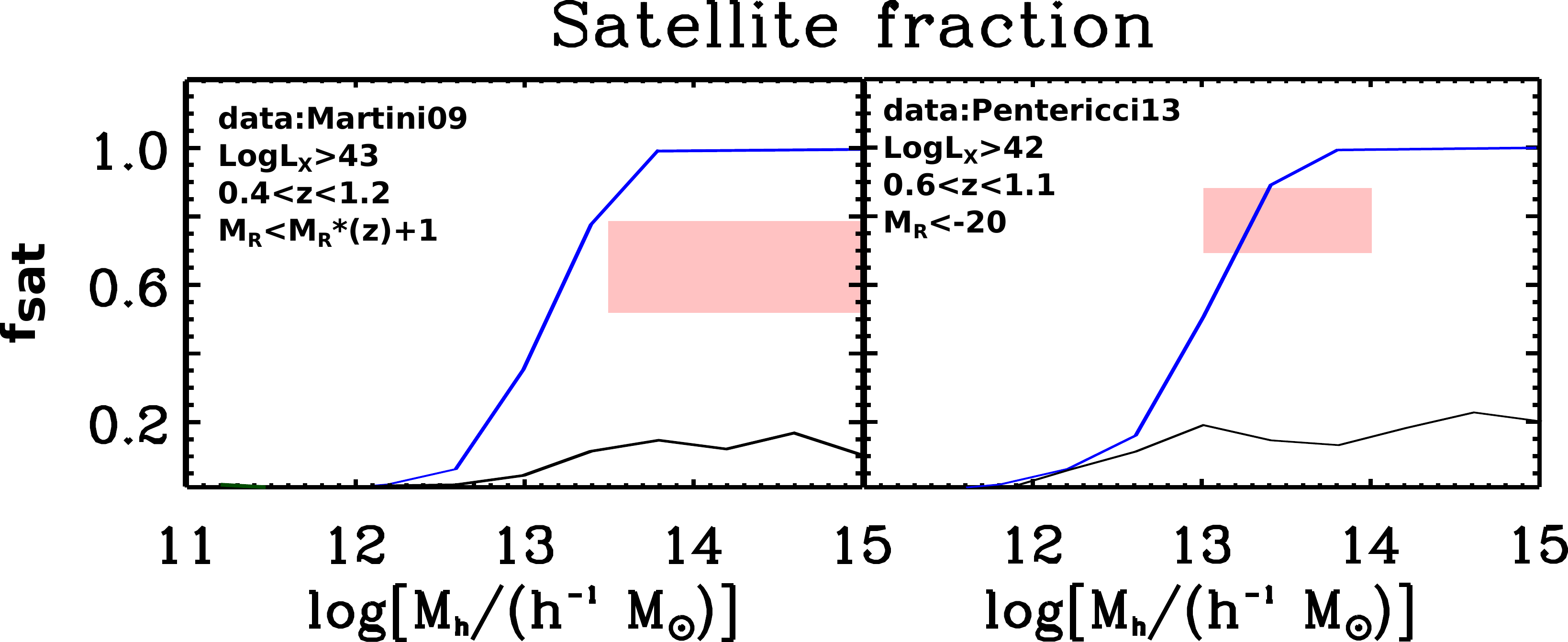}

\caption{\textit{Upper panels:}AGN satellite fraction as predicted by our models as a function of AGN luminosity and dark matter halo mass, for two redshift bins. For the DI scenario, only the prediction with normalization $\alpha=5$ is shown. \textit{Lower panels:} same as upper panels, but compared with data from \citet{Martini2009} and \citet{Pentericci2013}. The observed fraction (red box) has been obtained from the AGN radial distribution expressed in $R/R_{200}$, considering as satellites all the AGN not belonging to the innermost radial bin. The box represents the 1-$\sigma$ interval computed with the low number statistics estimator by \citet{Gerhels1986}.}
\label{sat_frac_ext}
\end{figure}

Fig. \ref{sat_frac} shows the AGN satellite fraction $f_{sat}$ as predicted by our models, at different redshift and for different luminosity cuts. In general, the AGN satellite fraction does not show any clear evolutionary trend with redshift, with its value oscillating in the range $0.05 \sim 0.15$. The only exception is for the DI scenario for $43.5<Log L_{bol}<44.5$, where the satellite fraction almost double with respect the low redshift case; however, we note that the impact of such increment on the $f_{sat}$ computed considering the whole luminosity interval ($43<Log L_{bol} <48$) is almost negligible. Galaxy interactions are characterized on average by a slightly lower satellite fraction ($f_{sat} \sim 0.05 - 0.1$) with respect to the DI scenario ($f_{sat} \sim 0.075 - 0.15$), except for the highest luminosity bin ($Log L_{bol} > 46$). The difference between the two models becomes even more evident when considering the satellite fraction as a function of the dark matter host halo mass (fig. \ref{sat_frac_ext}). Especially in the most massive hosts (groups and clusters), the fraction strongly increases and saturates to unity for the DI scenario, indicating that no AGN is triggered in central galaxies, while in the IT scenario the increment is not as pronounced and the fraction settles around lower values.

%The slightly lower AGN satellite fraction for the IT scenario can be explained by taking into account the cross section for galaxy interactions: its value depends on the mass and on the size of the host galaxy, and since central galaxies are more massive and larger in size with respect to satellite galaxies, they are more likely to become active. 

The low $f_{sat}$ in the IT scenario is induced by a very low probability of triggering satellite galaxies (the cross section for satellite interactions is small, see for instance \citealt{Angulo2009}), and the very high probability of triggering centrals. Conversely in the DI scenario central galaxies are generally less favored: at a fixed halo mass, central galaxies are more massive and with lower gas fraction with respect to satellites, leading to a higher probability of triggering satellites rather than centrals. We stress that while the exact value might be susceptible to the details of our modeling, the general prediction of a high satellite fraction for the DI scenario, whose main requirements are high disk and gas fractions, should be considered robust (see Sect. 6 for a further discussion). 
 
A first comparison with data from \citet{Martini2009} and \citet{Pentericci2013} (fig. \ref{sat_frac_ext} lower panels) seems to indicate that the IT scenario might not trigger enough intermediate-to-low luminosity AGN in satellite galaxies in massive environments (groups and clusters) at $z \lesssim 1$. In the IT scenario, the number of satellite AGN is constantly lower than those of centrals, largely irrespective of the luminosity/redshift interval. This could suggest that other mechanisms besides IT might contribute to the triggering of satellite AGN, at least in massive environments. 

We caution, however, that the comparison in fig. \ref{sat_frac_ext} should not be assumed as a conclusive test for the absolute predominance of the DI mode in triggering satellite AGN.  The fraction $f_{sat}$ provides only the relative number of satellite AGN, not the absolute abundance. In lack of any information concerning the abundance of central and satellite AGN, high $f_{sat}$ might be the result either of a high abundance of satellite AGN or of a lack of central AGN. In this respect, $f_{sat}$ values even higher than $10\%$ are not necessarily in disagreement with observational constraints, if a lack of central AGN occurs. We will better investigate the abundance of central and satellite AGN predicted by our models in the next sections.

\section{Results: Comparison with observations}
We present here the comparison between the outputs of our SAM concerning the DI and IT mode, and several 2PCF and MOF measurements.% We remind that the two feeding modes have been included in the SAM and compared with observations separately; by so doing, our aim is to further constrain the regimes where IT and DI might be effective in triggering AGN activity. 

%The measurements we compared with, besides being characterized by different redshift and luminosity ranges, are the result of both optical and X-ray surveys. and consists both in optical-selected bright quasars, as well in moderately luminous X-ray selected AGN.

The AGN samples we compare with are characterized by different redshift and luminosity ranges, and are extracted from both bright quasar optical surveys and moderately luminous AGN X-ray surveys. To broadly take into account the observational selections relative to the different types of surveys, we decided to make use of an observational absorption function \citep{Ueda2014}. The latter provides the fraction of AGN with a certain column density $N_H$ as a function of AGN luminosity and redshift. When comparing with measurements from bright quasar optical surveys, for example, we filtered the SAM mock AGN catalogs through our adopted absorption function to select only unobscured AGN with column density $N_H<10^{22} cm^{-2}$. On the other hand, we excluded AGN with $N_H>10^{24} cm^{-2}$ (CTK population) when we compared with measurements from X-ray surveys. We note however that correcting for absorption has a little impact on the exact shape of the AGN MOF predicted by our models while it mainly influences the normalization. 

Furthermore, we have taken into account all the luminosity and redshift cuts concerning different observational samples. In particular: a) we accounted for the redshift range spanned by the sample; b) in case of optical surveys, we selected AGN converting the flux limit of the survey in the \textit{i} band into a limit on the AGN bolometric luminosity (following \citealt{Richards2006}); c) for X-ray surveys, AGN have been selected according to their flux in the soft or hard X-ray band (obtained using the bolometric corrections of \citealt{Marconi2004}) and the limiting flux of the survey; d) if specified in the reference paper, we also applied an additional cut on the host galaxy magnitude in the \textit{i} band, so as to reproduce the selection bias related to the need of spectroscopic redshift.% Galaxies magnitude in the \textit{i} band for each time step of our simulation are computed using a magnitude-stellar mass relation as provided by our SAM. 

%In what follows we first show the comparison with three AGN HOD indirect measurements (that is, AGN HOD obtained from clustering analysis); then, we show the comparison with three AGN HOD measured directly, and ultimately the comparison with several 2PCF measurements.

\subsection{Comparison with AGN MOF from 2PCF measurements}

\begin{figure*}
    \centering
\includegraphics[scale=0.24]{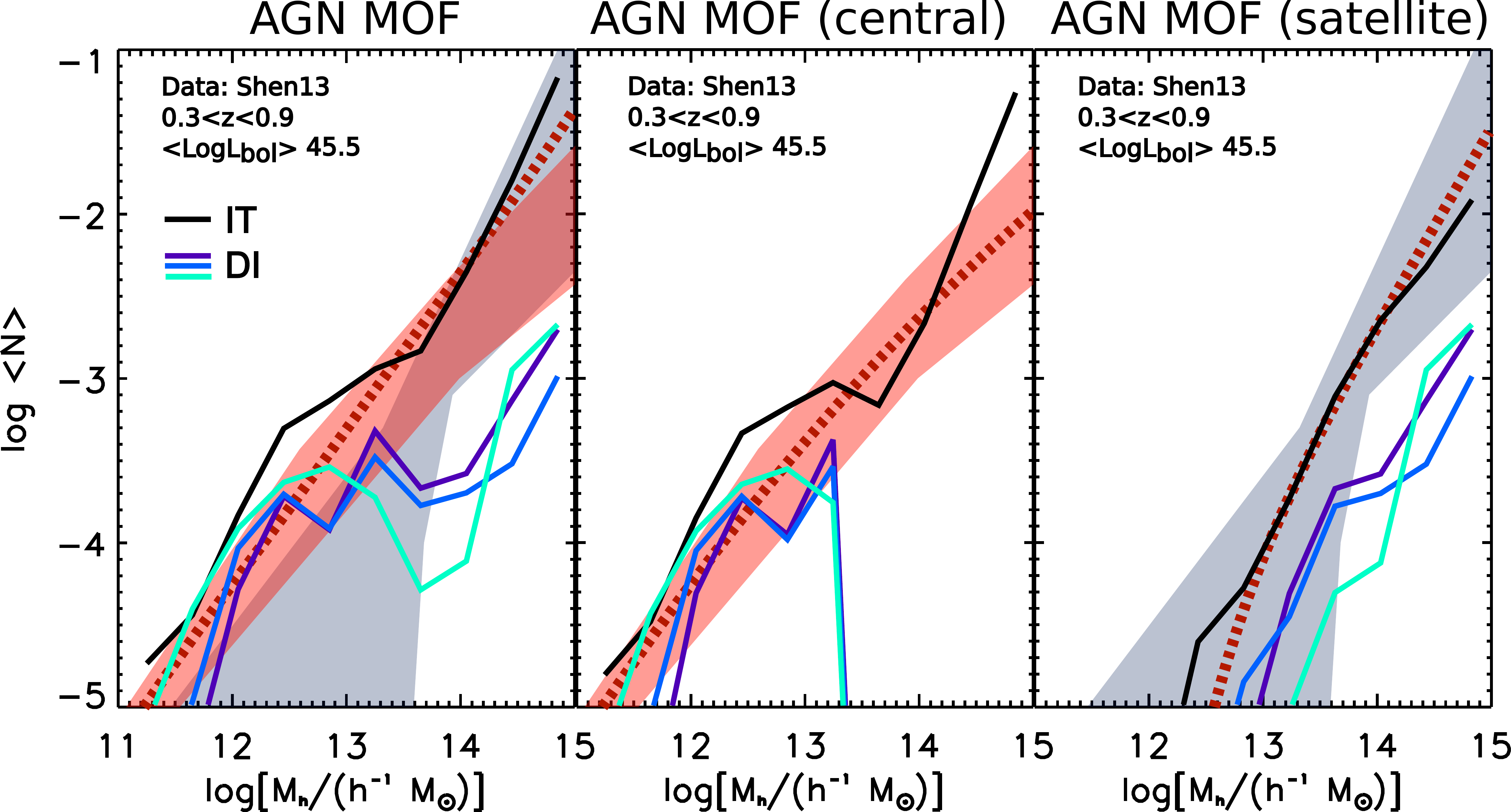}
\includegraphics[scale=0.24]{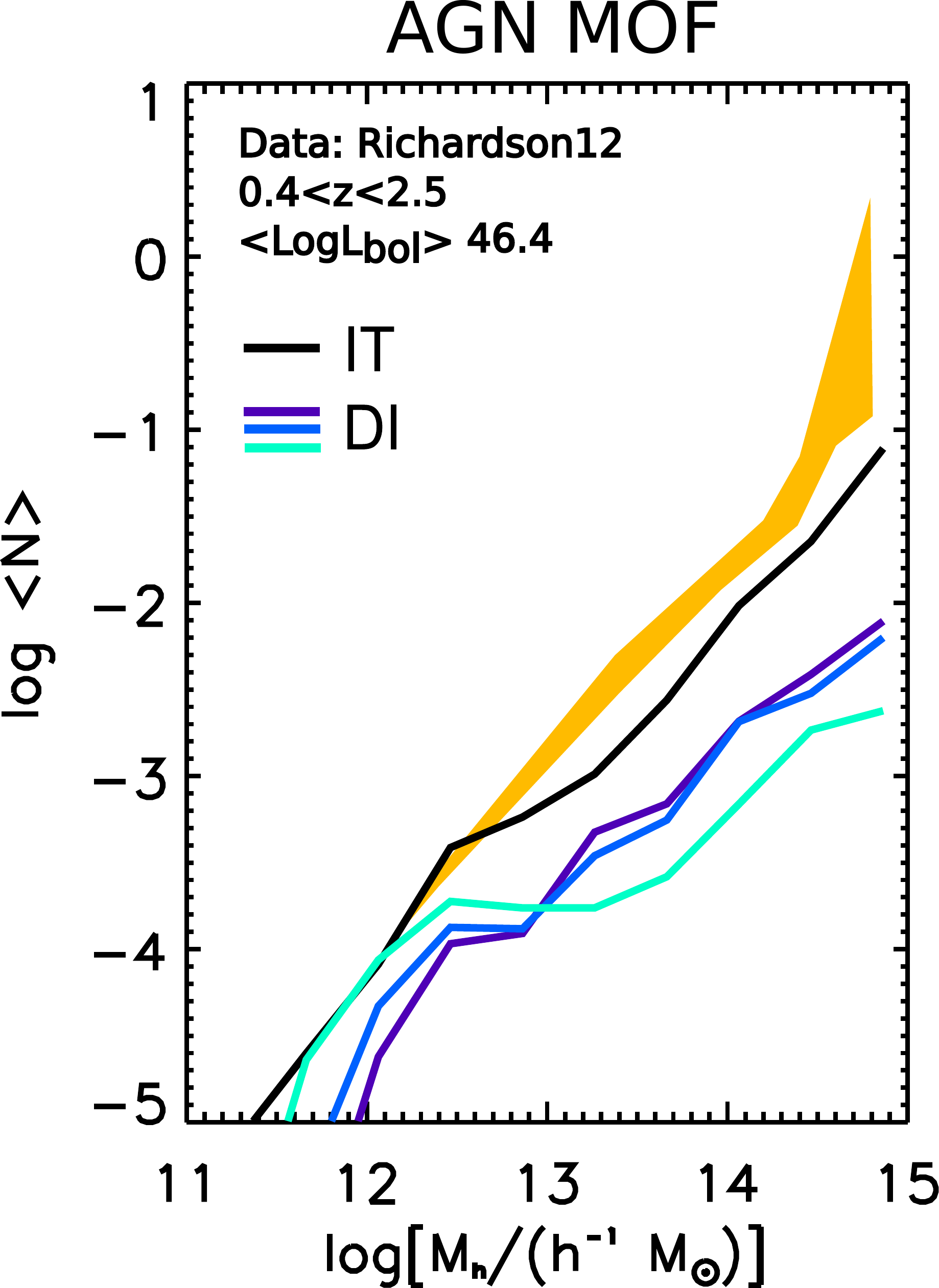}
\caption{AGN MOF as predicted by our SAM, compared with two different AGN MOF obtained from 2PCF measurements. The predictions of our model are represented by continuous lines: black line for the IT scenario, light blue, blue and purple for the DI scenario (the color code is the same of fig. \ref{dutycycle}). \textit{Left panels:} comparison with the AGN MOF from \citet{Shen2013}. The left, central and right panels refer to the total AGN MOF and the AGN MOF of central and satellite galaxies, respectively. The red dotted lines represent the AGN MOF best fit as obtained from \citet{Shen2013}; in every panel, the red-shaded region represents the 68.3$\%$ confidence interval obtained from the MCMC chain for central AGN, while the blue-shaded region is for satellite AGN. \textit{Right panel:} AGN MOF from \citet{Richardson2012}. The yellow-shaded region represents the 68.3$\%$ confidence interval for total AGN MOF as obtained from 2PCF measurements.}
%  \textit{Lower-right panel:} data from \citet{Richardson2013}. The continuous red line represents the the total AGN MOF as obtained from their 2PCF measurement, while the red-dotted lines are the AGN MOF for central and satellite AGN. The red-shaded is the 68.3$\%$ confidence interval for central AGN, while the blue-shaded region is for satellite AGN.}
\label{HOD2pcfa}
\end{figure*}
First, we compare with two AGN MOF $"$indirect$"$ measurements, that is, MOF obtained from clustering measurements. These AGN MOF have been obtained by the authors from the observed AGN 2PCF using the halo model framework: particularly, once having fixed a parametric expression for the AGN MOF, a Markov Chain Monte Carlo modeling of the AGN 2PCF is performed so as to probe the parameter space of the input AGN MOF.  These two AGN MOF measurements from \cite{Richardson2012} and \cite{Shen2013} are shown in fig. \ref{HOD2pcfa}, along with the predictions of our model.

%We provide in Appendix C a more accurate description of the different reference data sets considered in this work. 

The left three panels of fig. \ref{HOD2pcfa} represents the comparison with data from \citet{Shen2013}, who obtained the AGN MOF from a 2-point AGN-galaxy cross correlation function (CCF) measurement. They considered a subset of optically-selected quasars in the SDSS DR7 quasar catalog \citep{Abazajian2009,Schneider2010}, in the redshift range $0.3 < z < 0.9$ with median redshift $<z>=0.53$. The quasar sample is flux limited to $i<19.1$, with a median luminosity of $Log L_{bol} \approx 45.5$ erg $s^{-1}$ (roughly corresponding to $ \approx L_{knee}$). As for the galaxy sample, they considered the DR10 CMASS galaxy sample \citep{White2011} from the Baryonic Oscillation Spectroscopic Survey \citep{Schlegel2009}. 

The right panel represents the comparison with data taken from \citet{Richardson2012}. The authors firstly obtain the AGN 2PCF by combining an AGN sample from SDSS DR7 on large scale and a binary quasar sample take from \citet{Hennawi2006} relative to small scales; then, by applying the halo model formalism, the AGN MOF showed in fig \ref{HOD2pcfa} is obtained. The AGN large-scale sample span a redshift range $ 0.4 < z < 2.5$, with a median redshift of $<1.4>$, and is flux limited to $i < 19.1$. The binary quasar sample, relative to scales $r_p < 20$ h$^{-1}$kpc, has been obtained by detecting faint companions (flux limited to $i < 21.0$) around a parent sample of SDSS DR3 and 2QZ \citep{Croom2004} quasars, and is characterized by a slightly higher median redshift ($<z> =1.6$). The average luminosity of the sample is very high, approximately $log L_{bol} \approx 46.4$.

To compare with the AGN sample from \citet{Richardson2012}, we have chosen to use the redshift and luminosity cut associated with the large-scale sample ($0<z<2.5$ and $i<19.1$) for AGN in central galaxies, while AGN in satellites have been selected in the same redshift range but with a lower luminosity threshold ($i<21$), so as to mimic the selection cut of the small-scale sample. 

The comparison in fig. \ref{HOD2pcfa} broadly indicates that while the predictions for the IT scenario  agree generally well with data, the predictions for the DI scenario are able to match the observational constraints only in a limited range of dark matter halo masses, depending on the sample considered. 

In the three panels on the left we show the comparison with the AGN MOF from the optically selected AGN sample of \citet{Shen2013}. Both the total and the central and satellite components are displayed. Their sample has an AGN median luminosity $\approx L_{knee}$, thus it probes exactly the luminosity range where the DI scenario competes with the IT one in driving the bulk of AGN activity at such redshift. 

%The upper three panels are particularly useful, since they provide insights into the AGN MOF of both central and satellite AGN. Moreover, the optically selected AGN sample from \citet{Shen2013} has a median luminosity $\approx L_*$, thus probing exactly the luminosity range where our DI scenario competes with IT in driving the bulk of AGN activity at such redshift. 

Here, the DI scenario matches the total AGN MOF only up to $10^{13.5}$ h$^{-1} \msun$, then for higher halo masses it underpredicts the number of AGN. This is mainly due to the lack of AGN in central galaxies  in massive environments (central panel). Moreover, although DIs are able to trigger a discrete number of AGN in satellites, this is not enough to match the satellite AGN MOF, indicating that DIs might only contribute partially to the observed satellite AGN population.

The drop in central AGN directly follows from the results obtained in Sect. 4.1 and 4.3: DIs do not trigger AGN activity in high mass central galaxies, but rather preferentially occur in less massive satellite galaxies, with the fraction of AGN in satellite galaxies saturating to unity in massive clusters. The lack of central AGN in massive environments is present in all the comparisons we made, regardless of the luminosity and redshift cuts considered, thus it constitutes a robust feature of the DI model. 
 
 %the central dominant galaxy in massive environment (such as clusters) is not gas rich and disk-dominated, but it is rather gas poor and bulge-dominated, built through violent and frequent merging events through cosmic time. In these galaxies, our DI scenario is not able to efficiently trigger AGN activity (eq. \ref{efstathi} and \ref{hopkins}); 

%As for satellite AGN, the main problem is that at low redshift also the typical satellite galaxy residing in clusters is not completely gas rich and disk-dominated. 
At low redshift, binary aggregations between satellite galaxies and tidal stripping contribute to disrupt galaxy disks and consume gas reservoirs, lowering the probability of triggering  AGN activity by DIs. %(even if the effect is less dramatic with respect to central galaxies).

The comparison with \cite{Richardson2012} (right panel of fig. \ref{HOD2pcfa}) concerns an optically selected quasar sample with higher average bolometric luminosity with respect to the \cite{Shen2013} data.  The comparison favors the IT scenario, while for the DI mode only the prediction with the highest normalization (and hence highest AGN luminosity) reproduces the abundance of AGN for halo masses in the range $10^{11}-10^{12.5}$ h$^{-1} \msun$. The other two DI predictions, instead, underpredict the number of active galaxies for every dark matter halo mass. This follows from the results obtained in \cite{Menci2014}, where we showed that violent major mergers rather than DIs are the most likely mechanism for  triggering luminous quasars.

We remind that great attention must be paid in comparing our predictions with AGN MOF obtained from 2PCF measurements, especially when they involve wide redshift and luminosity intervals. %Indeed, this is a delicate process that relies on several assumptions that have to be validated. 
%As we stressed in Sect. 4.3, this approach basically ignores the effects of any possible redshift and luminosity evolution in the clustering strength of the AGN population. 
Indeed, the AGN MOF is usually interpreted as the MOF corresponding to the average luminosity and redshift of the sample; if present, any redshift and/or luminosity evolution of the AGN space density (see Sect. 4.2) might introduce systematic effects that might lead to a misleading interpretation of the results \citep{Richardson2012}.

%;moreover, since the halo model framework considers the mass function and the bias factor at the median redshift to obtain the AGN MOF, one might argue on the validity of such approach in case of wide redshift intervals \citep{Richardson2012}.% As noted before, this also hampers the comparison with theoretical predictions, making difficult to understand where (in terms of redshift and luminosity intervals) a particular scenario for AGN triggering fails in reproducing the observed data.

%clustering studies often report a single halo mass scale which may be difficult to interpret in the context of samples that span a wide range of halo masses

The second main problem concerns the assumed parametric expression used to obtain the AGN MOF from 2PCF measurements. The typical choice is to assume an AGN MOF similar to the galaxy MOF, motivated by the results of hydrodynamic cosmological simulations  \citep{DiMatteo2008,Chatterjee2012}: that is, a softened step function saturating to unity for central AGN and a rolling-off power law for satellite AGN. However, as pointed out by \citet{Shen2013}, there is a certain degree of degeneracy in the MOF parameterizations, especially at high halo masses and in the satellite MOF, which is mainly constrained by small scale clustering (difficult to probe) and affected by the assumptions concerning the distributions of satellite AGN inside dark matter halos.

Additional probes that could more directly than the 2PCF pin down the underlying host halo distribution will be very useful to set more secure constraints on the models. We discuss some of these probes in the next section.

\subsection{Comparison with AGN MOF - direct measurements and abundance matching approach}

\begin{figure*}
    \centering
\includegraphics[scale=0.23]{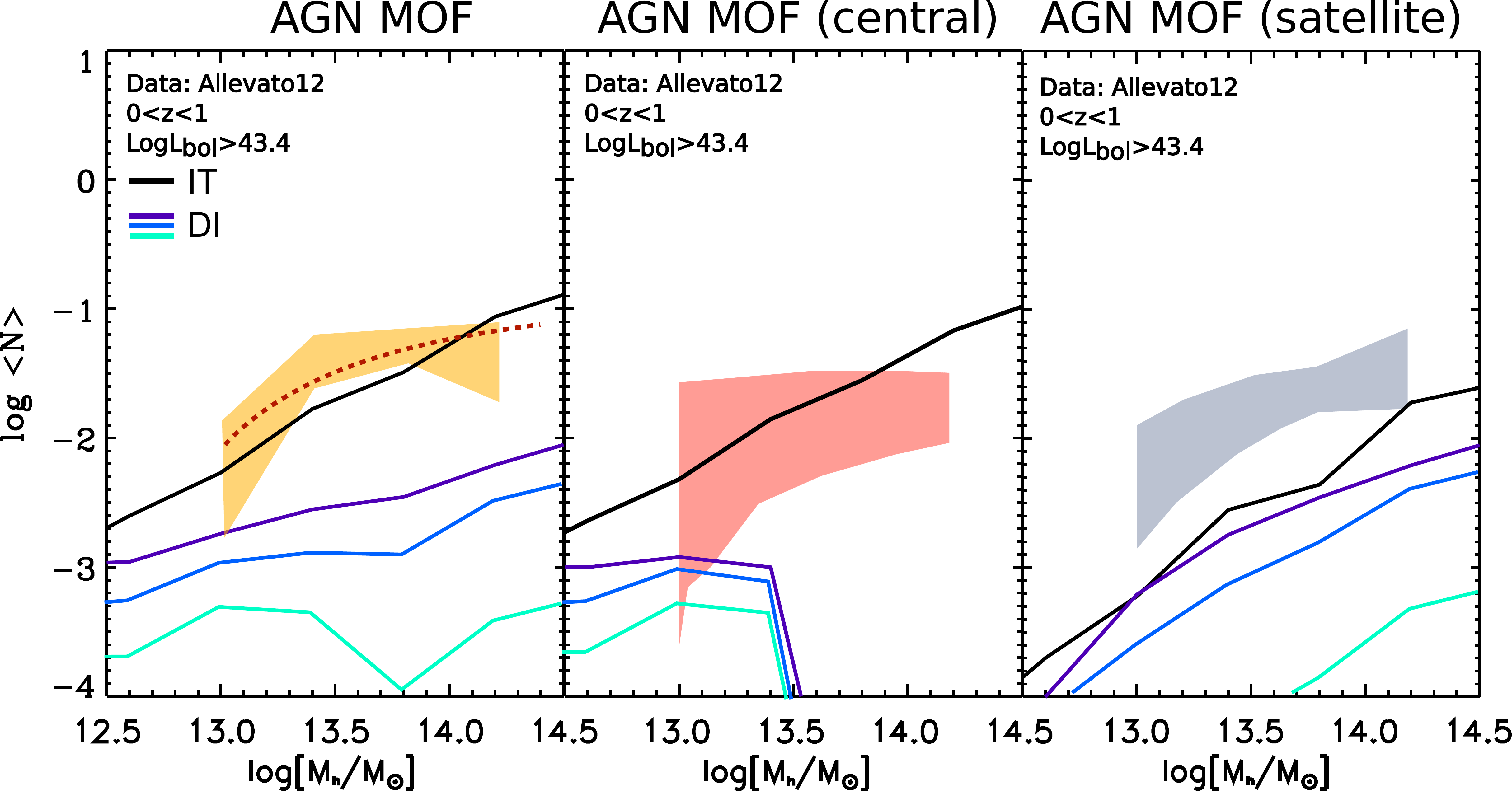}
\includegraphics[scale=0.23]{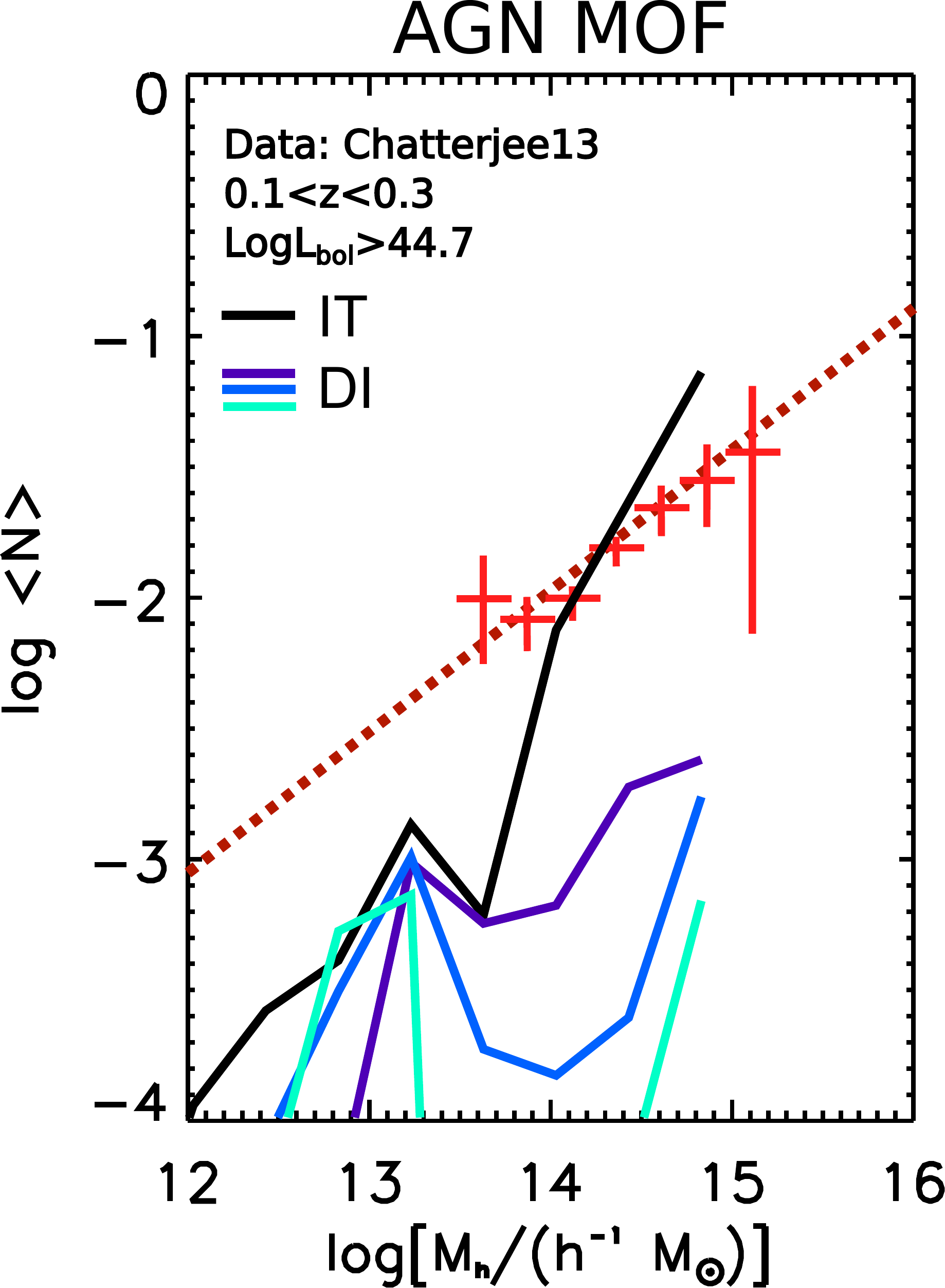}
\includegraphics[scale=0.23]{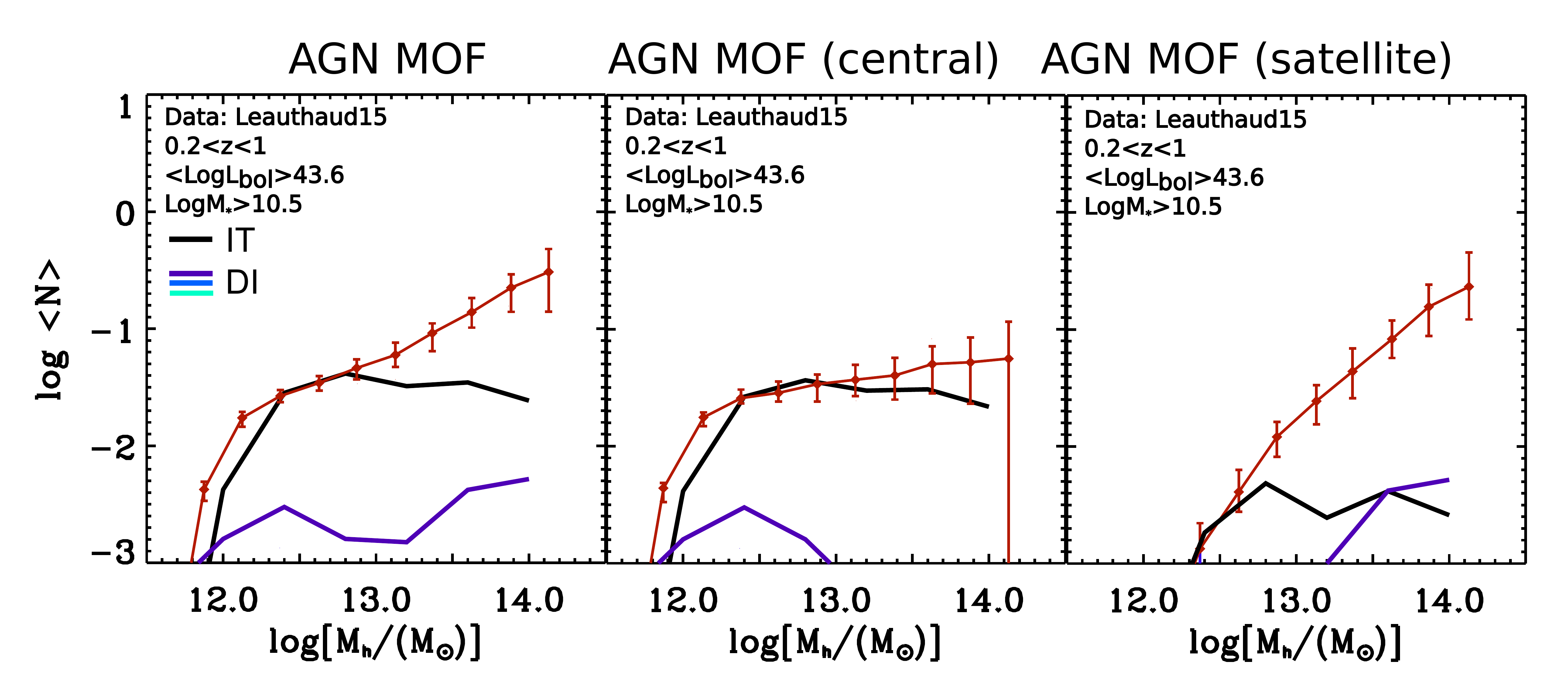}
\caption{AGN MOF as predicted by our SAM, compared with two AGN MOF directly measured, and with one obtained using an abundance matching approach. The predictions of our model are represented by continuous lines, with the same color code of fig. 1 .\textit{Upper-left panels:} comparison with the AGN MOF from \citet{Allevato2012}. The left, central and right panels refer to the total AGN MOF and the AGN MOF of central and satellite galaxies, respectively. The red dotted lines represent the total AGN MOF best fit as obtained from \citet{Allevato2012}; in every panel, the shaded region represents the 68.3$\%$ confidence interval obtained from their fit  (yellow for the total AGN MOF, red for centrals, blue for satellites). \textit{Upper-right panel:} AGN MOF from \citet{Chatterjee2012}. Red crosses represent the observational data, while the red dotted line their best fit power law model. \textit{Lower panels:} AGN MOF from \citet{Leauthaud2015}; The left, central and right panels refer to the total AGN MOF and the AGN MOF of central and satellite galaxies, respectively.}
\label{AGNHODInd}
\end{figure*}

In fig. \ref{AGNHODInd} we compare the predictions of our SAM with two direct AGN MOF measurements, taken from \citet{Chatterjee2013} and \citet{Allevato2012}, and with the AGN MOF obtained from \citet{Leauthaud2015} using an abundance matching approach. %The description concerning the samples and the selection cuts used to reproduce the observational data are given in Appendix C.2. % to compare with these AGN HOD we took into account all the selection cuts concerning the observed samples.

The upper left panels of fig. \ref{AGNHODInd} show data from \citet{Allevato2012}, who considered  41 XMM-COSMOS \citep{Salvato2011} and 17 C-COSMOS \citep{Elvis2009} AGN with photometric and spectroscopic redshift at $z \leq 1$, and associated them with member galaxies of X-ray detected galaxy groups in the COSMOS field \citep{Finoguenov2007} to obtain the AGN MOF. Group masses have been assigned from an empirical mass-luminosity relation \citep{Leauthaud2010}, and are in the range $Log M_h \sim 10^{13} - 10^{14.2}\msun $. AGN have been assigned to central and satellite galaxies by cross-matching the AGN sample with galaxy membership catalogs \citep{Leauthaud2007,George2011}. The authors also tried to take into account the luminosity and redshift dependence of their sample, by properly correcting the AGN MOF with a weight factor. Particularly, their weight factor considered both the fact that they were including AGN with luminosity lower than $L_{X(0.1-2.4KeV)} < 10^{42.2}$ ergs$^{-1}$ (which is the luminosity threshold at the highest redshift probed by their sample) and the redshift evolution of the AGN number density. To compare with their data, we selected AGN in the redshift range $0<z<1$ with $L_{X(0.1-2.4KeV)} > 10^{42.2}$ ergs$^{-1}$ and corrected the AGN number density at different redshifts with a weight factor accounting for the redshift evolution similar to the one adopted by \citet{Allevato2012}. Given the low redshift range and the high dark matter halo mass probed here, no additional cut on the host galaxy stellar mass have been applied, since it would not affect the AGN MOF.

\citet{Chatterjee2013} (upper right panel) used SDSS DR7 quasars and galaxy group catalogs from MaxBCG sample \citep{Koester2007} to empirically measure the MOF of quasars at low redshift. They used a subsample of SDSS quasars in the redshift range $0.1 <z< 0.3$ and absolute magnitude $M_i < -22$ ($Log L_{bol} \gtrsim 44.7$). The MaxBGC sample contains galaxy groups and clusters with velocity dispersion greater than $\sim$ 400 km/s, in the redshift range $0.1 <z< 0.3$. Groups and clusters are obtained by identifying first the brightest cluster galaxies (BCGs), and then by assigning group members by selecting those galaxies that lie within $R_{gal200}$ (defined as the radius within which the galaxy density is 200 times higher than background) of a BCG. The group/cluster mass is then computed using the modified optical richness method of \citet{Rykoff2012}, with a mass error equal to 33$\%$. Quasars are then associated with groups and clusters depending on their positions.

Ultimately, in the lower panels we show the comparison with \citet{Leauthaud2015}, who used a novel approach to obtain the AGN MOF, based on the stellar-to-halo mass relation (SHMR). The authors focused on a sample of moderately luminous AGN from the XMM-COSMOS and C-COSMOS X-ray catalogs. The sample span the redshift interval $0.2 < z < 1$, with a median redshift of $<z>=0.7$. Besides the luminosity cut due to the limiting flux of XMM-COSMOS and C-COSMOS surveys, the sample has been further constrained in luminosity to $10^{41.5}$ erg s$^{-1}< L_{X,0.5-10 keV } < 10^{43.5}$ erg  s$^{-1}$ and has a median luminosity of $<L_X> = 10^{42.7}$ erg s$^{-1}$ (corresponding roughly to $<L_{bol}> \sim 43.6$). The authors also imposed a lower host galaxy stellar mass limit of $Log M_* >10.5$ (with host stellar masses being obtained from a SED-fitting procedure), resulting in a median stellar host mass of $<M_*> = 1.3 10^{11} \msun$. The authors firstly showed the similarity in the host halo occupation of active and inactive galaxies of a given stellar mass, by using galaxy-galaxy lensing measurements; then, through the use of the SHMR and knowing the stellar mass of AGN host galaxy, accurate predictions of the AGN MOF have been made.

Similarly to the results obtained in the previous section, the IT scenario agrees well with data, while DIs generally provide a poor match.

In the comparison with the optically selected AGN sample from \citet{Chatterjee2013}, only galaxy interactions are able to trigger the observed abundance of AGN in massive halos with $M_h > 10^{13.5}$ h$^{-1}\msun$, with DIs being disfavored as main fueling mechanism, due to the lack of AGN in central galaxies. %Unfortunately, less massive halos ($M_h < 10^{13.5}$ h$^{-1}\msun$) are not probed here, even if the authors provide an extrapolation of the AGN MOF also at lower masses, once having fitted their data. It is nonetheless worth noting that the predictions of our SAM concerning halos $M_h < 10^{13.5}$ h$^{-1}\msun$ provides similar abundance for both the IT and DI scenario (even if the predictions are particularly noisy due to the paucity of AGN at such low redshift). This would suggest that in such luminosity and redshift ranges DIs might be as effective as IT in triggering AGN activity.

The other two comparisons concern relatively low luminosity ($L\lesssim L_{knee}$), X ray-selected AGN samples at $z<1$, therefore they are particularly useful to further test only the IT scenario. Indeed, in such luminosity and redshift range, our DI model does not provide the correct abundance of AGN to match the observed AGN luminosity function (they are most effective in triggering more luminous $L\approx L_{knee}$ AGN).We therefore expect an underprediction of the AGN MOF for the DI scenario at every halo mass.

In both the comparisons with data from \citet{Allevato2012} and \citet{Leauthaud2015}, our IT scenario generally provides a good match (i.e. within the $68.3\%$ confidence interval) with the observed central MOF over the full range of masses probed by data, but tends to underpredict the abundance of AGN in satellite galaxies.%, in particular when comparing with \citet{Leauthaud2015}. 

In the comparison with \citet{Allevato2012} data, despite the shape of the satellite MOF predicted by our SAM is similar to the observed one, the IT scenario is only able to account for $20\%$ of the observed abundance of AGN in satellite galaxies, with $f_{sat} \approx 10\%$ (to be compared to $f_{sat} \approx 60\%$ of the Allevato et al. sample). The mismatch with observational data is significant at $\sim 1.9 \sigma$ level (for reference, we have a $\sim 2.3 \sigma$ level  disagreement for the DI prediction with $\alpha=2$).  Concerning the comparison with \citet{Leauthaud2015},  the IT model predicts a $f_{sat} \approx 8\%$ against $f_{sat} \approx 18\%$ inferred by the authors. The discrepancy becomes particularly evident at high mass where the number of AGN satellite is underpredicted by more than one order of magnitude.

This represents a test to what we highlighted in Sect. 4.3: the low fraction of AGN in satellite galaxies of the IT scenario seems to conflict with observational results, which favor higher AGN satellite fractions. 
%($\sim 60\%$ in the sample of \citealt{Allevato2012}, $\sim 18\%$ in the sample from \citealt{Leauthaud2015}).
It is possible that other mechanisms (probably other $"$in-situ$"$ processes, such as stochastic accretion or cold flows) besides galaxy interactions might contribute to the triggering of AGN activity in satellite galaxies, at least for low-to-intermediate luminosity AGN.

However, it is worth noting that the lensing signal from which \citet{Leauthaud2015} obtain their AGN MOF is poorly sensitive to low satellite fractions: even if the fiducial satellite fraction inferred by the authors is equal to $f_{sat}=18\%$, they note that reducing it to $f_{sat}=0\%$ has only a little effect on the lensing signal, making difficult to firmly exclude values lower than the fiducial one. In particular, if there were some kind of difference in the clustering strength of normal galaxies and AGN (at least concerning satellite AGN in massive environments), this might affect their estimate of the satellite AGN MOF, possibly reconciling our predictions with data.  The low AGN satellite fraction for the IT scenario might also partially be the result of an incorrect estimate of satellite galaxies stellar mass from our SAM, since lowering the stellar mass selection cut would result into an increment of the fraction $f_{sat}$. Furthermore, the value of the satellite fraction inferred by \citet{Allevato2012} and \citet{Leauthaud2015} are very different from one another, suggesting that $f_{sat}$ might be sensible to selection effects (AGN luminosity, host galaxy stellar mass, redshift) and to the environment.  Even if the comparisons presented in this section represent a more reliable test of the AGN satellite MOF since data are not biased by parameterization degeneracy, more data are needed to accurately test the efficiency of our IT scenario in triggering AGN activity in satellite galaxies.

\begin{figure*}
    \centering
    
\includegraphics[scale=0.25]{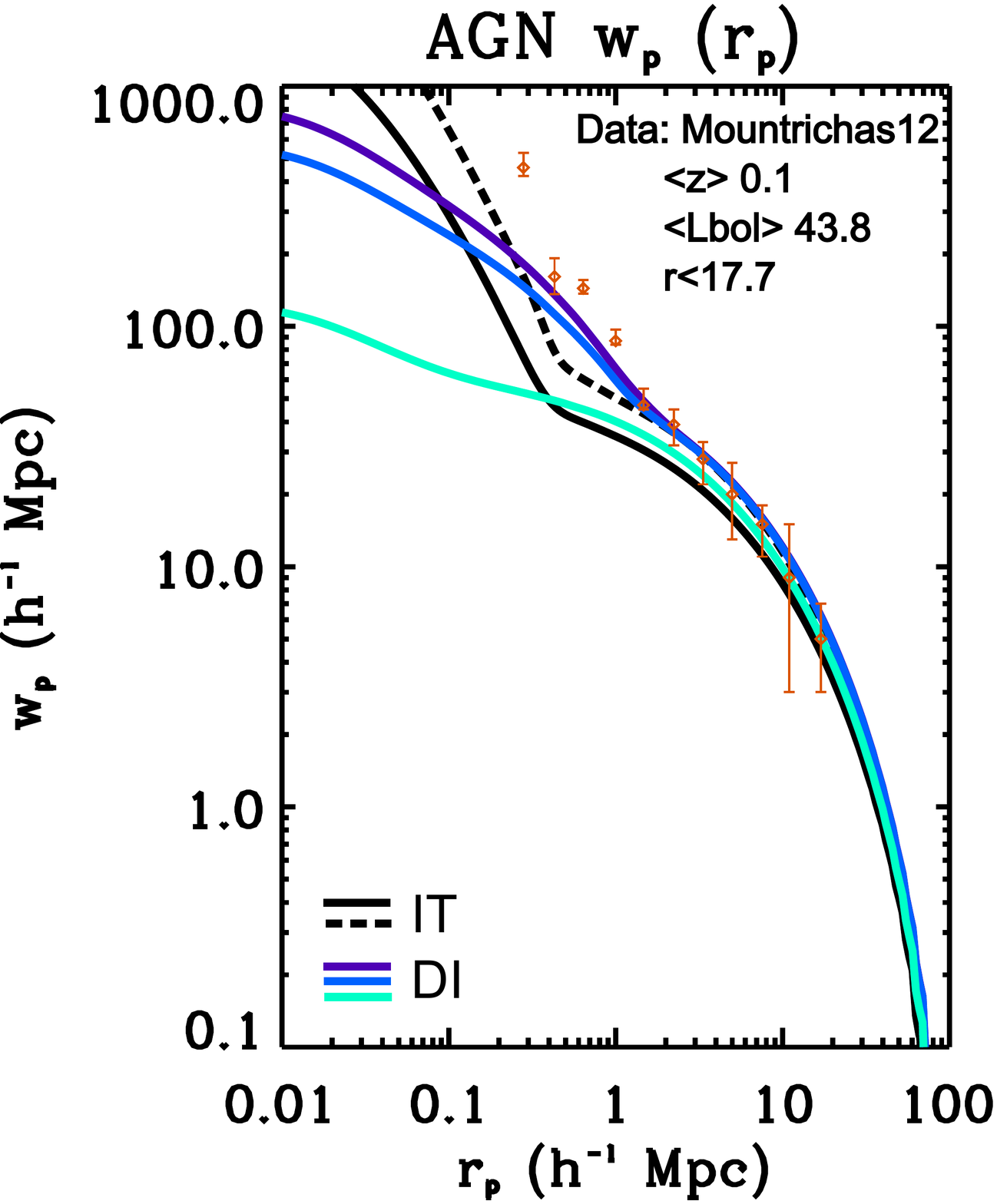}
\includegraphics[scale=0.25]{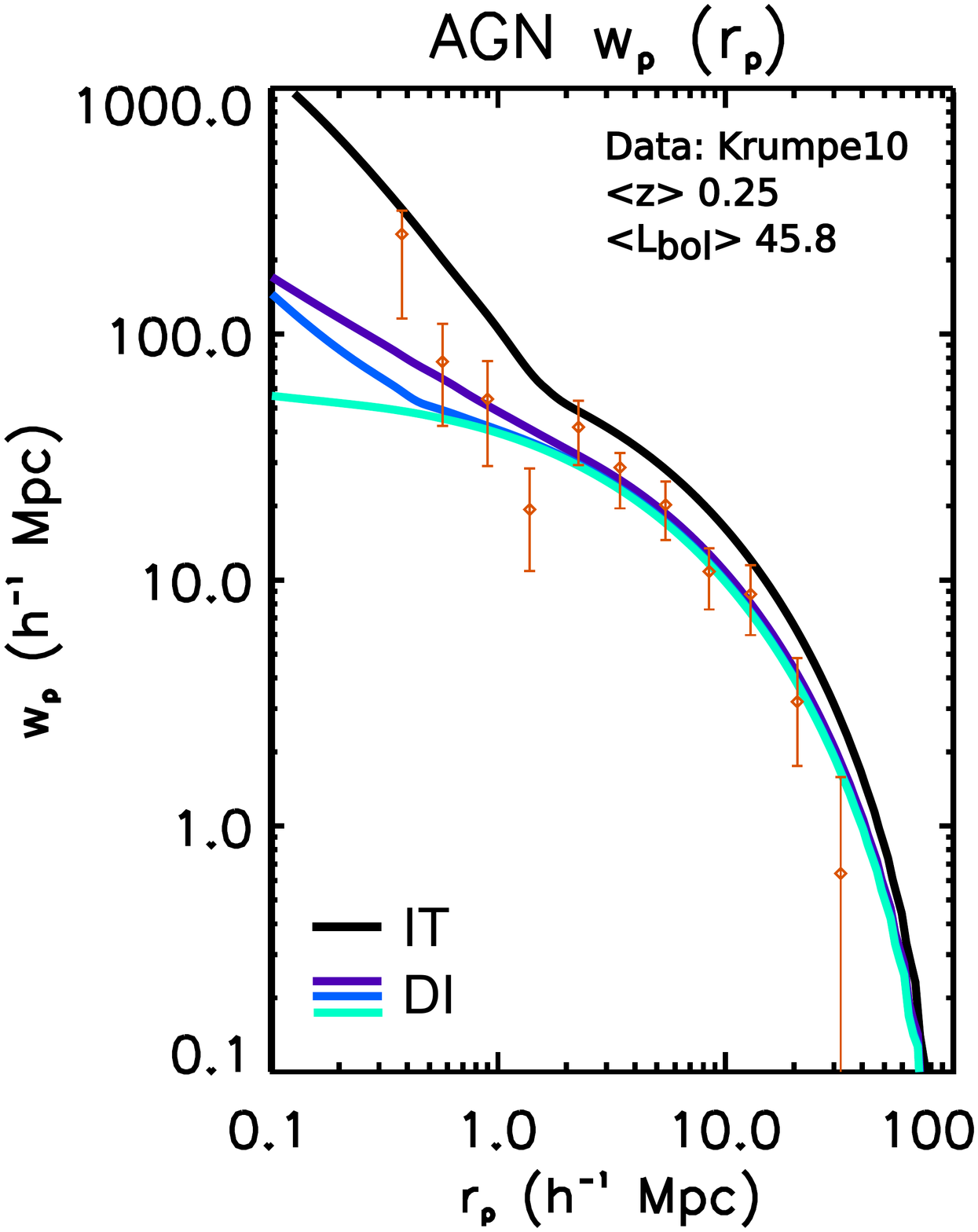}
\includegraphics[scale=0.25]{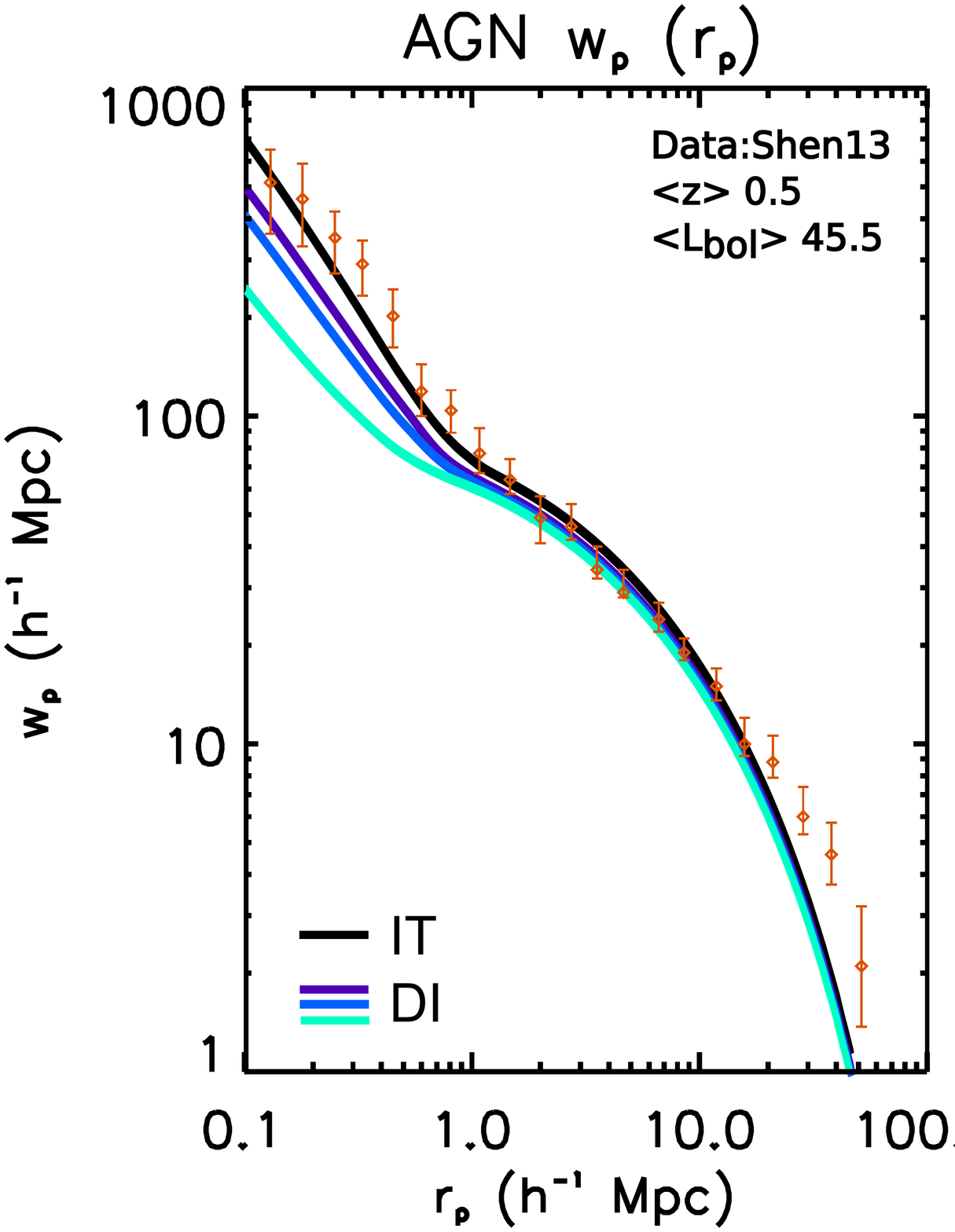}
\includegraphics[scale=0.25]{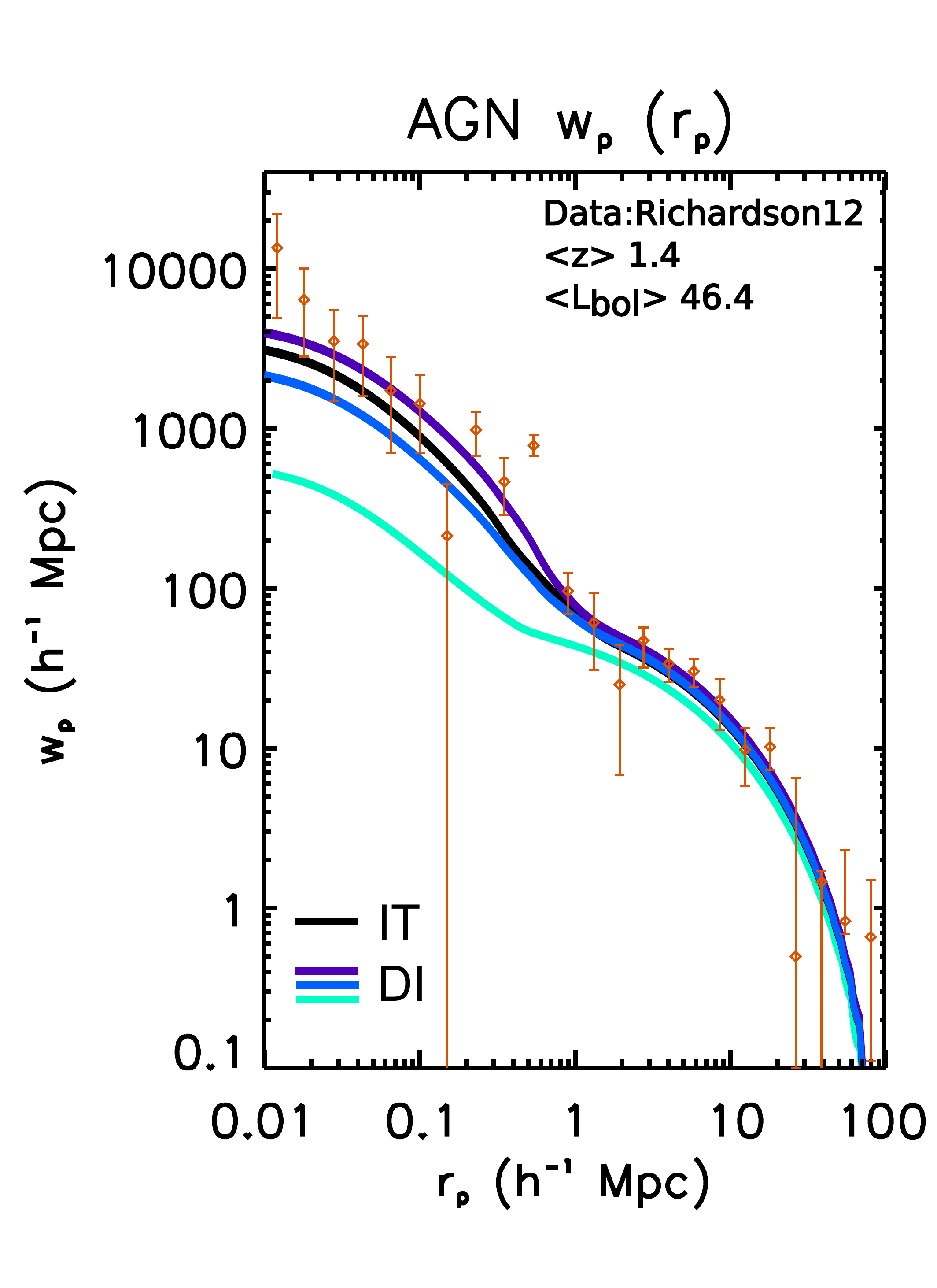}
\includegraphics[scale=0.25]{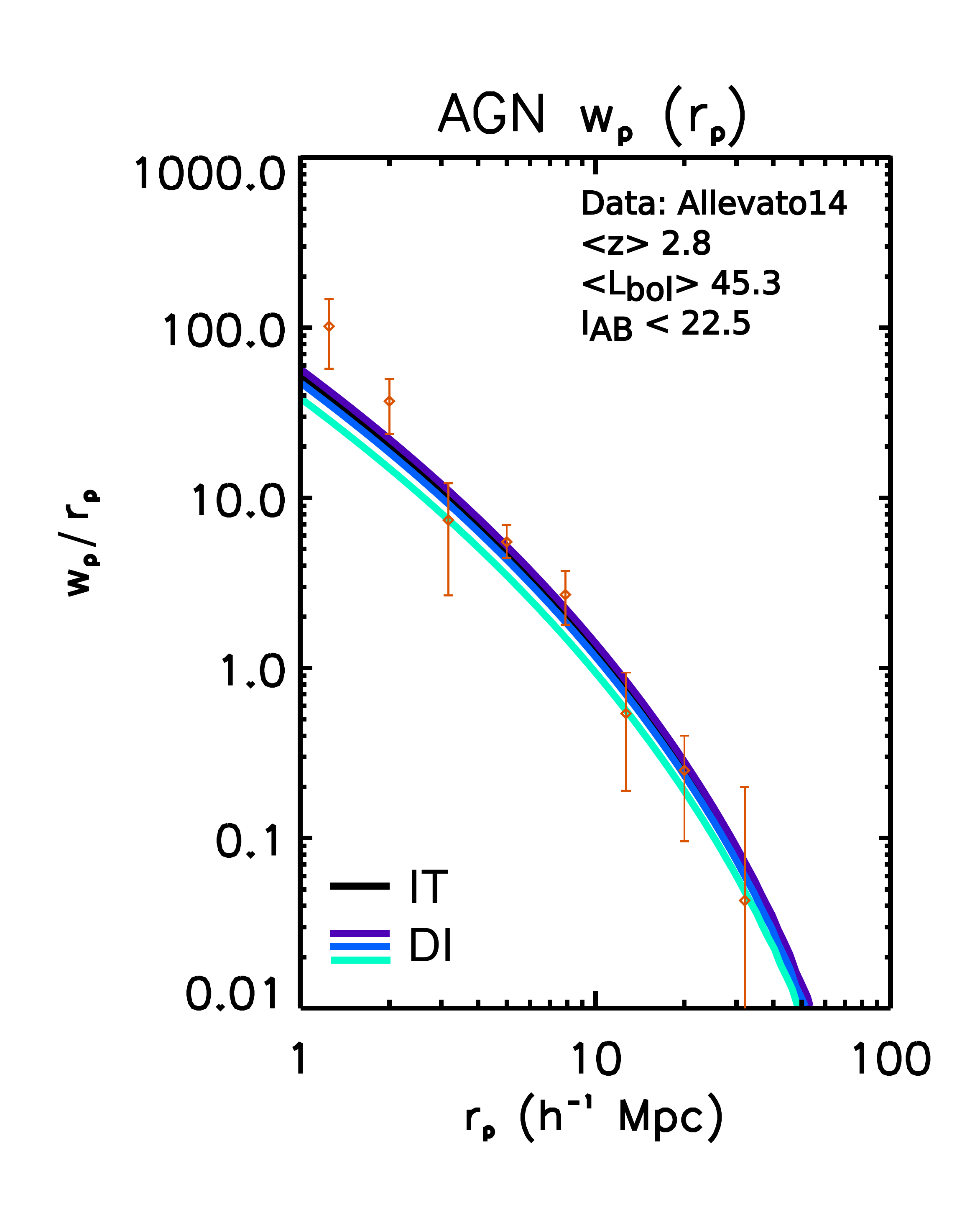}
\includegraphics[scale=0.25]{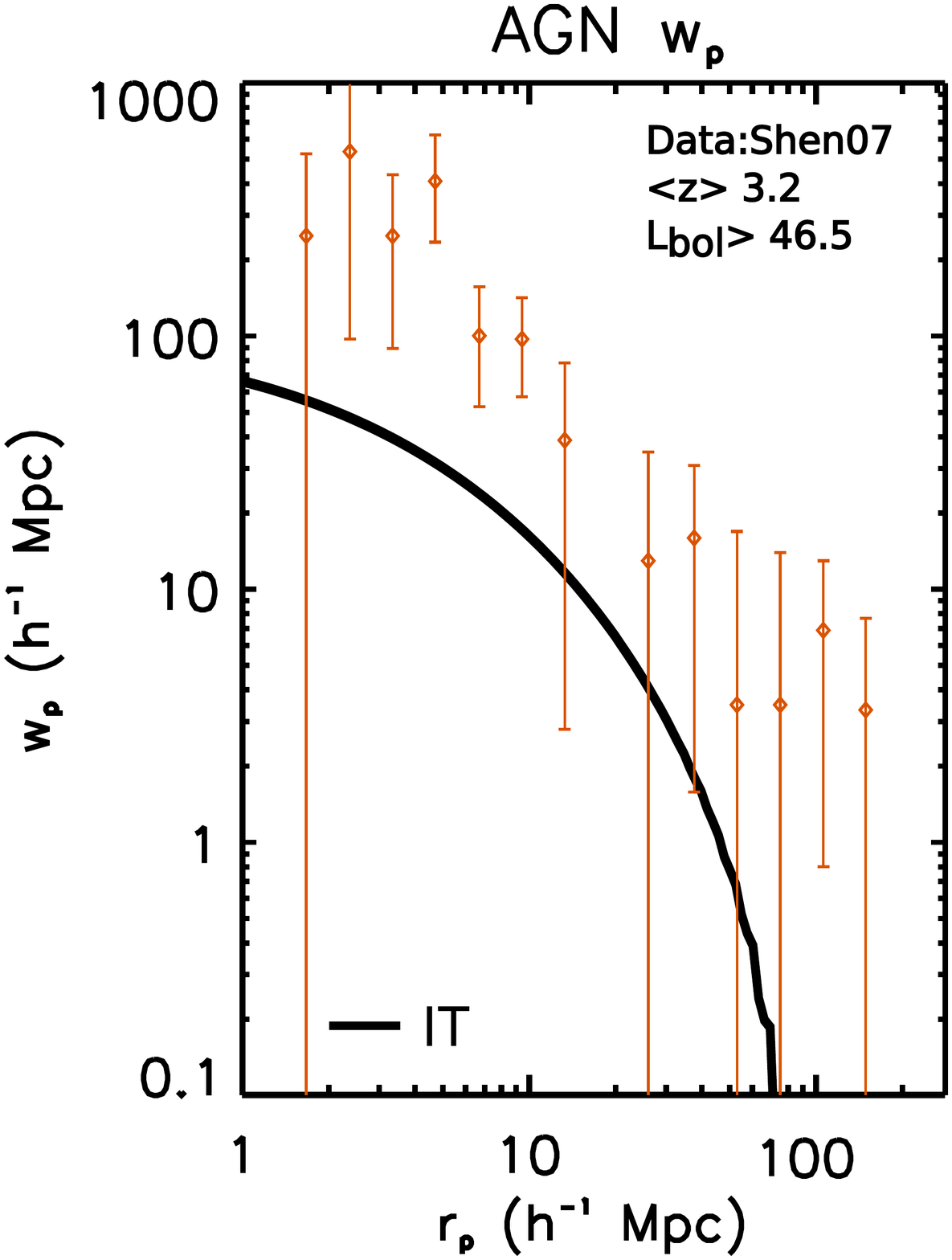}

\caption{AGN ACF and CCF as predicted by our SAM, compared with several 2PCF measurements from  \citet{Shen2007,Krumpe2010,Mountrichas2012, Richardson2012,Richardson2013,Shen2013,Chatterjee2013,Allevato2014} (see text for further details). The predictions of our model are represented by continuous lines, with the same color code of fig. 1. In the comparison with \citet{Mountrichas2012}, the black dashed line refers to the prediction for the IT scenario with a slightly higher threshold ($L_{X(2-10 KeV)} > 10^{42}$ erg s$^{-1}$) on the AGN luminosity (see text for more details).}
\label{AGNHODCF}
\end{figure*}

\subsection{Comparison with AGN correlation function}
Last, we proceed with the comparison with a number of observed AGN 2PCF at different redshifts. The AGN 2PCF provides two pieces of information: the study of the bias factor, implicit in the 2-halo term, pinpoints the average halo mass where the AGN sample resides, while the 1-halo term  provides insights into the small-scale clustering, which is related to the number of AGN pairs and satellite fraction inside dark matter halos.
Note that in this section we start from a model predicted MOF that\textit{ uniquely }specifies the implied 2PCF.
%Even if the piece of information concerning the AGN distribution over the full range of DM halo masses provided by the AGN HOD is lost with the 2PCF, there are at least two main advantages in comparing with such clustering measurements. Firstly, there exist a substantial number of measured AGN 2PCF in literature, allowing to test the predictions of our SAM over a wide range of different redshifts and luminosities; secondly, 2PCF measurements - differently from the AGN MOF extrapolated from the 2PCF - are univocal and do not suffer from any parameterization degeneracy.  

In fig. \ref{AGNHODCF} we show the comparison of our models with five AGN auto correlation functions and one AGN-galaxy cross correlation function. The AGN-galaxy cross correlation function \citep{Shen2013} and one AGN auto correlation function \citep{Richardson2012} concern the same data sets relative to the AGN MOFs discussed in Sect. 5.1. To these we added the comparisons with four AGN auto-correlation function from \citet{Krumpe2010}, \citet{Mountrichas2012}, \citet{Allevato2014}, and  \citet{Shen2007}. To obtain the 2PCF displayed in fig. \ref{AGNHODCF}, we made use of the bias factor and the halo mass function at the median redshift of the sample. 
% The other four AGN auto correlation functions are based on two low redshift X-ray selected AGN samples from \citet{Krumpe2010} and \citet{Mountrichas2012}, a high redshift X-ray selected AGN sample from \citet{Allevato2014}, and a high redshift luminous quasars sample from \citet{Shen2007}. More details about each data set we compared with are given in Appendix C.3. 

%The comparisons displayed in fig. \ref{AGNHODCF} concern 4 AGN auto-correlation functions and one AGN-galaxy cross-correlation function. The AGN-galaxy cross-correlation function \citep{Shen2013} and one AGN auto-correlation function \citep{Richardson2012} concern the same data sets relative to the AGN MOFs discussed in Sect. 5.1. and Appendix C1. 
%To these we added the comparisons with four AGN auto-correlation function from \citet{Krumpe2010}, \citet{Mountrichas2012}, \citet{Allevato2014}, and  \citet{Shen2007}. 

\citet{Mountrichas2012} investigated the clustering properties of very low redshift ($0.03<z<0.2$, $<z> \sim 0.1$) X-ray AGN from the XMM/SDSS survey \citep{Georgakakis2011}. The sample includes low luminosity AGN with $L_{X(2-10 KeV)}>10^{41}$ erg s$^{-1}$ and has a median luminosity of $<L_{X(2-10 KeV)}>10^{42.6}$ erg s$^{-1}$ (corresponding to $<L_{bol}> \sim 43.8$). The optical identification from the SDSS sample introduces an additional cut on the host magnitude of $r < 17.7$.  The sample from \citet{Krumpe2010} concerns a low redshift ($0.16 < z < 0.36$) sample of X-ray selected broad line AGN from the ROSAT All-Sky Survey (RASS, \citealt{Voges1999}) with SDSS optical counterparts. The survey presents a typical flux limit of $\sim 10^{13}$ erg cm$^{-2}$ s$^{-1}$ (0.1-2.4 keV), with a median luminosity of the sample is $<L_{X(0.1-2.4 KeV)}> =1.5 10^{44}$ erg s$^{-1}$ ($<Log L_{bol}> \approx 45.8$). The RASS/SDSS AGN sample is reasonably complete for magnitude 15 < i < 19. %The AGN 2PCF shown in fig. \ref{AGNHODInd} has been extrapolated by the authors from the measured cross-correlation function with a sample of Luminous Red Galaxies (LRGs) from the SDSS DR7. 
In this comparison, we have selected only AGN with $n_H<10^{22} cm^{-2}$.

The other two comparisons concern two high redshift clustering measurements, from \citet{Allevato2014} and \citet{Shen2007}. \citet{Allevato2014} studied the clustering properties of a sample of 252 C-COSMOS AGN and 94 XMM-Newton AGN detected in the soft band with spectroscopic or photometric redshift.  The redshift interval considered here is $2.2<z<6.8$, with median redshift $<z>=2.8$; the median luminosity is equal to $<L_{bol}> \approx 10^{45.3}$ ergs$^{-1}$ , hence corresponding to the intermediate-luminosity regime. To compare with this data set, we have selected AGN according to the Chandra soft band flux limit. Since the C-COSMOS sample is 83$\%$ spectroscopically complete at $I_{AB} <22.5$ \citep{Civano2012}, we have applied this other cut to the magnitude of our AGN host galaxies. \citet{Shen2007} considered instead an optically-selected high redshift sample of luminous AGN from SDSS DR5 \citep{Adelman-McCarthy2007}, in the redshift interval $2.9 < z < 5.4$ (median redshift $<z> = 3.2$).  The whole AGN sample is flux limited to $i < 20.2$ (which corresponds to $Log L_{bol}>46.5$ at $z \approx 2.9$).

The data are generally reproduced quite well by our models, except for \citet{Shen2007}, where we have an underprediction of the normalization of the 2-halo term. We note here that only the comparison with the IT scenario is shown: this sample concerns high luminosity, high redshift AGN, and we know from \citet{Menci2014} that the DI scenario is not able to trigger such AGN. The high correlation length predicted by \citet{Shen2007} measurements, together with the rareness of massive halos at high redshift, have profound implications on the quasars properties (e.g.,\citealt{White2008,Bonoli2010,Shankar2010}). In particular, \citet{Shen2007} data would imply a very high AGN duty cycle and AGN radiation efficiency, together with a tight $L_{bol}-M_h$ relation, which are rather extreme requirements that are difficultly met by our SAM. Given the low number of luminous quasars probed by their sample, it is also possible that their correlation length is overestimated, or its statistical error underestimated. 

Recently, \citet{Eftekharzadeh2015}, studying a sample of spectroscopically confirmed optically selected quasars from the BOSS survey, performed a more precise estimation of the quasars bias at high redshift ($2.2<z<2.8$). They obtain a bias of $b=3.55\pm0.10$, corresponding to $Log M_h \sim 12.3$, which is sensibly lower than the value obtained by \citet{Shen2007} ($b\sim8$ at $z\sim3$), although it is important to keep in mind the two samples do not completely overlap in terms of redshift and luminosity intervals (\citealt{Eftekharzadeh2015} data probe lower luminosities). Accounting for \citet{Eftekharzadeh2015} data selection cuts following the basic prescriptions described at the beginning of Sect. 5, we verified that our models (both IT and DI) predict a value for the bias of $b \sim 2.8$, still lower than $b=3.55\pm0.10$. It might be that given the high redshift range and the luminosity of their sample, the prescriptions we assumed to mimic observational biases in case of optical surveys might be inadequate (above all, stellar contamination becomes relevant at these redshifts), and more precise selection cuts on the hosts might be needed.

In all the other cases, the AGN correlation functions are reproduced quite well by our predictions; more importantly, there is basically no appreciable difference in the 2-halo term predicted for the two scenarios. A small difference between the two models only appears in the low redshift comparison with \citet{Krumpe2010}, with the IT scenario being characterized by a slightly higher bias factor.  Even if the AGN MOF for the DI scenario always shows a drop at high halo masses, this has little impact on the average halo mass and bias factor, mainly  because the more massive halos are rare, especially at high redshift, and their contribution to the normalization of the 2-halo term is almost negligible. We note that a possible drop in the central AGN MOF in massive halos might have little or no effect on the AGN 2PCF at high redshift has already been anticipated by other authors \citep{Richardson2012,Kayo2012}.

The results obtained in this section and in Sect. 4.2 seem to indicate that the sole analysis of the large scale clustering (i.e. bias factor) constitutes a poor constraint of the DI and IT modes considered here,  especially at high redshift. Even if the AGN MOF for the two scenarios show some differences, this does not translate in an appreciable difference in the AGN 2PCF. The use of large intervals in terms of redshift and AGN luminosity to infer the AGN 2PCF further contributes to make the situation more complex. 

It is also interesting to note that the AGN 2PCF measurements (as well as the AGN MOF in previous sections) we compared with concern both X-ray and optically selected AGN samples. Differences in the bias factor inferred from optical and X-ray AGN surveys have been often interpreted as a clear sign of different triggering mechanisms at play. While it is true that different triggering mechanisms might be characterized by a different clustering strength, our analysis suggests that the differences in the bias factor of surveys carried out at different wavelengths might be driven mainly by different selection cuts in terms of AGN luminosity, redshift range, host galaxy properties (e.g. stellar mass), which overcome a possible signature of different triggering mechanisms at play. In this respect, clearer selections on the host galaxy properties and smaller luminosity/redshift intervals might emphasize any difference in the clustering strength of separate SMBH feeding modes.

A similar result has been obtained by \citet{Hopkins2014}: the authors compared semi-empirical models for AGN fueling based on both mergers and stochastic accretion, in which the fueling in the latter is essentially a random process arising whenever dense gas clouds reach the nucleus. They found that the stochastic fueling dominates AGN by number, though it accounts for just $10\%$ per cent of BH mass growth at masses  $10^8 \msun$. In total, fueling in disky hosts accounts for $30\%$ of the total AGN luminosity density/BH mass density. They also argue that the large scale clustering is not a sensitive probes of BH fueling mechanisms, in agreement with our results.
 
Concerning the comparison with the small scale clustering, the situation is a little complex: small scale clustering is difficult to probe, and also the procedure we used to obtain the 1-halo term might be susceptible to the small number of AGN pairs (see also Appendix A). Nevertheless, we might expect some differences between the DI and IT scenarios. Due to the shortage of central active galaxies, DIs should be characterized on average by a lower relative number of $N_{cen}N_{sat}$ pairs, and unless they are  replaced by enough $N_{sat}N_{sat}$ pairs, this should imply less power at small scales.

The data set that probes the smallest scales is represented by \citet{Richardson2012}. Here only the DI prediction with the highest normalization of the inflow shows a clear departure from small-scale data and from the IT prediction; however, the good match of the two other DI predictions should not considered as conclusive. As shown by the AGN MOF in Sect 5.1, these two models underpredict the number density of luminous AGN at every halo mass, also missing the very luminous AGN in central galaxies in $10^{12}\sim10^{13}h^{-1}\msun$ dark matter halos. This gives more power to small scales, but at the same time implies that the two predictions can only account for a small fraction of the luminous AGN sample we compared with. 
 
Similarly, the better match of the two DI predictions with $\alpha=2$ and $\alpha=5$ in the comparison with \citet{Mountrichas2012} is not decisive: here the observational sample includes extremely faint AGN (down to $L_{X(2-10 KeV)}\sim10^{41}$ erg s$^{-1}$), thus probing the luminosity regime where the DI scenario  does not trigger enough AGN activity to match the AGN luminosity function. As for the IT scenario, we note it shows less power at small scales than what we should have expected from observational data. This might be due to the excess of very faint AGN for the IT scenario at low reshift (see Sect. 4.1), which mainly reside as centrals in small dark matter halos (see Sect 4.2). In fig. \ref{AGNHODCF} we show that selecting slightly more luminous AGN ($L_{X(2-10 KeV)} > 10^{42}$ erg s$^{-1}$, rather than $L_{X(2-10 KeV)} > 10^{41}$ erg s$^{-1}$) improves the match at small scales, while little affecting the clustering at large scales.

In the two other comparisons that partially probe the small scales regime (\citealt{Krumpe2010} and \citealt{Shen2013}) concerning moderately luminous AGN ($L \sim L_{knee}$), the DI mode shows less power at small scales with respect to the IT model, especially in the former comparison. Indeed, at low redshift the drop in the high-mass end of central active galaxies of the DI scenario becomes more relevant, since massive environments are more common, substantially affecting the small scale clustering. In the comparison with \citet{Shen2013}, the cross-correlation with the galaxy sample partially mitigate the differences between the two models. % though DIs still display slightly less power at small scales.

Lastly, we note that in order to compute the 1-halo term, we have always assumed the spatial distribution of satellite AGN to follow the NFW profile \citep{Navarro1997} with concentration-mass relation described as \citet{Zheng2007}. We checked that assuming a higher concentration parameter (up to 5 time larger), as suggested by some observational studies (e.g., \citealt{Lin2007}), though gives slightly more power at small scales, do not drastically affect our predictions.

More data are needed to further test our models in reproducing small scales clustering, also at higher redshift (where DIs increase their efficiency and might reverse the trend with the IT mode). The 1-halo term can provide information on the efficiency of distinct triggering modes, since it is quite sensible to differences in the relative distribution of central and satellite AGN. However, its analysis should always be followed by the study of other complimentary observables (e.g., MOF directly measured, pairwise velocity distribution), in order to give more stringent constraints on the AGN satellite fraction and to break the degeneracies in the HOD modeling.

\section{Discussion}
%In this section we further discuss our results, focusing on the AGN HOD parametric form provided by our models and on any redshift or luminosity dependence of AGN clustering. % and and test their robustness against a number of different assumptions.

\subsection{DI scenario: robustness of results}
One of the main prediction we obtained for the DI scenario is that DIs fall short in triggering central AGN in massive halos, with AGN in satellites not being enough to match the observational constraints. As a first test, we check whether the lack of DI AGN in massive halos might be due to observational statistical errors.
%We know that the DI scenario is particularly effective in triggering $L_*$ AGN: DIs are not able to reproduce the bright end of the luminosity function (LF), especially at high redshift. Similarly, the DI scenario underpredicts the number density of low luminosity AGN ($Log L_{bol} <44$).
The DI models have in fact the tendency to underproduce the bright end of the AGN luminosity function \citep{Menci2014}, and in turn the distribution of luminous AGN in massive halos. At least part of this shortfall in these type of models may be attributable to random errors in the observed luminosities. To check for this, we convolved our mock AGN catalogs with Gaussian errors. Being interested in the maximal effect of this procedure, we have forced the bright end of the predicted LF to match the observed one, regardless of the assumed width of the Gaussian errors. For reference purposes, we have used for the DI prediction with $\alpha=10$ ($\alpha=2$) a dispersion of width $\delta Log L_{1450}=0.$2 mag ($\delta Log L_{1450}=0.8$ mag) at $z \lesssim 1$, increasing to $\delta Log L_{1450}=0.8$ mag ($\delta Log L_{1450}=1.8$ mag) at $z \sim 2.5$.

%However, as expected, including a scatter in the luminosities also tends to reduce the clustering signal at all scales, as shown in Fig. 10.

\begin{figure}
    \centering
\includegraphics[scale=0.22]{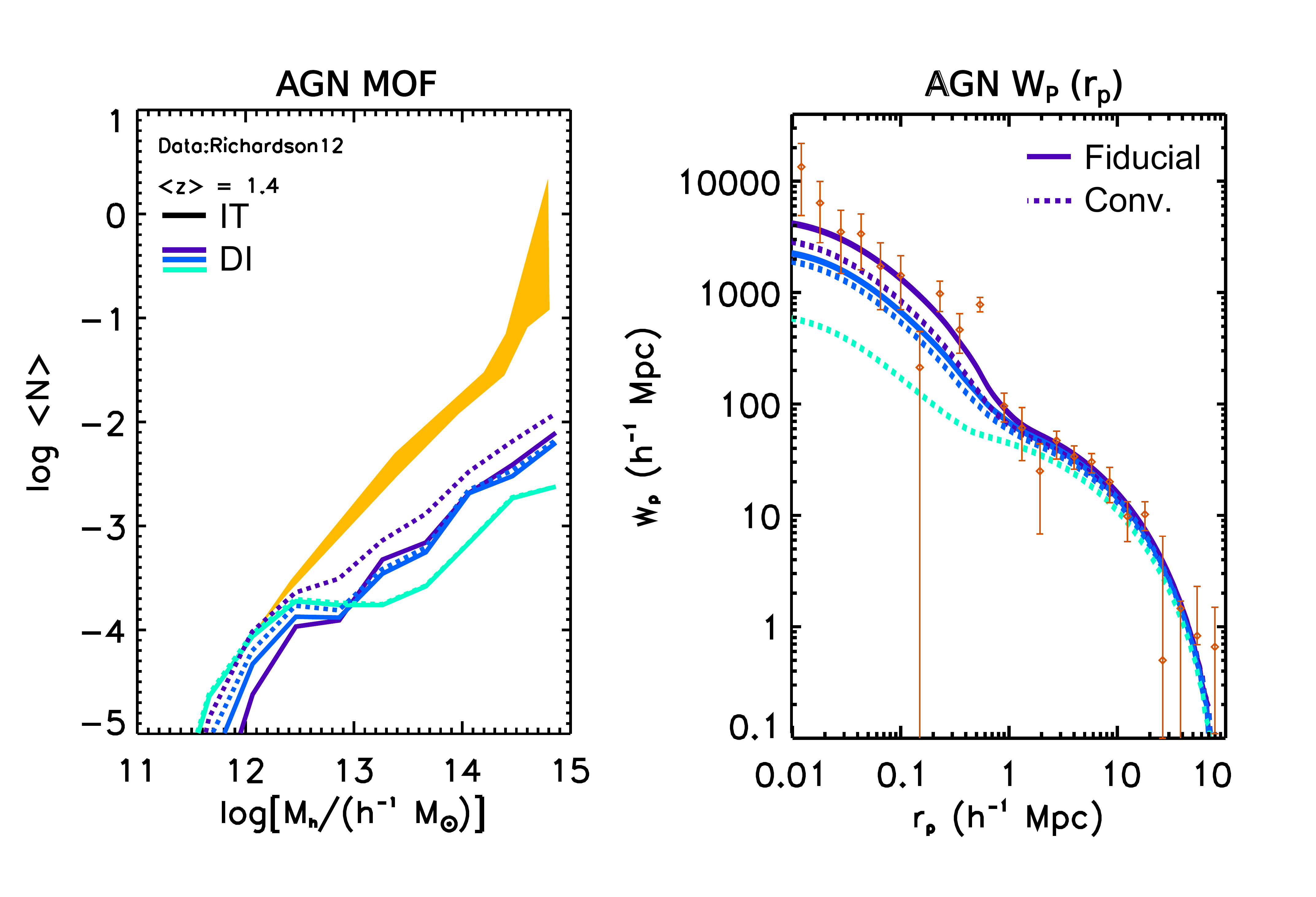}

\caption{AGN MOF and 2PCF for the DI scenario after having matched the AGN luminosity function bright-end (dotted lines). Continuous lines represent the fiducial predictions. Data from \citet{Richardson2012}}
\label{Conv}
\end{figure}

Fig. \ref{Conv} shows the effects of such procedure on the AGN MOF and 2PCF. The global effect of including a scatter in the AGN luminosities on the AGN MOF and on the 2PCF are minor, though not negligible, with the predicted clustering strength decreasing with increasing scatter, as expected (e.g., \citealt{Haiman2001,Martini2001,Shankar2010}). Overall, we believe that the underproduction of the DI models in producing luminous AGN is a true physical effect, caused by the low gas and disk fractions in massive galaxies. %Nevertheless, not being able to match the bright end of the luminosity function does not affect any of our previous results concerning the DI scenario.

%Overall, we believe that the underproduction of the the DI models in producing luminous AGN is a true physical effect, as well , caused by the low gas and disk fractions in massive galaxies. 

The lack of AGN in central galaxies residing in massive dark matter halos, as well as the high AGN satellite fraction, are caused mainly by the triggering criterion and to a less extent by the model for the mass inflow, but should be considered general features of the DI scenario regardless of its precise modeling.  %(eq. 5, although strongly depends on the disk and gas fraction of the galaxy, only regulates the AGN luminosity, not the onset of the AGN activity). 
The key point is that central galaxies are always more massive than satellites at a fixed dark matter halo mass. Since in massive galaxies the disk and gas fractions decrease, at a fixed halo mass the ideal conditions to trigger DIs should be found in satellites rather than in centrals. This is also supported by observations. For instance \citet{Bluck2014} find that the fraction of passive SDSS galaxies at low redshift is higher in centrals than in the satellites of the same dark matter halo.

%DI model driven by these properties should favor the onset of AGN activity in satellites (especially at low redshift). This view is also supported by observations: for instance, bluck et al, obtain a higher fraction of passive galaxies (and hence with low gas fraction) in centrals rather than satellite as a fixed DM halo mass.

%Anyway, our DI scenario is quite extreme in this respect and completely fails in triggering AGN activity in   galaxies with $M_* > 10^{11.5} \msun$, therefore causing the lack of AGN in centrals in massive environments. The main reason is that the critical mass needed for a disk to become unstable is too high in massive hosts, due to the stabilizing effect of the hot component (eq.4) and the relatively high B/T fraction. 

We stress that our modeling needs a triggering criterion: the type of gas inflows taken into account here (eq. 5) are meant to arise specifically in response to large-scale strong torques on gas from non-axisymmetric perturbations to the stellar gravitational potential \citep{Hopkins2011}. It could be interesting to explore a different modeling of the inflow triggered by DIs: indeed, in literature the DI scenario has been modeled and investigated in several fashions often leading to quite different (if not contrasting) results.

For instance, in the Durham SAM \citep{Fanidakis2011} and in the Somerville SAM \citep{Hirschmann2012} the implementation of the DI scenario takes into account a criterion for the stability of the disk similar to the one assumed in our modeling, but substantially differs into the modeling of the inflow onto the central SMBH. In particular, the former authors assume that in case of an unstable disk, all the mass of the disk accretes onto the SMBH, leading to a quite dramatic effectiveness of the DI scenario; in the latter case, the authors assume that only a fixed (and tunable) fraction of the mass in excess accretes onto the SMBH. Our modeling is somewhat less dramatic than these two cases: although the criterion for the stability is similar, the accretion is regulated at each time step by eq. \ref{hopkins}, which suppresses the inflow in case of low gas and disk fractions; moreover, the additional starburst following the AGN activity further contributes to deplete gas reservoirs and to stabilize the disk.

In high redshift, gas-rich violently unstable disk, migration of gas clouds \citep{Gabor2013}, possibly in combination with smoother cold inflows \citep{Dekel2009,Bournaud2011}, might lead to a substantial growth of $z\sim2$ SMBHs. Clump accretion might persist also at lower redshifts, though with a lower impact on the SMBH population also depending on the details of the modeling \citep{Hopkins2006,Gabor2013}. In these models the inflow is triggered under a set of different conditions (e.g., continuous refueling of the gas reservoirs in the inner galactic region, formation and survival of gas clumps, etc.) which are not necessarily related to large-scale disk instabilities (in the sense of eq. \ref{efstathi}). The inflows might be highly variable, being clump accretion a stochastic process, which might lead to violent accretion episodes and dramatic morphological transformations. We will explore the impact of different modeling of the DI scenario (e.g., impact of clump accretion) in future work.

\subsection{IT scenario: robustness of results}

%Our IT scenario, on the other hand, provides a good match to the observed AGN luminosity function, as well  to the observed AGN HOD. As we showed in the previous section, including observational error in our estimates do not drastically change any of our prediction.

One of the major uncertainty in our modeling of the IT scenario concerns the fate of the gas destabilized during a galaxy interaction. 

In our SAM, we assume that a fixed fraction ($1/4$, see Sect 2.2.1) accretes onto the SMBH, while the remaining part feeds a circumnuclear starburst, which adds to the $"$normal$"$ star formation activity presents in the galaxy (see Sect.2.1). Our fiducial value comes from \citet{Sanders1996}, and its validity has also been verified \textit{a posteriori} through the comparison with a number of observables (such as the$ M_{BH}-M_*$ relation, in \citealt{Menci2014},  or the star forming properties of AGN host galaxies in \citealt{Lamastra2013a,Lamastra2013}). 

Observationally, linking the star forming properties of AGN host galaxies to AGN luminosity / SMBH accretion rate has never been an easy task, and many observational studies have faced a challenging situation, mainly due to observational bias and/or AGN variability (see, for instance, \citealt{Rosario2012,Chen2013,Hickox2014}). In case of galaxy interactions, the lack of a reliable diagnostic able to effectively distinguish between the $"$normal$"$ component of star formation activity (normally present in every galaxy depending on the cold gas reservoirs) and the additional $"$burst$"$ of star formation triggered during the interaction has further contributed to make the situation more complex.

Despite the complex observational situation, we note that our results are quite insensitive to slight changes in the exact value of the fraction.  First of all, the contribution to the total galaxy stellar mass at z=0 owning to the burst phase following a galaxy interaction (i.e., $3/4$ of the destabilized gas) is less than $10\%$ \citep{Lamastra2013a}, indicating that assigning a higher/lower fraction should not drastically affect host galaxy properties.

Secondly, assuming a slightly different values for the fraction of gas accreting onto the SMBH should also little affect the SMBH population properties, thanks to the AGN feedback. A lower fraction indicating less gas onto the SMBH implies lower AGN luminosities, but at the same time, it implies lower AGN feedback and less gas expelled from the galaxy, encouraging further AGN activity at a later time. On the contrary, higher fraction indicating higher luminosities would imply stronger AGN feedback, inhibiting further AGN activity. 

To be more quantitative, at $z \sim 0.5$, assuming that a fraction $f=1/2$ of the destabilized gas accretes onto the SMBH causes the AGN luminosity to increase of a factor $\sim 0.8 mag$, which leads to a slight overestimate of the AGN luminosity function (particularly effective for bright AGN). Conversely, reducing the fraction down to $f=1/10$ leads to a slight underestimate of the AGN luminosity function, even if the effect is less noticeable. As for the AGN MOF, it almost does not change its shape (which implies unaltered clustering strength and AGN satellite fraction),  while it only differs in terms of normalization, depending on the luminosity range, at maximum of a factor $\sim 2-3$.

%To be more quantitative, at $z \sim 0.5$, assuming that a fraction $f=1/2$ (0.05) of the destabilized gas accretes onto the SMBH causes the normalization of the luminosity function to increase (decrease) of a factor $\sim 2$ ($\sim3$). As for the AGN MOF, it almost does not change its shape (which implies unaltered clustering strength and AGN satellite fraction),  while it only differs in terms of normalization, dependending on the luminosity range, at maximum of a factor $\sim 2$ ($\sim 3$). 

Having said this, it might be nonetheless possible that extremely weak interactions - responsible for the triggering of low luminosity AGN ($Log L_{bol} < 44$) in our SAM - might not directly trigger an inflow onto the SMBH at all, or that our treatment of the inflow in these cases might be inadequate. This is the regime where weak fly-by events dominate the statistics, usually concerning objects with quite different sizes and high relative velocities with respect galaxy circular velocity (otherwise a merger would occur), capable of inducing only small variations in the galaxy potential/ angular momentum.  In these very cases, weak interactions might only trigger a perturbation to the disk potential at large galactic scales that at a later time might trigger an inflow onto the SMBH in a  more complex form of DI (see, e.g., \citealt{Hopkins2010,Hopkins2011}). This might represent another possible improvement worth to be explored in a future work, in light of a possible scenario where the two modes (IT and DI) might co-exist at the same time (see also Sect. 6.5).

\subsection{Obscured vs. unobscured AGN clustering}
Whether type I (unobscured) AGN clusters more or less strongly than type II(obscured) AGN is still a matter of debate. Some authors find evidence of statistically significant discrepancy in the clustering strength of obscured and unobscured X-ray selected AGN, both at low \citep{Cappelluti2010,Allevato2011} and high redshift \citep{Allevato2014}. Mild evidence ($\sim 2 \sigma$) is also pointed out by \citet{Hickox2011}, who find that obscured mid-IR selected quasars at $0.7<z<1.8$ cluster more strongly than unobscured quasars. \citet{Donoso2014} obtain a similar result for WISE (Wide-field Infrared
Survey Explorer, \citealt{Wright2010}) mid-IR selected AGN at $z \sim 1.1$.

On the contrary, \citet{Gilli2009} suggest that obscured and unobscured XMM-COSMOS AGN inhabit similar environment. No difference in the clustering strength of unobscured radio-loud quasars and obscured radio-galaxies is found by \citet{Wylezalek2013}. Mendez et al. (2015) find no significant difference in the clustering of obscured and unobscured WISE-selected AGN, and further suggest that the results obtained by \citet{Donoso2014} might be driven by the difference in the redshift distributions of their obscured and unobscured samples. 

Any difference in the clustering strength of obscured and unobscured AGN would have strong implications into the physics of AGN obscuration. Indeed,  it would call into question the simplest unified scheme for AGN, according to which obscuration is purely a geometrical effect due to the AGN orientation with respect to the line-of-sight.

From the point of view of our SAM, since we are accounting for AGN obscuration through the empirical absorption function from \citet{Ueda2014}, we are implicitly assuming the "luminosity and redshift dependent unified scheme" of AGN, neglecting any explicit dependence of the covering fraction due to environmental effects.

Hence, in our SAM any obscured and unobscured AGN sample sharing the same redshift and luminosity distributions must be naturally characterized by the same clustering strength. In case the luminosity and redshift distributions were not perfectly overlapping,  it might be possible to observe some slight differences between the obscured and unobscured population due to any implicit luminosity and/or redshift dependence of the AGN clustering.

In this paper we have not explicitly focused on the differences in the clustering of the obscured and unobscured population; the use of the absorption function was only limited to roughly reproducing the observational selection bias of different types of survey (optical vs. X-ray). However, If the luminosity and redshift dependent unified scheme were not correct, but there were some kind of environmental dependence of the covering fraction, this might slightly alter some of our predictions. For instance, assuming a model were the black hole growth lags behind that of the halo \citep{Peng2006,Alexander2008, Hopkins2008,Decarli2010,King2010,Hickox2011}, with young obscured quasars being characterized by an early phase of rapid growth to "catch up" their final black hole mass, our X-ray AGN clustering predictions should be biased towards higher $M_{h}$ dark matter halo mass.

\subsection{Effects of selection cuts}

 \begin{figure}
\includegraphics[scale=0.2]{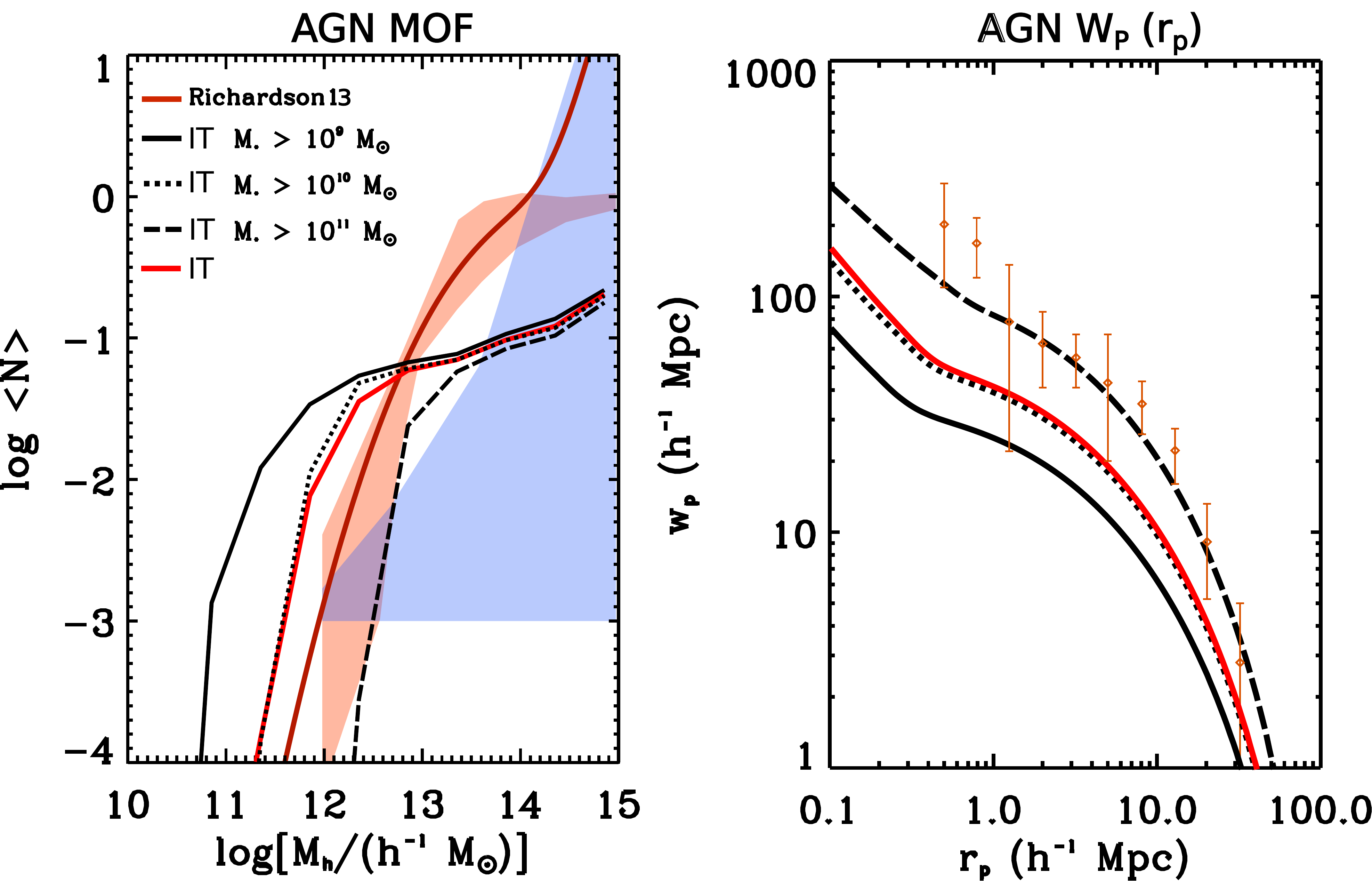}
\includegraphics[scale=0.2]{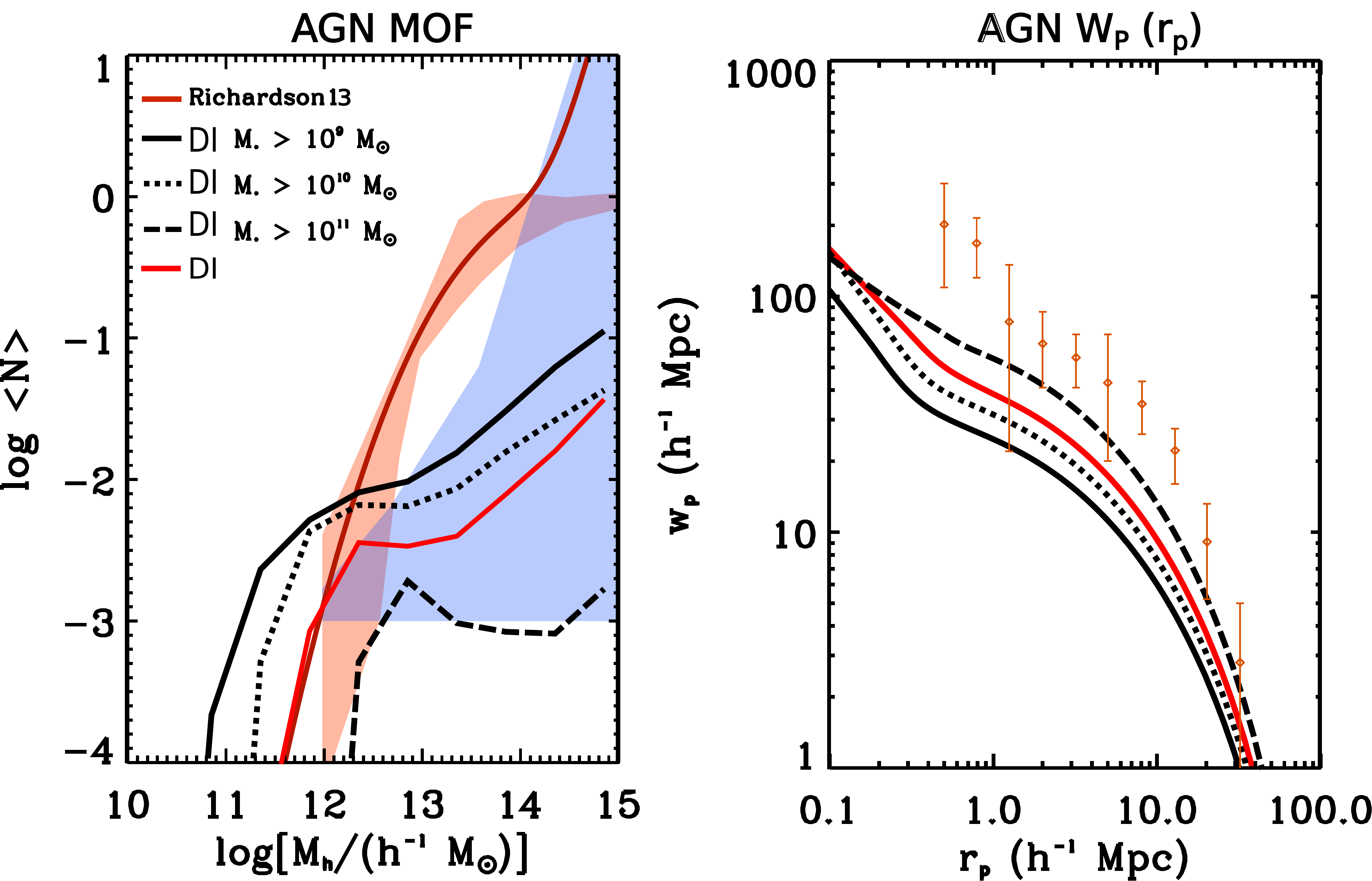}

\caption{AGN MOF and 2PCF for the IT and DI scenarios concerning the comparison with \citet{Richardson2013} data. %The continuous red line represents the the total AGN MOF as obtained from their 2PCF measurement, while the red-dotted lines are the AGN MOF for central and satellite AGN. The red-shaded is the 68.3$\%$ confidence interval for central AGN, while the blue-shaded region is for satellite AGN.
The continuous red lines represent the fiducial predictions for the IT and DI scenarios obtained considering the full redshift and luminosity interval of the sample, as well as the cut on the host galaxy magnitude of $I_{AB}<23$. The continuous, dashed and dotted black lines have been obtained instead substituting the condition on the host galaxy magnitude ($I_{AB} <23$) with a cut on the host galaxy stellar mass of $Log M_*>9,10,11 \msun$.} %The red continuous lines (IT v2 and DI v2) have instead been obtained considering AGN around the median redshift and luminosity of the sample ($1<z<1.4$, $43.2<Log L_X<44$) and with the standard cut in the host galaxy magnitude of $I_{AB}<23$.}
\label{massrich}
\end{figure}

We also test the effect of using different selection cuts on the AGN MOF and 2PCF. 

In order to do so, we compare our predictions with the AGN 2PCF from \citet{Richardson2013} and the associated AGN MOF obtained using the HOD modeling. The authors considered a sample of XMM-COSMOS AGN in the redshift range of $0<z<4$ ($<z>=1.2$), with typical luminosities in the range of $L_{X} \sim 41-45$, and with a magnitude cut for AGN host galaxies of $I_{AB} < 23$. These broad selection intervals are ideal to test the impact of diverse observational cuts on the models outputs.

At face value, the IT and DI scenario fail in reproducing both the 2PCF and the AGN MOF inferred by the authors. In particular, the average halo mass predicted by our models is in the range $Log M_h \sim 12.2 - 12.4$, at odds with the $Log M_h \sim 13$ derived by
\citet{Richardson2013}.
 
Given the large redshift and luminosity intervals spanned by their sample, the discrepancy might at least in part be induced by not properly accounting for the non-trivial selection functions inherent in the observational sample.  In fig. \ref{massrich} we check how the AGN MOF and 2PCF are influenced by different selection cuts specifically in terms of host galaxy stellar mass. As displayed by fig. \ref{massrich}, considering only the most massive host galaxies affects the AGN MOF and 2PCF (as already highlighted in Sect 4.2), resulting into an increment of the AGN bias and a better match to the observed 2PCF. The match is improved in the IT model when selecting galaxies with stellar mass $M_*>10^{11} \msun$ . The effect is slightly less pronounced in the DI scenario, since DIs are effective only in a limited range of host galaxy stellar masses ($Log M_* <11.5 \msun$). 

%In fig. \ref{massrich} we have also plotted for a qualitatively comparison the AGN MOF and 2PCF obtained by considering a small redshift and luminosity intervals around the median values of the sample ($1<z<1.4$, $43.2<L_X<44$, $I_{AB}<23$). The normalization of the AGN MOF is sensibly reduced, due to the smaller luminosity interval

%Also in this case, the AGN MOFs obtained using smaller intervals (fig. \ref{massrich}, red lines) are quite different with respect the one obtained in Sect 4.4.1, being characterized by a lower normalization and being skewed toward high DM halo masses. Also the AGN 2PCF is different, nonetheless providing a better match to the observed one. The change in the normalization is due to the fact that by using a smaller luminosity interval we are excluding the less luminous AGN, which are more numerous. We note that a change in the normalization as a function of redshift should have expected also for limited luminosity intervals, due to the strong dependence of the AGN duty cycle by redshift shown by observations (see Sect. 4.1).

%This comparison is particularly relevant as the higher X-ray clustering signal has been often interpreted as an additional channel for AGN triggering.

Fig. \ref{massrich} clearly highlights the importance of properly taking into account selection cuts in the host galaxy population (such as stellar mass), especially when comparing results concerning AGN selected at different wavelengths. Indeed, the higher X-ray clustering signal, which has been often interpreted as an additional channel for AGN triggering, might be understood in terms of selection biases and in terms of the properties of the underlying host galaxy population.

This view is also supported by the recent paper of Mendez et al. (2015). Studying the clustering properties of X-ray, radio, and mid-IR AGN from the PRIMUS and DEEP2 redshift surveys, they found a higher clustering signal for X-ray and radio AGN with respect to mid-IR selected AGN. Nonetheless, the differences disappeared when comparing the clustering of each AGN sample with matched galaxy samples with similar properties in terms of stellar mass, star formation, redshift. %The authors further claimed that AGN clustering could be entirely interpreted in terms of galaxy clustering and AGN selection effects.

%Ultimately, we note that interpreting the differences in the AGN clustering strengths as the signature of differing feeding modes might also depend on the type of triggering mechanism considered, whose definition is clearly model dependent. For instance, our IT scenario includes both strong interactions (such as major mergers, responsible for the most luminous quasars) and weak interactions (minor mergers, weak fly-by events, responsible for less luminous AGN), being the difference between the two regimes somewhat arbitrary.

% This plot highlights the importance of taking into account any selection cut in the AGN host galaxy, due to the relation between the galaxy stellar mass and the DM halo mass. In a scenario such a the IT, which reproduce the propertie sof the AGn population over a wide range of host masse, redshift and AGN luminosity..
 
\subsection{DI+IT scenario}
Although studying the separate effects of the two modes is more effective
in highlighting their role in determining the clustering properties of AGN (the goal of the present paper), considering a scenario where both mechanisms simultaneously have a role in the triggering of AGN activity is undoubtedly an intriguing possibility.
However, such a development needs some caution.

Indeed, the two modes have to co-exist and be implemented in a physically justified way. Simple considering a tunable fraction of the AGN population to be triggered by one of the two modes would not provide any physically useful information, though it might produce a better fit to observations. For instance, a possible way to mix the two scenarios could be the one sketched at the end of Sect. 6.2: that is, a scenario where our fiducial IT implementation is retained for the strongest interactions, along with the DI scenario for isolated galaxies, while a $"$mixed$"$ description accounts for the case of weak interactions (which should be responsible for disturbing the galaxy potential and causing a DI-driven inflow at a later time).

While the development of such a model clearly goes beyond the aim of this paper, it might be nonetheless interesting to discuss - at least from a qualitative point of view - to what extent a combined DI+IT scenario might affect the results presented in this paper.

It is clear from the past sections that while our IT scenario provides a quite good match to observational data, the effectiveness of our DI scenario is quite limited (e.g., in terms of luminosity and redshift intervals).  So, we might ask  whether a possible DI+IT scenario might improve the match with observational data where the IT scenario seems to fail.

For instance, the IT scenario seems not to trigger enough low-intermediate luminosity AGN in satellite galaxies, preferring centrals. This is highlighted by the comparison with \citet{Allevato2012} and \citet{Leauthaud2015} in fig. \ref{AGNHODInd}, but also by the comparison with \citet{Pentericci2013} and \citet{Martini2009} data in fig. \ref{sat_frac_ext}. Hence it might be possible that at least a fraction of the observed satellite AGN are triggered by DIs. A rough estimate of the maximum effect of a combined DI+IT scenario can be obtained by simply adding to the prediction of the MOF for the IT scenario the satellite MOF predicted for the DI model with $\alpha=2$. This represents a rough estimate of the maximum effect because, in a realistic situation, the two modes would compete against each other in consuming the galaxy gas reservoirs and the total effect would be fainter than simply adding the two MOFs.

By applying this procedure to the aforementioned data sets, the match with observational data improves, but still remains statistically significant: in particular, the DI+IT scenario shows a disagreement with the satellite fractions of \citet{Pentericci2013} and \citet{Martini2009} at$\sim  3 \sigma$ level, while   the disagreement with satellite MOFs of \citet{Allevato2012} and \citet{Leauthaud2015} remains at $\sim 1.3 \sigma$ and $\sim 3 \sigma$ level ($f_{sat}$ increases respectively from $10\%$ to $17\%$ and from  $8\%$ to $10\%$). This means that if the discrepancy with observational results is not due to any observational selection bias (as discussed in Sect 5.2), our DI scenario cannot supply the missing fraction of low-intermediate luminosity AGN in satellite galaxy at low redshift, and hence a possible DI+IT scenario would not provide a good match with data.

Lastly, we could also ask whether a possible DI+IT scenario might alter our predictions concerning the luminosity and redshift dependence of AGN clustering. As for the redshift dependence,  both modes are characterized by a similar average bias factor at different redshifts (as shown also by the comparison with the observed 2PCFs), and a similar shift towards lower halo masses for increasing redshift, driven by an increasing lack of massive halos. On the other hand, the mild luminosity dependence shown by DI and IT is not perfectly the same for the two scenarios and it is also quite susceptible to selection cuts (such as host stellar mass). Hence, in a more realistic case where the two modes might co-exist, while the redshift dependence of the AGN clustering highlighted in Sect 4.3. should be preserved, the same might not be guaranteed for the luminosity dependence.

\section{conclusions}
   
Using an advanced semi analytic model (SAM) for galaxy formation \citep{Menci2014,Gatti2015}, coupled to accurate halo occupation distribution modeling, we have investigated the imprint left by different AGN triggering mechanisms on the clustering strength of the AGN population at small and large scales. Two fueling mechanisms have been considered: a first accretion mode where AGN activity is triggered by disk instabilities (DI scenario) in isolated galaxies, and a second feeding mode where galaxy mergers and fly-by events (IT scenario) are responsible for producing a sudden destabilization of large quantities of gas, causing the mass inflow onto the central SMBH. The final goal of this paper was to highlight key features in the clustering properties of the two modes that might constitute robust probes to pin down the dominant SMBH fueling mechanism. We obtained the following results:

%Our main results are as follows: i) DIs, irrespective of their exact implementation in the SAM, tend to fall short in triggering AGN activity in galaxies at the centre of halos with $M_h>10^{13.5} h^{-1}\msun$ (at all redshifts), triggering mainly satellite galaxies in massive environments. On the contrary, IT scenario generally agrees well with observations at every halo mass. ii)  Both scenarios are quite degenerate in matching large-scale clustering measurements, suggesting that the sole average bias might not be an effective observational constraint. iii) DIs, being $"$in-situ$"$ processes, naturally predict fractions of active galaxies that do not depend much on host halo mass, while in the IT scenario the majority of the AGN population is hosted by halos with $M_h \sim 10^{12} h^{-1}\msun$.  The relative number of satellites in DIs at intermediate-to-low luminosities is always significantly higher than in IT models. The latter basic predictions could act as fundamental probes for setting more stringent constraints on AGN triggering mechanisms. In light of our advanced models, we also discuss possible improvements on current AGN semi-empirical halo occupation modelling, as well as the impact of different observational selection cuts in measuring AGN clustering, including possible discrepancies between optical and X-ray surveys and redshift/luminosity effects.

\begin{enumerate}
\item  DIs, irrespective of their exact implementation in the SAM, tend to fall short in triggering AGN activity in galaxies at the center of halos with $M_h>10^{13.5} h^{-1}\msun$ (at all redshifts). For centrals in less massive environments, DI are particularly effective in triggering $L \sim L_{knee}$ AGN. On the contrary, the IT scenario predicts abundance of active, central galaxies that generally agrees well with observations at every halo mass, for a wide range of AGN luminosities and redshift.

\item The relative number of satellites in DIs at intermediate-to-low luminosities is always significantly higher than in IT models, especially in groups and clusters, indicating a preference of DI AGN to inhabit satellite galaxies. However, the absolute abundance of satellite AGN at $z \lesssim 1$ is still underpredicted by DI models. 

\item Similarly, the low-satellite fraction predicted for the IT scenario would suggest that different feeding modes might partially contribute to the triggering of satellite AGN in groups and clusters, at least for intermediate-to-low luminosities at $z \lesssim 1$.

%The relative number of satellites in DIs at intermediate-to-low luminosities is always significantly higher than in IT models, especially in groups and clusters, where DIs trigger only satellite AGN. The low-satellite fraction predicted for the IT scenario would suggest that other feeding modes might partially contribute to the triggering of satellite AGN in groups and clusters, at least for intermediate-to-low luminosities at $z \lesssim 1$.

\item Both scenarios are quite degenerate in matching large-scale clustering measurements, concerning both X-ray and optically selected AGN surveys, with different average luminosity and redshift. This seems to indicate that the sole analysis of the large scale clustering (i.e. bias factor) constitutes a poor constraint of the DI and IT modes.

\item Selection cuts in terms of AGN luminosity, host galaxy properties, redshift interval might have a more relevant role in driving the differences often observed in the bias factor inferred from surveys carried out at different wavelengths.

\item Our analysis suggests the presence of both a mild luminosity and a more consistent redshift dependence of the AGN clustering: at $z\sim 0.5$ luminous AGN are hosted by halos with mass $10^{12}-10^{13} \msun$, while moderately luminous AGN occupy a wide range of dark matter halos of different mass. Less luminous AGN are biased towards lower dark matter halo masses. At high redshift, the average halo mass sensibly moves towards lower values.

 \end{enumerate}

%The claim that different types of 2PCF measurements (e.g., X-ray vs. optical) might probe different AGN triggering mechanisms is definitely a non trivial topic. Besides all the selection biases discussed above, we note that this might also depend on the type of triggering mechanism considered, whose definition is clearly model dependent: for instance, our IT scenario includes both strong interactions (such as major mergers, responsible for the most luminous QSOs) and weak interactions (minor mergers, weak fly-by events), being the difference between the two regimes somewhat arbitrary. 

% Secondly, a mixture of selection effects (luminosity and redshift intervals considered, as well stellar mass cuts) and luminosity and redshift dependence of AGN clustering might have a relevant role in driving the differences in the 2PCF of different AGN samples. 

Our analysis suggests the need of new AGN host halo mass distributions possibly directly probed via, e.g., lensing or dynamical measurements, and possibly with clearer selections on the host galaxy properties, where possible. \newline

\textit{Acknowledgement.} The authors thank the anonymous referee for several helpful suggestions that improved the presentation of this work.

%Keeping this in mind, our analysis suggests the need of new, directly measured AGN HOD, concerning AGN sample relative to limited redshift and luminosity intervals, with known morphological properties. Only then, with an accurate comparison with existing hydrodynamic cosmological simulations or semi analytic models, different triggering mechanisms for the AGN activity can be more effectively tested and eventually validated.

%Future \textit{direct} measurements of AGN HOD capable of probing \textit{different redshift and luminosity ranges} are needed to effectively pin down AGN triggering mechanisms

\bibliographystyle{mnras}
\bibliography{research}

\begin{thebibliography}{}
\makeatletter
\relax
\def\mn@urlcharsother{\let\do\@makeother \do\$\do\&\do\#\do\^\do\_\do\%\do\~}
\def\mn@doi{\begingroup\mn@urlcharsother \@ifnextchar [ {\mn@doi@}
  {\mn@doi@[]}}
\def\mn@doi@[#1]#2{\def\@tempa{#1}\ifx\@tempa\@empty \href
  {http://dx.doi.org/#2} {doi:#2}\else \href {http://dx.doi.org/#2} {#1}\fi
  \endgroup}
\def\mn@eprint#1#2{\mn@eprint@#1:#2::\@nil}
\def\mn@eprint@arXiv#1{\href {http://arxiv.org/abs/#1} {{\tt arXiv:#1}}}
\def\mn@eprint@dblp#1{\href {http://dblp.uni-trier.de/rec/bibtex/#1.xml}
  {dblp:#1}}
\def\mn@eprint@#1:#2:#3:#4\@nil{\def\@tempa {#1}\def\@tempb {#2}\def\@tempc
  {#3}\ifx \@tempc \@empty \let \@tempc \@tempb \let \@tempb \@tempa \fi \ifx
  \@tempb \@empty \def\@tempb {arXiv}\fi \@ifundefined
  {mn@eprint@\@tempb}{\@tempb:\@tempc}{\expandafter \expandafter \csname
  mn@eprint@\@tempb\endcsname \expandafter{\@tempc}}}

\bibitem[\protect\citeauthoryear{{Abazajian} et~al.,}{{Abazajian}
  et~al.}{2009}]{Abazajian2009}
{Abazajian} K.~N.,  et~al., 2009, \mn@doi [\apjs]
  {10.1088/0067-0049/182/2/543}, \href
  {http://adsabs.harvard.edu/abs/2009ApJS..182..543A} {182, 543}

\bibitem[\protect\citeauthoryear{{Adelman-McCarthy} et~al.,}{{Adelman-McCarthy}
  et~al.}{2007}]{Adelman-McCarthy2007}
{Adelman-McCarthy} J.~K.,  et~al., 2007, \mn@doi [\apjs] {10.1086/518864},
  \href {http://adsabs.harvard.edu/abs/2007ApJS..172..634A} {172, 634}

\bibitem[\protect\citeauthoryear{{Aird} et~al.,}{{Aird}
  et~al.}{2012}]{Aird2012}
{Aird} J.,  et~al., 2012, \mn@doi [\apj] {10.1088/0004-637X/746/1/90}, \href
  {http://adsabs.harvard.edu/abs/2012ApJ...746...90A} {746, 90}

\bibitem[\protect\citeauthoryear{{Alexander} et~al.,}{{Alexander}
  et~al.}{2008}]{Alexander2008}
{Alexander} D.~M.,  et~al., 2008, \mn@doi [\aj] {10.1088/0004-6256/135/5/1968},
  \href {http://adsabs.harvard.edu/abs/2008AJ....135.1968A} {135, 1968}

\bibitem[\protect\citeauthoryear{{Allevato} et~al.,}{{Allevato}
  et~al.}{2011}]{Allevato2011}
{Allevato} V.,  et~al., 2011, \mn@doi [\apj] {10.1088/0004-637X/736/2/99},
  \href {http://adsabs.harvard.edu/abs/2011ApJ...736...99A} {736, 99}

\bibitem[\protect\citeauthoryear{{Allevato} et~al.,}{{Allevato}
  et~al.}{2012}]{Allevato2012}
{Allevato} V.,  et~al., 2012, \mn@doi [\apj] {10.1088/0004-637X/758/1/47},
  \href {http://adsabs.harvard.edu/abs/2012ApJ...758...47A} {758, 47}

\bibitem[\protect\citeauthoryear{{Allevato} et~al.,}{{Allevato}
  et~al.}{2014}]{Allevato2014}
{Allevato} V.,  et~al., 2014, \mn@doi [\apj] {10.1088/0004-637X/796/1/4}, \href
  {http://adsabs.harvard.edu/abs/2014ApJ...796....4A} {796, 4}

\bibitem[\protect\citeauthoryear{{Angulo}, {Lacey}, {Baugh}  \&
  {Frenk}}{{Angulo} et~al.}{2009}]{Angulo2009}
{Angulo} R.~E.,  {Lacey} C.~G.,  {Baugh} C.~M.,   {Frenk} C.~S.,  2009, \mn@doi
  [\mnras] {10.1111/j.1365-2966.2009.15333.x}, \href
  {http://adsabs.harvard.edu/abs/2009MNRAS.399..983A} {399, 983}

\bibitem[\protect\citeauthoryear{{Arp}}{{Arp}}{1970}]{Arp1970}
{Arp} H.,  1970, \mn@doi [\aj] {10.1086/110932}, \href
  {http://adsabs.harvard.edu/abs/1970AJ.....75....1A} {75, 1}

\bibitem[\protect\citeauthoryear{{Baugh}}{{Baugh}}{2006}]{Baugh2006}
{Baugh} C.~M.,  2006, \mn@doi [Reports on Progress in Physics]
  {10.1088/0034-4885/69/12/R02}, \href
  {http://adsabs.harvard.edu/abs/2006RPPh...69.3101B} {69, 3101}

\bibitem[\protect\citeauthoryear{{Berlind} et~al.,}{{Berlind}
  et~al.}{2003}]{Berlind2003}
{Berlind} A.~A.,  et~al., 2003, \mn@doi [\apj] {10.1086/376517}, \href
  {http://adsabs.harvard.edu/abs/2003ApJ...593....1B} {593, 1}

\bibitem[\protect\citeauthoryear{{Bessiere}, {Tadhunter}, {Ramos Almeida}  \&
  {Villar Mart{\'{\i}}n}}{{Bessiere} et~al.}{2012}]{Bessiere2012}
{Bessiere} P.~S.,  {Tadhunter} C.~N.,  {Ramos Almeida} C.,   {Villar
  Mart{\'{\i}}n} M.,  2012, \mn@doi [\mnras]
  {10.1111/j.1365-2966.2012.21701.x}, \href
  {http://adsabs.harvard.edu/abs/2012MNRAS.426..276B} {426, 276}

\bibitem[\protect\citeauthoryear{{Bluck}, {Mendel}, {Ellison}, {Moreno},
  {Simard}, {Patton}  \& {Starkenburg}}{{Bluck} et~al.}{2014}]{Bluck2014}
{Bluck} A.~F.~L.,  {Mendel} J.~T.,  {Ellison} S.~L.,  {Moreno} J.,  {Simard}
  L.,  {Patton} D.~R.,   {Starkenburg} E.,  2014, \mn@doi [\mnras]
  {10.1093/mnras/stu594}, \href
  {http://adsabs.harvard.edu/abs/2014MNRAS.441..599B} {441, 599}

\bibitem[\protect\citeauthoryear{{Bond}, {Cole}, {Efstathiou}  \&
  {Kaiser}}{{Bond} et~al.}{1991}]{Bond1991}
{Bond} J.~R.,  {Cole} S.,  {Efstathiou} G.,   {Kaiser} N.,  1991, \mn@doi
  [\apj] {10.1086/170520}, \href
  {http://adsabs.harvard.edu/abs/1991ApJ...379..440B} {379, 440}

\bibitem[\protect\citeauthoryear{{Bongiorno} et~al.,}{{Bongiorno}
  et~al.}{2012}]{Bongiorno2012}
{Bongiorno} A.,  et~al., 2012, \mn@doi [\mnras]
  {10.1111/j.1365-2966.2012.22089.x}, \href
  {http://adsabs.harvard.edu/abs/2012MNRAS.427.3103B} {427, 3103}

\bibitem[\protect\citeauthoryear{{Bonoli}, {Shankar}, {White}, {Springel}  \&
  {Wyithe}}{{Bonoli} et~al.}{2010}]{Bonoli2010}
{Bonoli} S.,  {Shankar} F.,  {White} S.~D.~M.,  {Springel} V.,   {Wyithe}
  J.~S.~B.,  2010, \mn@doi [\mnras] {10.1111/j.1365-2966.2010.16285.x}, \href
  {http://adsabs.harvard.edu/abs/2010MNRAS.404..399B} {404, 399}

\bibitem[\protect\citeauthoryear{{Bournaud}, {Dekel}, {Teyssier}, {Cacciato},
  {Daddi}, {Juneau}  \& {Shankar}}{{Bournaud} et~al.}{2011}]{Bournaud2011}
{Bournaud} F.,  {Dekel} A.,  {Teyssier} R.,  {Cacciato} M.,  {Daddi} E.,
  {Juneau} S.,   {Shankar} F.,  2011, \mn@doi [\apjl]
  {10.1088/2041-8205/741/2/L33}, \href
  {http://adsabs.harvard.edu/abs/2011ApJ...741L..33B} {741, L33}

\bibitem[\protect\citeauthoryear{{Bruzual} \& {Charlot}}{{Bruzual} \&
  {Charlot}}{2003}]{Bruzual2003}
{Bruzual} G.,  {Charlot} S.,  2003, \mn@doi [\mnras]
  {10.1046/j.1365-8711.2003.06897.x}, \href
  {http://adsabs.harvard.edu/abs/2003MNRAS.344.1000B} {344, 1000}

\bibitem[\protect\citeauthoryear{{Cappelluti}, {Ajello}, {Burlon}, {Krumpe},
  {Miyaji}, {Bonoli}  \& {Greiner}}{{Cappelluti} et~al.}{2010}]{Cappelluti2010}
{Cappelluti} N.,  {Ajello} M.,  {Burlon} D.,  {Krumpe} M.,  {Miyaji} T.,
  {Bonoli} S.,   {Greiner} J.,  2010, \mn@doi [\apjl]
  {10.1088/2041-8205/716/2/L209}, \href
  {http://adsabs.harvard.edu/abs/2010ApJ...716L.209C} {716, L209}

\bibitem[\protect\citeauthoryear{{Cappelluti}, {Allevato}  \&
  {Finoguenov}}{{Cappelluti} et~al.}{2012}]{Cappelluti2012}
{Cappelluti} N.,  {Allevato} V.,   {Finoguenov} A.,  2012, \mn@doi [Advances in
  Astronomy] {10.1155/2012/853701}, \href
  {http://adsabs.harvard.edu/abs/2012AdAst2012....1C} {2012, 1}

\bibitem[\protect\citeauthoryear{{Cavaliere}, {Lapi}  \& {Menci}}{{Cavaliere}
  et~al.}{2002}]{Cavaliere2002}
{Cavaliere} A.,  {Lapi} A.,   {Menci} N.,  2002, \mn@doi [\apjl]
  {10.1086/345890}, \href {http://adsabs.harvard.edu/abs/2002ApJ...581L...1C}
  {581, L1}

\bibitem[\protect\citeauthoryear{{Chatterjee}, {Degraf}, {Richardson}, {Zheng},
  {Nagai}  \& {Di Matteo}}{{Chatterjee} et~al.}{2012}]{Chatterjee2012}
{Chatterjee} S.,  {Degraf} C.,  {Richardson} J.,  {Zheng} Z.,  {Nagai} D.,
  {Di Matteo} T.,  2012, \mn@doi [\mnras] {10.1111/j.1365-2966.2011.19917.x},
  \href {http://adsabs.harvard.edu/abs/2012MNRAS.419.2657C} {419, 2657}

\bibitem[\protect\citeauthoryear{{Chatterjee}, {Nguyen}, {Myers}  \&
  {Zheng}}{{Chatterjee} et~al.}{2013}]{Chatterjee2013}
{Chatterjee} S.,  {Nguyen} M.~L.,  {Myers} A.~D.,   {Zheng} Z.,  2013, \mn@doi
  [\apj] {10.1088/0004-637X/779/2/147}, \href
  {http://adsabs.harvard.edu/abs/2013ApJ...779..147C} {779, 147}

\bibitem[\protect\citeauthoryear{{Chen} et~al.,}{{Chen}
  et~al.}{2013}]{Chen2013}
{Chen} C.-T.~J.,  et~al., 2013, \mn@doi [\apj] {10.1088/0004-637X/773/1/3},
  \href {http://adsabs.harvard.edu/abs/2013ApJ...773....3C} {773, 3}

\bibitem[\protect\citeauthoryear{{Civano} et~al.,}{{Civano}
  et~al.}{2012}]{Civano2012}
{Civano} F.,  et~al., 2012, \mn@doi [\apjs] {10.1088/0067-0049/201/2/30}, \href
  {http://adsabs.harvard.edu/abs/2012ApJS..201...30C} {201, 30}

\bibitem[\protect\citeauthoryear{{Coil}, {Hennawi}, {Newman}, {Cooper}  \&
  {Davis}}{{Coil} et~al.}{2007}]{Coil2007}
{Coil} A.~L.,  {Hennawi} J.~F.,  {Newman} J.~A.,  {Cooper} M.~C.,   {Davis} M.,
   2007, \mn@doi [\apj] {10.1086/509099}, \href
  {http://adsabs.harvard.edu/abs/2007ApJ...654..115C} {654, 115}

\bibitem[\protect\citeauthoryear{{Coil} et~al.,}{{Coil}
  et~al.}{2009}]{Coil2009}
{Coil} A.~L.,  et~al., 2009, \mn@doi [\apj] {10.1088/0004-637X/701/2/1484},
  \href {http://adsabs.harvard.edu/abs/2009ApJ...701.1484C} {701, 1484}

\bibitem[\protect\citeauthoryear{{Cole}, {Lacey}, {Baugh}  \& {Frenk}}{{Cole}
  et~al.}{2000}]{Cole2000}
{Cole} S.,  {Lacey} C.~G.,  {Baugh} C.~M.,   {Frenk} C.~S.,  2000, \mn@doi
  [\mnras] {10.1046/j.1365-8711.2000.03879.x}, \href
  {http://adsabs.harvard.edu/abs/2000MNRAS.319..168C} {319, 168}

\bibitem[\protect\citeauthoryear{{Combes} et~al.,}{{Combes}
  et~al.}{2009}]{Combes2009}
{Combes} F.,  et~al., 2009, \mn@doi [\aap] {10.1051/0004-6361/200912181}, \href
  {http://adsabs.harvard.edu/abs/2009A%26A...503...73C} {503, 73}

\bibitem[\protect\citeauthoryear{{Cooray} \& {Sheth}}{{Cooray} \&
  {Sheth}}{2002}]{Cooray2002}
{Cooray} A.,  {Sheth} R.,  2002, \mn@doi [\physrep]
  {10.1016/S0370-1573(02)00276-4}, \href
  {http://adsabs.harvard.edu/abs/2002PhR...372....1C} {372, 1}

\bibitem[\protect\citeauthoryear{{Cox}, {Jonsson}, {Somerville}, {Primack}  \&
  {Dekel}}{{Cox} et~al.}{2008}]{Cox2008}
{Cox} T.~J.,  {Jonsson} P.,  {Somerville} R.~S.,  {Primack} J.~R.,   {Dekel}
  A.,  2008, \mn@doi [\mnras] {10.1111/j.1365-2966.2007.12730.x}, \href
  {http://adsabs.harvard.edu/abs/2008MNRAS.384..386C} {384, 386}

\bibitem[\protect\citeauthoryear{{Croom}, {Smith}, {Boyle}, {Shanks}, {Miller},
  {Outram}  \& {Loaring}}{{Croom} et~al.}{2004}]{Croom2004}
{Croom} S.~M.,  {Smith} R.~J.,  {Boyle} B.~J.,  {Shanks} T.,  {Miller} L.,
  {Outram} P.~J.,   {Loaring} N.~S.,  2004, \mn@doi [\mnras]
  {10.1111/j.1365-2966.2004.07619.x}, \href
  {http://adsabs.harvard.edu/abs/2004MNRAS.349.1397C} {349, 1397}

\bibitem[\protect\citeauthoryear{{Daddi} et~al.,}{{Daddi}
  et~al.}{2007}]{Daddi2007}
{Daddi} E.,  et~al., 2007, \mn@doi [\apj] {10.1086/521818}, \href
  {http://adsabs.harvard.edu/abs/2007ApJ...670..156D} {670, 156}

\bibitem[\protect\citeauthoryear{{Decarli}, {Falomo}, {Treves}, {Labita},
  {Kotilainen}  \& {Scarpa}}{{Decarli} et~al.}{2010}]{Decarli2010}
{Decarli} R.,  {Falomo} R.,  {Treves} A.,  {Labita} M.,  {Kotilainen} J.~K.,
  {Scarpa} R.,  2010, \mn@doi [\mnras] {10.1111/j.1365-2966.2009.16049.x},
  \href {http://adsabs.harvard.edu/abs/2010MNRAS.402.2453D} {402, 2453}

\bibitem[\protect\citeauthoryear{{Dekel} et~al.,}{{Dekel}
  et~al.}{2009}]{Dekel2009}
{Dekel} A.,  et~al., 2009, \mn@doi [\nat] {10.1038/nature07648}, \href
  {http://adsabs.harvard.edu/abs/2009Natur.457..451D} {457, 451}

\bibitem[\protect\citeauthoryear{{Di Matteo}, {Springel}  \& {Hernquist}}{{Di
  Matteo} et~al.}{2005}]{DiMatteo2005}
{Di Matteo} T.,  {Springel} V.,   {Hernquist} L.,  2005, \mn@doi [\nat]
  {10.1038/nature03335}, \href
  {http://adsabs.harvard.edu/abs/2005Natur.433..604D} {433, 604}

\bibitem[\protect\citeauthoryear{{Di Matteo}, {Colberg}, {Springel},
  {Hernquist}  \& {Sijacki}}{{Di Matteo} et~al.}{2008}]{DiMatteo2008}
{Di Matteo} T.,  {Colberg} J.,  {Springel} V.,  {Hernquist} L.,   {Sijacki} D.,
   2008, \mn@doi [\apj] {10.1086/524921}, \href
  {http://adsabs.harvard.edu/abs/2008ApJ...676...33D} {676, 33}

\bibitem[\protect\citeauthoryear{{Donoso}, {Yan}, {Stern}  \& {Assef}}{{Donoso}
  et~al.}{2014}]{Donoso2014}
{Donoso} E.,  {Yan} L.,  {Stern} D.,   {Assef} R.~J.,  2014, \mn@doi [\apj]
  {10.1088/0004-637X/789/1/44}, \href
  {http://adsabs.harvard.edu/abs/2014ApJ...789...44D} {789, 44}

\bibitem[\protect\citeauthoryear{{Efstathiou}, {Lake}  \&
  {Negroponte}}{{Efstathiou} et~al.}{1982}]{Efstathiou1982}
{Efstathiou} G.,  {Lake} G.,   {Negroponte} J.,  1982, \mnras, \href
  {http://adsabs.harvard.edu/abs/1982MNRAS.199.1069E} {199, 1069}

\bibitem[\protect\citeauthoryear{{Eftekharzadeh} et~al.,}{{Eftekharzadeh}
  et~al.}{2015}]{Eftekharzadeh2015}
{Eftekharzadeh} S.,  et~al., 2015, \mn@doi [\mnras] {10.1093/mnras/stv1763},
  \href {http://adsabs.harvard.edu/abs/2015MNRAS.453.2779E} {453, 2779}

\bibitem[\protect\citeauthoryear{{Elvis} et~al.,}{{Elvis}
  et~al.}{2009}]{Elvis2009}
{Elvis} M.,  et~al., 2009, \mn@doi [\apjs] {10.1088/0067-0049/184/1/158}, \href
  {http://adsabs.harvard.edu/abs/2009ApJS..184..158E} {184, 158}

\bibitem[\protect\citeauthoryear{{Fanidakis}, {Baugh}, {Benson}, {Bower},
  {Cole}, {Done}  \& {Frenk}}{{Fanidakis} et~al.}{2011}]{Fanidakis2011}
{Fanidakis} N.,  {Baugh} C.~M.,  {Benson} A.~J.,  {Bower} R.~G.,  {Cole} S.,
  {Done} C.,   {Frenk} C.~S.,  2011, \mn@doi [\mnras]
  {10.1111/j.1365-2966.2010.17427.x}, \href
  {http://adsabs.harvard.edu/abs/2011MNRAS.410...53F} {410, 53}

\bibitem[\protect\citeauthoryear{{Finoguenov} et~al.,}{{Finoguenov}
  et~al.}{2007}]{Finoguenov2007}
{Finoguenov} A.,  et~al., 2007, \mn@doi [\apjs] {10.1086/516577}, \href
  {http://adsabs.harvard.edu/abs/2007ApJS..172..182F} {172, 182}

\bibitem[\protect\citeauthoryear{{Fontanot}, {De Lucia}, {Monaco}, {Somerville}
   \& {Santini}}{{Fontanot} et~al.}{2009}]{Fontanot2009}
{Fontanot} F.,  {De Lucia} G.,  {Monaco} P.,  {Somerville} R.~S.,   {Santini}
  P.,  2009, \mn@doi [\mnras] {10.1111/j.1365-2966.2009.15058.x}, \href
  {http://adsabs.harvard.edu/abs/2009MNRAS.397.1776F} {397, 1776}

\bibitem[\protect\citeauthoryear{{Gabor} \& {Bournaud}}{{Gabor} \&
  {Bournaud}}{2013}]{Gabor2013}
{Gabor} J.~M.,  {Bournaud} F.,  2013, \mn@doi [\mnras] {10.1093/mnras/stt1046},
  \href {http://adsabs.harvard.edu/abs/2013MNRAS.434..606G} {434, 606}

\bibitem[\protect\citeauthoryear{{Gatti}, {Lamastra}, {Menci}, {Bongiorno}  \&
  {Fiore}}{{Gatti} et~al.}{2015}]{Gatti2015}
{Gatti} M.,  {Lamastra} A.,  {Menci} N.,  {Bongiorno} A.,   {Fiore} F.,  2015,
  \mn@doi [\aap] {10.1051/0004-6361/201425094}, \href
  {http://adsabs.harvard.edu/abs/2015A%26A...576A..32G} {576, A32}

\bibitem[\protect\citeauthoryear{{Gehrels}}{{Gehrels}}{1986}]{Gerhels1986}
{Gehrels} N.,  1986, \mn@doi [\apj] {10.1086/164079}, \href
  {http://adsabs.harvard.edu/abs/1986ApJ...303..336G} {303, 336}

\bibitem[\protect\citeauthoryear{{Georgakakis} \& {Nandra}}{{Georgakakis} \&
  {Nandra}}{2011}]{Georgakakis2011}
{Georgakakis} A.,  {Nandra} K.,  2011, \mn@doi [\mnras]
  {10.1111/j.1365-2966.2011.18387.x}, \href
  {http://adsabs.harvard.edu/abs/2011MNRAS.414..992G} {414, 992}

\bibitem[\protect\citeauthoryear{{George} et~al.,}{{George}
  et~al.}{2011}]{George2011}
{George} M.~R.,  et~al., 2011, \mn@doi [\apj] {10.1088/0004-637X/742/2/125},
  \href {http://adsabs.harvard.edu/abs/2011ApJ...742..125G} {742, 125}

\bibitem[\protect\citeauthoryear{{Gilli} et~al.,}{{Gilli}
  et~al.}{2009}]{Gilli2009}
{Gilli} R.,  et~al., 2009, \mn@doi [\aap] {10.1051/0004-6361:200810821}, \href
  {http://adsabs.harvard.edu/abs/2009A%26A...494...33G} {494, 33}

\bibitem[\protect\citeauthoryear{{Gruppioni} et~al.,}{{Gruppioni}
  et~al.}{2015}]{Gruppioni2015}
{Gruppioni} C.,  et~al., 2015, \mn@doi [\mnras] {10.1093/mnras/stv1204}, \href
  {http://adsabs.harvard.edu/abs/2015MNRAS.451.3419G} {451, 3419}

\bibitem[\protect\citeauthoryear{{Haiman}, {Mohr}  \& {Holder}}{{Haiman}
  et~al.}{2001}]{Haiman2001}
{Haiman} Z.,  {Mohr} J.~J.,   {Holder} G.~P.,  2001, \mn@doi [\apj]
  {10.1086/320939}, \href {http://adsabs.harvard.edu/abs/2001ApJ...553..545H}
  {553, 545}

\bibitem[\protect\citeauthoryear{{Hennawi} et~al.,}{{Hennawi}
  et~al.}{2006}]{Hennawi2006}
{Hennawi} J.~F.,  et~al., 2006, \mn@doi [\aj] {10.1086/498235}, \href
  {http://adsabs.harvard.edu/abs/2006AJ....131....1H} {131, 1}

\bibitem[\protect\citeauthoryear{{Henriques} \& {Thomas}}{{Henriques} \&
  {Thomas}}{2010}]{Henriques2010}
{Henriques} B.~M.~B.,  {Thomas} P.~A.,  2010, \mn@doi [\mnras]
  {10.1111/j.1365-2966.2009.16151.x}, \href
  {http://adsabs.harvard.edu/abs/2010MNRAS.403..768H} {403, 768}

\bibitem[\protect\citeauthoryear{{Hickox} et~al.,}{{Hickox}
  et~al.}{2009}]{Hickox2009}
{Hickox} R.~C.,  et~al., 2009, \mn@doi [\apj] {10.1088/0004-637X/696/1/891},
  \href {http://adsabs.harvard.edu/abs/2009ApJ...696..891H} {696, 891}

\bibitem[\protect\citeauthoryear{{Hickox} et~al.,}{{Hickox}
  et~al.}{2011}]{Hickox2011}
{Hickox} R.~C.,  et~al., 2011, \mn@doi [\apj] {10.1088/0004-637X/731/2/117},
  \href {http://adsabs.harvard.edu/abs/2011ApJ...731..117H} {731, 117}

\bibitem[\protect\citeauthoryear{{Hickox}, {Mullaney}, {Alexander}, {Chen},
  {Civano}, {Goulding}  \& {Hainline}}{{Hickox} et~al.}{2014}]{Hickox2014}
{Hickox} R.~C.,  {Mullaney} J.~R.,  {Alexander} D.~M.,  {Chen} C.-T.~J.,
  {Civano} F.~M.,  {Goulding} A.~D.,   {Hainline} K.~N.,  2014, \mn@doi [\apj]
  {10.1088/0004-637X/782/1/9}, \href
  {http://adsabs.harvard.edu/abs/2014ApJ...782....9H} {782, 9}

\bibitem[\protect\citeauthoryear{{Hirschmann}, {Somerville}, {Naab}  \&
  {Burkert}}{{Hirschmann} et~al.}{2012}]{Hirschmann2012}
{Hirschmann} M.,  {Somerville} R.~S.,  {Naab} T.,   {Burkert} A.,  2012,
  \mn@doi [\mnras] {10.1111/j.1365-2966.2012.21626.x}, \href
  {http://adsabs.harvard.edu/abs/2012MNRAS.426..237H} {426, 237}

\bibitem[\protect\citeauthoryear{{Hopkins} \& {Hernquist}}{{Hopkins} \&
  {Hernquist}}{2006}]{Hopkins2006}
{Hopkins} P.~F.,  {Hernquist} L.,  2006, \mn@doi [\apjs] {10.1086/505753},
  \href {http://adsabs.harvard.edu/abs/2006ApJS..166....1H} {166, 1}

\bibitem[\protect\citeauthoryear{{Hopkins} \& {Quataert}}{{Hopkins} \&
  {Quataert}}{2010}]{Hopkins2010}
{Hopkins} P.~F.,  {Quataert} E.,  2010, \mn@doi [\mnras]
  {10.1111/j.1365-2966.2010.17064.x}, \href
  {http://adsabs.harvard.edu/abs/2010MNRAS.407.1529H} {407, 1529}

\bibitem[\protect\citeauthoryear{{Hopkins} \& {Quataert}}{{Hopkins} \&
  {Quataert}}{2011}]{Hopkins2011}
{Hopkins} P.~F.,  {Quataert} E.,  2011, \mn@doi [\mnras]
  {10.1111/j.1365-2966.2011.18542.x}, \href
  {http://adsabs.harvard.edu/abs/2011MNRAS.415.1027H} {415, 1027}

\bibitem[\protect\citeauthoryear{{Hopkins}, {Hernquist}, {Cox}  \& {Kere{\v
  s}}}{{Hopkins} et~al.}{2008}]{Hopkins2008}
{Hopkins} P.~F.,  {Hernquist} L.,  {Cox} T.~J.,   {Kere{\v s}} D.,  2008,
  \mn@doi [\apjs] {10.1086/524362}, \href
  {http://adsabs.harvard.edu/abs/2008ApJS..175..356H} {175, 356}

\bibitem[\protect\citeauthoryear{{Hopkins}, {Kocevski}  \& {Bundy}}{{Hopkins}
  et~al.}{2014}]{Hopkins2014}
{Hopkins} P.~F.,  {Kocevski} D.~D.,   {Bundy} K.,  2014, \mn@doi [\mnras]
  {10.1093/mnras/stu1736}, \href
  {http://adsabs.harvard.edu/abs/2014MNRAS.445..823H} {445, 823}

\bibitem[\protect\citeauthoryear{{Kauffmann}, {Nusser}  \&
  {Steinmetz}}{{Kauffmann} et~al.}{1997}]{Kauffmann1997}
{Kauffmann} G.,  {Nusser} A.,   {Steinmetz} M.,  1997, \mnras, \href
  {http://adsabs.harvard.edu/abs/1997MNRAS.286..795K} {286, 795}

\bibitem[\protect\citeauthoryear{{Kayo} \& {Oguri}}{{Kayo} \&
  {Oguri}}{2012}]{Kayo2012}
{Kayo} I.,  {Oguri} M.,  2012, \mn@doi [\mnras]
  {10.1111/j.1365-2966.2012.21321.x}, \href
  {http://adsabs.harvard.edu/abs/2012MNRAS.424.1363K} {424, 1363}

\bibitem[\protect\citeauthoryear{{King}}{{King}}{2010}]{King2010}
{King} A.~R.,  2010, \mn@doi [\mnras] {10.1111/j.1365-2966.2009.16013.x}, \href
  {http://adsabs.harvard.edu/abs/2010MNRAS.402.1516K} {402, 1516}

\bibitem[\protect\citeauthoryear{{Koester} et~al.,}{{Koester}
  et~al.}{2007}]{Koester2007}
{Koester} B.~P.,  et~al., 2007, \mn@doi [\apj] {10.1086/509599}, \href
  {http://adsabs.harvard.edu/abs/2007ApJ...660..239K} {660, 239}

\bibitem[\protect\citeauthoryear{{Komatsu} et~al.,}{{Komatsu}
  et~al.}{2009}]{Komatsu2009a}
{Komatsu} E.,  et~al., 2009, \mn@doi [\apjs] {10.1088/0067-0049/180/2/330},
  \href {http://adsabs.harvard.edu/abs/2009ApJS..180..330K} {180, 330}

\bibitem[\protect\citeauthoryear{{Kormendy} \& {Ho}}{{Kormendy} \&
  {Ho}}{2013}]{Kormendy2013}
{Kormendy} J.,  {Ho} L.~C.,  2013, \mn@doi [\araa]
  {10.1146/annurev-astro-082708-101811}, \href
  {http://adsabs.harvard.edu/abs/2013ARA%26A..51..511K} {51, 511}

\bibitem[\protect\citeauthoryear{{Koss}, {Mushotzky}, {Veilleux}  \&
  {Winter}}{{Koss} et~al.}{2010}]{Koss2010}
{Koss} M.,  {Mushotzky} R.,  {Veilleux} S.,   {Winter} L.,  2010, \mn@doi
  [\apjl] {10.1088/2041-8205/716/2/L125}, \href
  {http://adsabs.harvard.edu/abs/2010ApJ...716L.125K} {716, L125}

\bibitem[\protect\citeauthoryear{{Koutoulidis}, {Plionis}, {Georgantopoulos}
  \& {Fanidakis}}{{Koutoulidis} et~al.}{2013}]{Koutoulidis2013}
{Koutoulidis} L.,  {Plionis} M.,  {Georgantopoulos} I.,   {Fanidakis} N.,
  2013, \mn@doi [\mnras] {10.1093/mnras/sts119}, \href
  {http://adsabs.harvard.edu/abs/2013MNRAS.428.1382K} {428, 1382}

\bibitem[\protect\citeauthoryear{{Krumpe}, {Miyaji}  \& {Coil}}{{Krumpe}
  et~al.}{2010}]{Krumpe2010}
{Krumpe} M.,  {Miyaji} T.,   {Coil} A.~L.,  2010, \mn@doi [\apj]
  {10.1088/0004-637X/713/1/558}, \href
  {http://adsabs.harvard.edu/abs/2010ApJ...713..558K} {713, 558}

\bibitem[\protect\citeauthoryear{{Krumpe}, {Miyaji}, {Coil}  \&
  {Aceves}}{{Krumpe} et~al.}{2012}]{Krumpe2012}
{Krumpe} M.,  {Miyaji} T.,  {Coil} A.~L.,   {Aceves} H.,  2012, \mn@doi [\apj]
  {10.1088/0004-637X/746/1/1}, \href
  {http://adsabs.harvard.edu/abs/2012ApJ...746....1K} {746, 1}

\bibitem[\protect\citeauthoryear{{Krumpe}, {Miyaji}, {Husemann}, {Fanidakis},
  {Coil}  \& {Aceves}}{{Krumpe} et~al.}{2015}]{krumpe2015}
{Krumpe} M.,  {Miyaji} T.,  {Husemann} B.,  {Fanidakis} N.,  {Coil} A.~L.,
  {Aceves} H.,  2015, preprint, \href
  {http://adsabs.harvard.edu/abs/2015arXiv150901261K} {} (\mn@eprint {arXiv}
  {1509.01261})

\bibitem[\protect\citeauthoryear{{Lacey} \& {Cole}}{{Lacey} \&
  {Cole}}{1993}]{Lacey1993}
{Lacey} C.,  {Cole} S.,  1993, \mnras, \href
  {http://adsabs.harvard.edu/abs/1993MNRAS.262..627L} {262, 627}

\bibitem[\protect\citeauthoryear{{Lamastra}, {Menci}, {Fiore}  \&
  {Santini}}{{Lamastra} et~al.}{2013a}]{Lamastra2013a}
{Lamastra} A.,  {Menci} N.,  {Fiore} F.,   {Santini} P.,  2013a, \mn@doi [\aap]
  {10.1051/0004-6361/201220754}, \href
  {http://adsabs.harvard.edu/abs/2013A%26A...552A..44L} {552, A44}

\bibitem[\protect\citeauthoryear{{Lamastra}, {Menci}, {Fiore}, {Santini},
  {Bongiorno}  \& {Piconcelli}}{{Lamastra} et~al.}{2013b}]{Lamastra2013}
{Lamastra} A.,  {Menci} N.,  {Fiore} F.,  {Santini} P.,  {Bongiorno} A.,
  {Piconcelli} E.,  2013b, \mn@doi [\aap] {10.1051/0004-6361/201322667}, \href
  {http://adsabs.harvard.edu/abs/2013A%26A...559A..56L} {559, A56}

\bibitem[\protect\citeauthoryear{{Landy} \& {Szalay}}{{Landy} \&
  {Szalay}}{1993}]{Landy1993a}
{Landy} S.~D.,  {Szalay} A.~S.,  1993, \mn@doi [\apj] {10.1086/172900}, \href
  {http://adsabs.harvard.edu/abs/1993ApJ...412...64L} {412, 64}

\bibitem[\protect\citeauthoryear{{Lapi}, {Cavaliere}  \& {Menci}}{{Lapi}
  et~al.}{2005}]{Lapi2005}
{Lapi} A.,  {Cavaliere} A.,   {Menci} N.,  2005, \mn@doi [\apj]
  {10.1086/426376}, \href {http://adsabs.harvard.edu/abs/2005ApJ...619...60L}
  {619, 60}

\bibitem[\protect\citeauthoryear{{L{\"a}sker}, {Ferrarese}, {van de Ven}  \&
  {Shankar}}{{L{\"a}sker} et~al.}{2014}]{Laesker2014}
{L{\"a}sker} R.,  {Ferrarese} L.,  {van de Ven} G.,   {Shankar} F.,  2014,
  \mn@doi [\apj] {10.1088/0004-637X/780/1/70}, \href
  {http://adsabs.harvard.edu/abs/2014ApJ...780...70L} {780, 70}

\bibitem[\protect\citeauthoryear{{Leauthaud} et~al.,}{{Leauthaud}
  et~al.}{2007}]{Leauthaud2007}
{Leauthaud} A.,  et~al., 2007, \mn@doi [\apjs] {10.1086/516598}, \href
  {http://adsabs.harvard.edu/abs/2007ApJS..172..219L} {172, 219}

\bibitem[\protect\citeauthoryear{{Leauthaud} et~al.,}{{Leauthaud}
  et~al.}{2010}]{Leauthaud2010}
{Leauthaud} A.,  et~al., 2010, \mn@doi [\apj] {10.1088/0004-637X/709/1/97},
  \href {http://adsabs.harvard.edu/abs/2010ApJ...709...97L} {709, 97}

\bibitem[\protect\citeauthoryear{{Leauthaud} et~al.,}{{Leauthaud}
  et~al.}{2015}]{Leauthaud2015}
{Leauthaud} A.,  et~al., 2015, \mn@doi [\mnras] {10.1093/mnras/stu2210}, \href
  {http://adsabs.harvard.edu/abs/2015MNRAS.446.1874L} {446, 1874}

\bibitem[\protect\citeauthoryear{{Lidz}, {Hopkins}, {Cox}, {Hernquist}  \&
  {Robertson}}{{Lidz} et~al.}{2006}]{Lidz2006}
{Lidz} A.,  {Hopkins} P.~F.,  {Cox} T.~J.,  {Hernquist} L.,   {Robertson} B.,
  2006, \mn@doi [\apj] {10.1086/500444}, \href
  {http://adsabs.harvard.edu/abs/2006ApJ...641...41L} {641, 41}

\bibitem[\protect\citeauthoryear{{Lin} \& {Mohr}}{{Lin} \&
  {Mohr}}{2007}]{Lin2007}
{Lin} Y.-T.,  {Mohr} J.~J.,  2007, \mn@doi [\apjs] {10.1086/513565}, \href
  {http://adsabs.harvard.edu/abs/2007ApJS..170...71L} {170, 71}

\bibitem[\protect\citeauthoryear{{Lutz} et~al.,}{{Lutz}
  et~al.}{2010}]{Lutz2010}
{Lutz} D.,  et~al., 2010, \mn@doi [\apj] {10.1088/0004-637X/712/2/1287}, \href
  {http://adsabs.harvard.edu/abs/2010ApJ...712.1287L} {712, 1287}

\bibitem[\protect\citeauthoryear{{Magorrian} et~al.,}{{Magorrian}
  et~al.}{1998}]{Magorrian1998}
{Magorrian} J.,  et~al., 1998, \mn@doi [\aj] {10.1086/300353}, \href
  {http://adsabs.harvard.edu/abs/1998AJ....115.2285M} {115, 2285}

\bibitem[\protect\citeauthoryear{{Marconi} \& {Hunt}}{{Marconi} \&
  {Hunt}}{2003}]{Marconi2003}
{Marconi} A.,  {Hunt} L.~K.,  2003, \mn@doi [\apjl] {10.1086/375804}, \href
  {http://adsabs.harvard.edu/abs/2003ApJ...589L..21M} {589, L21}

\bibitem[\protect\citeauthoryear{{Marconi}, {Risaliti}, {Gilli}, {Hunt},
  {Maiolino}  \& {Salvati}}{{Marconi} et~al.}{2004}]{Marconi2004}
{Marconi} A.,  {Risaliti} G.,  {Gilli} R.,  {Hunt} L.~K.,  {Maiolino} R.,
  {Salvati} M.,  2004, \mn@doi [\mnras] {10.1111/j.1365-2966.2004.07765.x},
  \href {http://adsabs.harvard.edu/abs/2004MNRAS.351..169M} {351, 169}

\bibitem[\protect\citeauthoryear{{Martini} \& {Weinberg}}{{Martini} \&
  {Weinberg}}{2001}]{Martini2001}
{Martini} P.,  {Weinberg} D.~H.,  2001, \mn@doi [\apj] {10.1086/318331}, \href
  {http://adsabs.harvard.edu/abs/2001ApJ...547...12M} {547, 12}

\bibitem[\protect\citeauthoryear{{Martini}, {Sivakoff}  \&
  {Mulchaey}}{{Martini} et~al.}{2009}]{Martini2009}
{Martini} P.,  {Sivakoff} G.~R.,   {Mulchaey} J.~S.,  2009, \mn@doi [\apj]
  {10.1088/0004-637X/701/1/66}, \href
  {http://adsabs.harvard.edu/abs/2009ApJ...701...66M} {701, 66}

\bibitem[\protect\citeauthoryear{{McConnell} \& {Ma}}{{McConnell} \&
  {Ma}}{2013}]{McConnell2013}
{McConnell} N.~J.,  {Ma} C.-P.,  2013, \mn@doi [\apj]
  {10.1088/0004-637X/764/2/184}, \href
  {http://adsabs.harvard.edu/abs/2013ApJ...764..184M} {764, 184}

\bibitem[\protect\citeauthoryear{{McIntosh}, {Guo}, {Mo}, {van den Bosch}  \&
  {Yang}}{{McIntosh} et~al.}{2009}]{McIntosh2009}
{McIntosh} D.~H.,  {Guo} Y.,  {Mo} H.~J.,  {van den Bosch} F.,   {Yang} X.,
  2009, in American Astronomical Society Meeting Abstracts \#213. p. 423.09

\bibitem[\protect\citeauthoryear{{Menci}, {Cavaliere}, {Fontana}, {Giallongo},
  {Poli}  \& {Vittorini}}{{Menci} et~al.}{2004}]{Menci2004}
{Menci} N.,  {Cavaliere} A.,  {Fontana} A.,  {Giallongo} E.,  {Poli} F.,
  {Vittorini} V.,  2004, \mn@doi [\apj] {10.1086/381522}, \href
  {http://adsabs.harvard.edu/abs/2004ApJ...604...12M} {604, 12}

\bibitem[\protect\citeauthoryear{{Menci}, {Fontana}, {Giallongo}, {Grazian}  \&
  {Salimbeni}}{{Menci} et~al.}{2006}]{Menci2006}
{Menci} N.,  {Fontana} A.,  {Giallongo} E.,  {Grazian} A.,   {Salimbeni} S.,
  2006, \mn@doi [\apj] {10.1086/505528}, \href
  {http://adsabs.harvard.edu/abs/2006ApJ...647..753M} {647, 753}

\bibitem[\protect\citeauthoryear{{Menci}, {Fiore}, {Puccetti}  \&
  {Cavaliere}}{{Menci} et~al.}{2008}]{Menci2008}
{Menci} N.,  {Fiore} F.,  {Puccetti} S.,   {Cavaliere} A.,  2008, \mn@doi
  [\apj] {10.1086/591438}, \href
  {http://adsabs.harvard.edu/abs/2008ApJ...686..219M} {686, 219}

\bibitem[\protect\citeauthoryear{{Menci}, {Gatti}, {Fiore}  \&
  {Lamastra}}{{Menci} et~al.}{2014}]{Menci2014}
{Menci} N.,  {Gatti} M.,  {Fiore} F.,   {Lamastra} A.,  2014, \mn@doi [\aap]
  {10.1051/0004-6361/201424217}, \href
  {http://adsabs.harvard.edu/abs/2014A%26A...569A..37M} {569, A37}

\bibitem[\protect\citeauthoryear{{Miyaji}, {Krumpe}, {Coil}  \&
  {Aceves}}{{Miyaji} et~al.}{2011}]{Miyaji2011}
{Miyaji} T.,  {Krumpe} M.,  {Coil} A.~L.,   {Aceves} H.,  2011, \mn@doi [\apj]
  {10.1088/0004-637X/726/2/83}, \href
  {http://adsabs.harvard.edu/abs/2011ApJ...726...83M} {726, 83}

\bibitem[\protect\citeauthoryear{{Mo}, {Mao}  \& {White}}{{Mo}
  et~al.}{1998}]{Mo1998}
{Mo} H.~J.,  {Mao} S.,   {White} S.~D.~M.,  1998, \mn@doi [\mnras]
  {10.1046/j.1365-8711.1998.01227.x}, \href
  {http://adsabs.harvard.edu/abs/1998MNRAS.295..319M} {295, 319}

\bibitem[\protect\citeauthoryear{{Moster}, {Naab}  \& {White}}{{Moster}
  et~al.}{2013}]{Moster2013}
{Moster} B.~P.,  {Naab} T.,   {White} S.~D.~M.,  2013, \mn@doi [\mnras]
  {10.1093/mnras/sts261}, \href
  {http://adsabs.harvard.edu/abs/2013MNRAS.428.3121M} {428, 3121}

\bibitem[\protect\citeauthoryear{{Mountrichas} \& {Georgakakis}}{{Mountrichas}
  \& {Georgakakis}}{2012}]{Mountrichas2012}
{Mountrichas} G.,  {Georgakakis} A.,  2012, \mn@doi [\mnras]
  {10.1111/j.1365-2966.2011.20059.x}, \href
  {http://adsabs.harvard.edu/abs/2012MNRAS.420..514M} {420, 514}

\bibitem[\protect\citeauthoryear{{Mullaney} et~al.,}{{Mullaney}
  et~al.}{2012}]{Mullaney2012a}
{Mullaney} J.~R.,  et~al., 2012, \mn@doi [\mnras]
  {10.1111/j.1365-2966.2011.19675.x}, \href
  {http://adsabs.harvard.edu/abs/2012MNRAS.419...95M} {419, 95}

\bibitem[\protect\citeauthoryear{{Myers} et~al.,}{{Myers}
  et~al.}{2006}]{Myers2006}
{Myers} A.~D.,  et~al., 2006, \mn@doi [\apj] {10.1086/499093}, \href
  {http://adsabs.harvard.edu/abs/2006ApJ...638..622M} {638, 622}

\bibitem[\protect\citeauthoryear{{Navarro}, {Frenk}  \& {White}}{{Navarro}
  et~al.}{1997}]{Navarro1997}
{Navarro} J.~F.,  {Frenk} C.~S.,   {White} S.~D.~M.,  1997, \apj, \href
  {http://adsabs.harvard.edu/abs/1997ApJ...490..493N} {490, 493}

\bibitem[\protect\citeauthoryear{{Padmanabhan}, {White}, {Norberg}  \&
  {Porciani}}{{Padmanabhan} et~al.}{2009}]{Padmanabhan2009}
{Padmanabhan} N.,  {White} M.,  {Norberg} P.,   {Porciani} C.,  2009, \mn@doi
  [\mnras] {10.1111/j.1365-2966.2008.14071.x}, \href
  {http://adsabs.harvard.edu/abs/2009MNRAS.397.1862P} {397, 1862}

\bibitem[\protect\citeauthoryear{{Peng}, {Impey}, {Rix}, {Kochanek}, {Keeton},
  {Falco}, {Leh{\'a}r}  \& {McLeod}}{{Peng} et~al.}{2006}]{Peng2006}
{Peng} C.~Y.,  {Impey} C.~D.,  {Rix} H.-W.,  {Kochanek} C.~S.,  {Keeton} C.~R.,
   {Falco} E.~E.,  {Leh{\'a}r} J.,   {McLeod} B.~A.,  2006, \mn@doi [\apj]
  {10.1086/506266}, \href {http://adsabs.harvard.edu/abs/2006ApJ...649..616P}
  {649, 616}

\bibitem[\protect\citeauthoryear{{Pentericci} et~al.,}{{Pentericci}
  et~al.}{2013}]{Pentericci2013}
{Pentericci} L.,  et~al., 2013, \mn@doi [\aap] {10.1051/0004-6361/201219759},
  \href {http://adsabs.harvard.edu/abs/2013A%26A...552A.111P} {552, A111}

\bibitem[\protect\citeauthoryear{{Porciani}, {Magliocchetti}  \&
  {Norberg}}{{Porciani} et~al.}{2004}]{Porciani2004}
{Porciani} C.,  {Magliocchetti} M.,   {Norberg} P.,  2004, \mn@doi [\mnras]
  {10.1111/j.1365-2966.2004.08408.x}, \href
  {http://adsabs.harvard.edu/abs/2004MNRAS.355.1010P} {355, 1010}

\bibitem[\protect\citeauthoryear{{Press} \& {Schechter}}{{Press} \&
  {Schechter}}{1974}]{Press1974}
{Press} W.~H.,  {Schechter} P.,  1974, \mn@doi [\apj] {10.1086/152650}, \href
  {http://adsabs.harvard.edu/abs/1974ApJ...187..425P} {187, 425}

\bibitem[\protect\citeauthoryear{{Richards} et~al.,}{{Richards}
  et~al.}{2006}]{Richards2006}
{Richards} G.~T.,  et~al., 2006, \mn@doi [\aj] {10.1086/503559}, \href
  {http://adsabs.harvard.edu/abs/2006AJ....131.2766R} {131, 2766}

\bibitem[\protect\citeauthoryear{{Richardson}, {Zheng}, {Chatterjee}, {Nagai}
  \& {Shen}}{{Richardson} et~al.}{2012}]{Richardson2012}
{Richardson} J.,  {Zheng} Z.,  {Chatterjee} S.,  {Nagai} D.,   {Shen} Y.,
  2012, \mn@doi [\apj] {10.1088/0004-637X/755/1/30}, \href
  {http://adsabs.harvard.edu/abs/2012ApJ...755...30R} {755, 30}

\bibitem[\protect\citeauthoryear{{Richardson}, {Chatterjee}, {Zheng}, {Myers}
  \& {Hickox}}{{Richardson} et~al.}{2013}]{Richardson2013}
{Richardson} J.,  {Chatterjee} S.,  {Zheng} Z.,  {Myers} A.~D.,   {Hickox} R.,
  2013, \mn@doi [\apj] {10.1088/0004-637X/774/2/143}, \href
  {http://adsabs.harvard.edu/abs/2013ApJ...774..143R} {774, 143}

\bibitem[\protect\citeauthoryear{{Richstone} et~al.,}{{Richstone}
  et~al.}{1998}]{Richstone1998}
{Richstone} D.,  et~al., 1998, \nat, \href
  {http://adsabs.harvard.edu/abs/1998Natur.395A..14R} {395, A14}

\bibitem[\protect\citeauthoryear{{Rosario} et~al.,}{{Rosario}
  et~al.}{2012}]{Rosario2012}
{Rosario} D.~J.,  et~al., 2012, \mn@doi [\aap] {10.1051/0004-6361/201219258},
  \href {http://adsabs.harvard.edu/abs/2012A%26A...545A..45R} {545, A45}

\bibitem[\protect\citeauthoryear{{Rosario} et~al.,}{{Rosario}
  et~al.}{2013}]{Rosario2013}
{Rosario} D.~J.,  et~al., 2013, \mn@doi [\apj] {10.1088/0004-637X/763/1/59},
  \href {http://adsabs.harvard.edu/abs/2013ApJ...763...59R} {763, 59}

\bibitem[\protect\citeauthoryear{{Rykoff} et~al.,}{{Rykoff}
  et~al.}{2012}]{Rykoff2012}
{Rykoff} E.~S.,  et~al., 2012, \mn@doi [\apj] {10.1088/0004-637X/746/2/178},
  \href {http://adsabs.harvard.edu/abs/2012ApJ...746..178R} {746, 178}

\bibitem[\protect\citeauthoryear{{Salvato} et~al.,}{{Salvato}
  et~al.}{2011}]{Salvato2011}
{Salvato} M.,  et~al., 2011, \mn@doi [\apj] {10.1088/0004-637X/742/2/61}, \href
  {http://adsabs.harvard.edu/abs/2011ApJ...742...61S} {742, 61}

\bibitem[\protect\citeauthoryear{{Sanders} \& {Mirabel}}{{Sanders} \&
  {Mirabel}}{1996}]{Sanders1996}
{Sanders} D.~B.,  {Mirabel} I.~F.,  1996, \mn@doi [\araa]
  {10.1146/annurev.astro.34.1.749}, \href
  {http://adsabs.harvard.edu/abs/1996ARA%26A..34..749S} {34, 749}

\bibitem[\protect\citeauthoryear{{Santini} et~al.,}{{Santini}
  et~al.}{2012}]{Santini2012}
{Santini} P.,  et~al., 2012, \mn@doi [\aap] {10.1051/0004-6361/201118266},
  \href {http://adsabs.harvard.edu/abs/2012A%26A...540A.109S} {540, A109}

\bibitem[\protect\citeauthoryear{{Saslaw}}{{Saslaw}}{1985}]{Saslaw1985}
{Saslaw} W.~C.,  1985, {Gravitational physics of stellar and galactic systems}

\bibitem[\protect\citeauthoryear{{Satyapal}, {Ellison}, {McAlpine}, {Hickox},
  {Patton}  \& {Mendel}}{{Satyapal} et~al.}{2014}]{Satyapal2014}
{Satyapal} S.,  {Ellison} S.~L.,  {McAlpine} W.,  {Hickox} R.~C.,  {Patton}
  D.~R.,   {Mendel} J.~T.,  2014, \mn@doi [\mnras] {10.1093/mnras/stu650},
  \href {http://adsabs.harvard.edu/abs/2014MNRAS.441.1297S} {441, 1297}

\bibitem[\protect\citeauthoryear{{Schlegel}, {White}  \&
  {Eisenstein}}{{Schlegel} et~al.}{2009}]{Schlegel2009}
{Schlegel} D.,  {White} M.,   {Eisenstein} D.,  2009, in astro2010: The
  Astronomy and Astrophysics Decadal Survey. p.~314 (\mn@eprint {arXiv}
  {0902.4680})

\bibitem[\protect\citeauthoryear{{Schneider} et~al.,}{{Schneider}
  et~al.}{2010}]{Schneider2010}
{Schneider} D.~P.,  et~al., 2010, \mn@doi [\aj] {10.1088/0004-6256/139/6/2360},
  \href {http://adsabs.harvard.edu/abs/2010AJ....139.2360S} {139, 2360}

\bibitem[\protect\citeauthoryear{{Shankar}, {Salucci}, {Granato}, {De Zotti}
  \& {Danese}}{{Shankar} et~al.}{2004}]{Shankar2004}
{Shankar} F.,  {Salucci} P.,  {Granato} G.~L.,  {De Zotti} G.,   {Danese} L.,
  2004, \mn@doi [\mnras] {10.1111/j.1365-2966.2004.08261.x}, \href
  {http://adsabs.harvard.edu/abs/2004MNRAS.354.1020S} {354, 1020}

\bibitem[\protect\citeauthoryear{{Shankar}, {Lapi}, {Salucci}, {De Zotti}  \&
  {Danese}}{{Shankar} et~al.}{2006}]{Shankar2006}
{Shankar} F.,  {Lapi} A.,  {Salucci} P.,  {De Zotti} G.,   {Danese} L.,  2006,
  \mn@doi [\apj] {10.1086/502794}, \href
  {http://adsabs.harvard.edu/abs/2006ApJ...643...14S} {643, 14}

\bibitem[\protect\citeauthoryear{{Shankar}, {Weinberg}  \&
  {Miralda-Escud{\'e}}}{{Shankar} et~al.}{2009}]{Shankar2009}
{Shankar} F.,  {Weinberg} D.~H.,   {Miralda-Escud{\'e}} J.,  2009, \mn@doi
  [\apj] {10.1088/0004-637X/690/1/20}, \href
  {http://adsabs.harvard.edu/abs/2009ApJ...690...20S} {690, 20}

\bibitem[\protect\citeauthoryear{{Shankar}, {Crocce}, {Miralda-Escud{\'e}},
  {Fosalba}  \& {Weinberg}}{{Shankar} et~al.}{2010}]{Shankar2010}
{Shankar} F.,  {Crocce} M.,  {Miralda-Escud{\'e}} J.,  {Fosalba} P.,
  {Weinberg} D.~H.,  2010, \mn@doi [\apj] {10.1088/0004-637X/718/1/231}, \href
  {http://adsabs.harvard.edu/abs/2010ApJ...718..231S} {718, 231}

\bibitem[\protect\citeauthoryear{{Shankar}, {Weinberg}  \&
  {Miralda-Escud{\'e}}}{{Shankar} et~al.}{2013}]{Shankar2013}
{Shankar} F.,  {Weinberg} D.~H.,   {Miralda-Escud{\'e}} J.,  2013, \mn@doi
  [\mnras] {10.1093/mnras/sts026}, \href
  {http://adsabs.harvard.edu/abs/2013MNRAS.428..421S} {428, 421}

\bibitem[\protect\citeauthoryear{{Shankar} et~al.,}{{Shankar}
  et~al.}{2014}]{Shankar2014}
{Shankar} F.,  et~al., 2014, \mn@doi [\apjl] {10.1088/2041-8205/797/2/L27},
  \href {http://adsabs.harvard.edu/abs/2014ApJ...797L..27S} {797, L27}

\bibitem[\protect\citeauthoryear{{Shen} et~al.,}{{Shen}
  et~al.}{2007}]{Shen2007}
{Shen} Y.,  et~al., 2007, \mn@doi [\aj] {10.1086/513517}, \href
  {http://adsabs.harvard.edu/abs/2007AJ....133.2222S} {133, 2222}

\bibitem[\protect\citeauthoryear{{Shen} et~al.,}{{Shen}
  et~al.}{2013}]{Shen2013}
{Shen} Y.,  et~al., 2013, \mn@doi [\apj] {10.1088/0004-637X/778/2/98}, \href
  {http://adsabs.harvard.edu/abs/2013ApJ...778...98S} {778, 98}

\bibitem[\protect\citeauthoryear{{Sheth} \& {Tormen}}{{Sheth} \&
  {Tormen}}{1999}]{Sheth1999}
{Sheth} R.~K.,  {Tormen} G.,  1999, \mn@doi [\mnras]
  {10.1046/j.1365-8711.1999.02692.x}, \href
  {http://adsabs.harvard.edu/abs/1999MNRAS.308..119S} {308, 119}

\bibitem[\protect\citeauthoryear{{Soltan}}{{Soltan}}{1982}]{Soltan1982}
{Soltan} A.,  1982, \mnras, \href
  {http://adsabs.harvard.edu/abs/1982MNRAS.200..115S} {200, 115}

\bibitem[\protect\citeauthoryear{{Starikova} et~al.,}{{Starikova}
  et~al.}{2011}]{Starikova2011}
{Starikova} S.,  et~al., 2011, \mn@doi [\apj] {10.1088/0004-637X/741/1/15},
  \href {http://adsabs.harvard.edu/abs/2011ApJ...741...15S} {741, 15}

\bibitem[\protect\citeauthoryear{{Tinker}, {Weinberg}, {Zheng}  \&
  {Zehavi}}{{Tinker} et~al.}{2005}]{Tinker2005}
{Tinker} J.~L.,  {Weinberg} D.~H.,  {Zheng} Z.,   {Zehavi} I.,  2005, \mn@doi
  [\apj] {10.1086/432084}, \href
  {http://adsabs.harvard.edu/abs/2005ApJ...631...41T} {631, 41}

\bibitem[\protect\citeauthoryear{{Treister}, {Schawinski}, {Urry}  \&
  {Simmons}}{{Treister} et~al.}{2012}]{Treister2012}
{Treister} E.,  {Schawinski} K.,  {Urry} C.~M.,   {Simmons} B.~D.,  2012,
  \mn@doi [\apjl] {10.1088/2041-8205/758/2/L39}, \href
  {http://adsabs.harvard.edu/abs/2012ApJ...758L..39T} {758, L39}

\bibitem[\protect\citeauthoryear{{Ueda}, {Akiyama}, {Hasinger}, {Miyaji}  \&
  {Watson}}{{Ueda} et~al.}{2014}]{Ueda2014}
{Ueda} Y.,  {Akiyama} M.,  {Hasinger} G.,  {Miyaji} T.,   {Watson} M.~G.,
  2014, \mn@doi [\apj] {10.1088/0004-637X/786/2/104}, \href
  {http://adsabs.harvard.edu/abs/2014ApJ...786..104U} {786, 104}

\bibitem[\protect\citeauthoryear{{Urrutia}, {Lacy}, {Spoon}, {Glikman},
  {Petric}  \& {Schulz}}{{Urrutia} et~al.}{2012}]{Urrutia2012}
{Urrutia} T.,  {Lacy} M.,  {Spoon} H.,  {Glikman} E.,  {Petric} A.,   {Schulz}
  B.,  2012, \mn@doi [\apj] {10.1088/0004-637X/757/2/125}, \href
  {http://adsabs.harvard.edu/abs/2012ApJ...757..125U} {757, 125}

\bibitem[\protect\citeauthoryear{{Vale} \& {Ostriker}}{{Vale} \&
  {Ostriker}}{2004}]{Vale2004}
{Vale} A.,  {Ostriker} J.~P.,  2004, \mn@doi [\mnras]
  {10.1111/j.1365-2966.2004.08059.x}, \href
  {http://adsabs.harvard.edu/abs/2004MNRAS.353..189V} {353, 189}

\bibitem[\protect\citeauthoryear{{Villforth} et~al.,}{{Villforth}
  et~al.}{2014}]{Villforth2014}
{Villforth} C.,  et~al., 2014, \mn@doi [\mnras] {10.1093/mnras/stu173}, \href
  {http://adsabs.harvard.edu/abs/2014MNRAS.439.3342V} {439, 3342}

\bibitem[\protect\citeauthoryear{{Voges} et~al.,}{{Voges}
  et~al.}{1999}]{Voges1999}
{Voges} W.,  et~al., 1999, \aap, \href
  {http://adsabs.harvard.edu/abs/1999A%26A...349..389V} {349, 389}

\bibitem[\protect\citeauthoryear{{White}, {Martini}  \& {Cohn}}{{White}
  et~al.}{2008}]{White2008}
{White} M.,  {Martini} P.,   {Cohn} J.~D.,  2008, \mn@doi [\mnras]
  {10.1111/j.1365-2966.2008.13817.x}, \href
  {http://adsabs.harvard.edu/abs/2008MNRAS.390.1179W} {390, 1179}

\bibitem[\protect\citeauthoryear{{White} et~al.,}{{White}
  et~al.}{2011}]{White2011}
{White} M.,  et~al., 2011, \mn@doi [\apj] {10.1088/0004-637X/728/2/126}, \href
  {http://adsabs.harvard.edu/abs/2011ApJ...728..126W} {728, 126}

\bibitem[\protect\citeauthoryear{{White} et~al.,}{{White}
  et~al.}{2012}]{White2012}
{White} M.,  et~al., 2012, \mn@doi [\mnras] {10.1111/j.1365-2966.2012.21251.x},
  \href {http://adsabs.harvard.edu/abs/2012MNRAS.424..933W} {424, 933}

\bibitem[\protect\citeauthoryear{{Wright} et~al.,}{{Wright}
  et~al.}{2010}]{Wright2010}
{Wright} E.~L.,  et~al., 2010, \mn@doi [\aj] {10.1088/0004-6256/140/6/1868},
  \href {http://adsabs.harvard.edu/abs/2010AJ....140.1868W} {140, 1868}

\bibitem[\protect\citeauthoryear{{Wylezalek} et~al.,}{{Wylezalek}
  et~al.}{2013}]{Wylezalek2013}
{Wylezalek} D.,  et~al., 2013, \mn@doi [\apj] {10.1088/0004-637X/769/1/79},
  \href {http://adsabs.harvard.edu/abs/2013ApJ...769...79W} {769, 79}

\bibitem[\protect\citeauthoryear{{Yu} \& {Tremaine}}{{Yu} \&
  {Tremaine}}{2002}]{Yu2002}
{Yu} Q.,  {Tremaine} S.,  2002, \mn@doi [\mnras]
  {10.1046/j.1365-8711.2002.05532.x}, \href
  {http://adsabs.harvard.edu/abs/2002MNRAS.335..965Y} {335, 965}

\bibitem[\protect\citeauthoryear{{Zheng} \& {Weinberg}}{{Zheng} \&
  {Weinberg}}{2007}]{Zheng2007}
{Zheng} Z.,  {Weinberg} D.~H.,  2007, \mn@doi [\apj] {10.1086/512151}, \href
  {http://adsabs.harvard.edu/abs/2007ApJ...659....1Z} {659, 1}

\makeatother
\end{thebibliography}

\appendix

\section{Comparison with DEUSS}
\begin{figure}
    \centering
\includegraphics[scale=0.6]{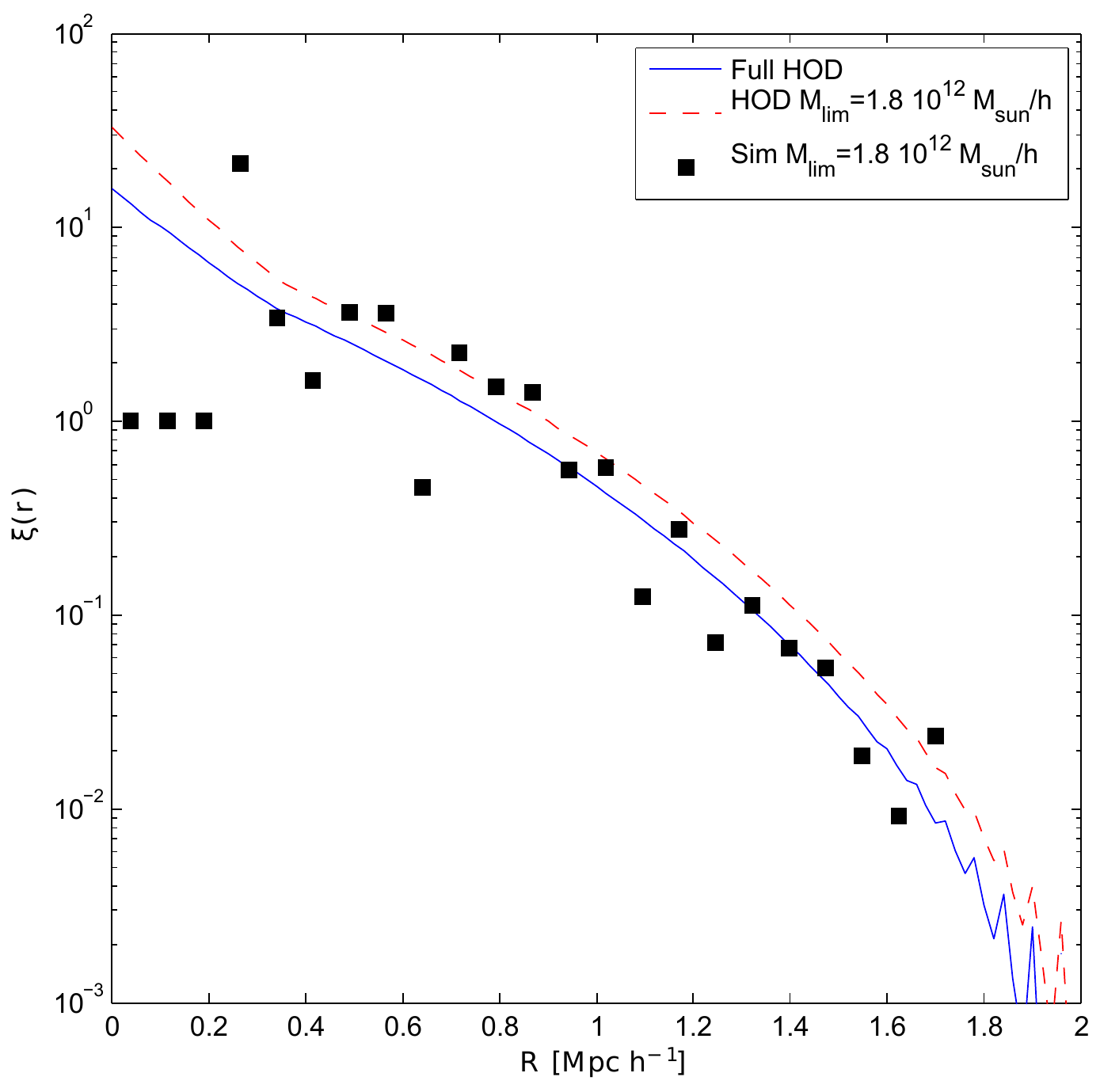}

\newpage
\caption{Comparison between the 2PCF obtained from our HOD modelling (blue and red lines) and with the one obtained using DEUSS (black dots). Both 2PCF have been obtained starting from the same AGN MOF.}
\end{figure}

In the following we used the Dark Energy Universe Simulation Series (DEUSS), a large ensemble of high performance cosmological Dark Matter (DM) simulations of realistic Dark Energy models that follows the gravitational evolution of billions of DM particles on volumes varying from inner halo scales to the size of the observable Universe\footnote{http://www.deus-consortium.org/}. DM halos are being detected using a Friends-of-Friends (FoF) algorithm, with a percolation factor $b=0.2$ widely used in the literature. In the specific, we make use of two simulations with different mass resolutions developed in the concordance $\Lambda$CDM model, with cosmological parameters obtained from WMAP-5 year data \citep{Komatsu2009a} $\Omega_{\Lambda}=0.74$, $\Omega_m=0.26$, $h=0.72$, $n=0.96$, $\Omega_b=0.04$, and $\sigma_8=0.79$. 

The numerical procedure to introduce luminous objects in a pure DM simulation is as follows. A central (resp. satellite) AGN is randomly assigned to each dark matter halo following the number density given by the AGN MOF. In order to beat the bridging effect inherent to FoF halos, the central AGN is placed at the minimum of the gravitational potential, the latter being computed from the particles detected with $b=0.3$ FoF. We found this positioning to be equivalent to using the center of mass in $95\%$ of DM halos. The satellites are positioned by picking randomly a DM particle belonging to the DM halo, thus ensuring we are following the right DM profile. 

From this distribution, the 2PCF is estimated using the \citet{Landy1993a} estimator. The statistical errors on the 2PCF are computed using a resampling technique based on 225 resamplings and the statistical errors drawn from a specific MOF distribution are computed through the average over 25 random MOF distributions.

Regarding the numerous systematics associated with the mass resolution and the limited volume of a simulation, we choose to use a simulation of 648 Mpc/h box length and $2048^3$ particles. This simulation ensure a volume large enough to have a sufficient enough density of AGN (i.e. the n,good normalization of the 2-halo term) while maximizing the mass resolution (i.e. the accuracy on the 1-one term). The choice on the mass resolution has been made by testing our results against a 648 Mpc/h box length and $1024^3$ particles simulation, which exhibits a poor resolution at small scale. The choice on the volume is obtained comparing a 2592 Mpc/h box length $2048^3$ particles simulation with a smaller 648 box length and $512^3$ particles simulation, at constant mass resolution. In the latter, the massive increase in volume is not changing significantly (more than a few percents) the results on the 2PCF.

%SDSS DR7 SHEN
%SDSS DR7 +SDSS3+2QZ rich12
%XMM COSMOS rich13
%XMM C COSMOS ALL12
%SDSS7 chatt13
%XMM C COSMOS LEAU
%RASS krumpe
%XMM C COSMOS ALL14
%SDSS DR5 shen

% Don't change these lines
\bsp	% typesetting comment
\label{lastpage}
\end{document}